\documentclass[11pt]{article}
\usepackage{amsfonts}

\usepackage{latexsym}
\begin{document}

\begin{flushright}
{ IFIC-07-65, \quad FTUV-07-2310, \quad hep-th/0710.4342 \\
published in \quad Nucl. Phys. {\bf B796}, 360--401 (2008) }
\end{flushright} \vspace{1.5cm}

\begin{center}

{\LARGE D=11 massless superparticle covariant  quantization, pure spinor BRST charge
and hidden symmetries}

\vskip 1.0cm

{\large Igor A. Bandos}

\vskip 1.0cm

{\it Departamento de F\'{\i}sica Te\'orica, Univ.~de Valencia and
IFIC (CSIC-UVEG), \\ 46100-Burjassot (Valencia), Spain
\\ and \\
 Institute for Theoretical Physics, NSC ``Kharkov Institute of
Physics  and Technology'',  \\ UA61108, Kharkov, Ukraine}

\date{{\large\bf  October 20, 2007}, Printed \today  }

\def\theequation{\arabic{section}.\arabic{equation}}

\bigskip

{\bf Abstract}
\end{center}

{We consider the covariant quantization of the D=11 massless superparticle (M0--brane)
in the spinor moving frame or twistor-like Lorentz harmonics formulation. The action
involves the set of $16$ constrained $32$ component Majorana spinors, the spinor
Lorentz harmonics $v_{\alpha q}^{\;\, -}$ parametrizing (as homogeneous coordinates,
modulo gauge symmetries) the celestial sphere $S^9$. There presence allows us to
separate covariantly the first and the second class constraints of the model. We show
that, after taking into account the second class constraints by means of Dirac brackets
and after further reducing the  first class constraints algebra, the system is
described in terms of a simple BRST charge ${\mathbb{Q}}^{susy}$ associated to the
$d=1$, $n=16$ supersymmetry algebra. The study of the cohomology of this BRST operator
requires a regularization by complexifying the bosonic ghosts for the
$\kappa$--symmetry, $\lambda_q$, and further reduction of the regularized cohomology
problem to the one for a simpler complex BRST charge $\tilde{\mathbb{Q}}^{susy}$. This
latter is essentially the pure spinor BRST charge by Berkovits, but with a composite
pure spinor constructed from the complex $d=9$ spinor with zero norm,
$\tilde{\lambda}_q$, and the spinorial harmonics $v_\alpha{}^-_q$. This exhibits a
possible origin of the complexity (non-hermiticity) characteristic of the Berkovits
pure spinor approach.

The simple structure of the nontrivial cohomology of the M0--brane BRST charge
$\mathbb{Q}^{susy}$ finds explanation in the properties that the superparticle action
exhibits in the so-called `covariantized light--cone' basis, where the M0-brane action
is expressed in terms of $\kappa$--symmetry invariant variables. The set of gauge
symmetries in this basis reduces to the $[SO(1,1)\times SO(9)]\subset\!\!\!\!\!\!\times
K_9$ Borel subgroup of $SO(1,10)$. Imposing their generators as conditions on the
superparticle wavefunctions, we arrive at the covariant quantization in terms of
physical degrees of freedom which hints possible hidden symmetries of $D=11$
supergravity. Besides $SO(16)$, which in the twistor like Lorentz harmonic formulation
is seen already at the classical level, we discuss also some indirect arguments in
favor of the possible $E_8$ symmetry.}
\\
{\rm PACs: 11.30.Pb, 11.25.-w, 04.65.+e, 11.10.Kk}

\thispagestyle{empty}

\setcounter{page}{0}

\tableofcontents

\renewcommand {\theequation} {\arabic{section}.\arabic{equation}}
\section{Introduction and summary}

\subsection{Introduction}
A covariant quantization of the massless $D$=11 superparticle (see
\cite{B+T=D-branes,Green+99}) has been recently considered \cite{BdAS2006} in its
 twistor-like Lorentz harmonics formulation \cite{BL98'} (see
also \cite{B90,BZ-str,BZ-strH,BZ-p}). This new, covariant {\it
supertwistor quantization} leds to the linearized $D$=11
supergravity multiplet in the spectrum (in agreement with the
light--cone results of \cite{Green+99}) and permitted to find a
possible origin of the hidden $SO(16)$ symmetry of the $D=11$
supergravity \cite{Nicolai87}. In this paper we study the BRST
quantization of the $D$=11 massless superparticle model in that
approach and then turn back to the covariant quantization of
physical degrees of freedom (different from the supertwistor one in
\cite{BdAS2006}) to search for an explanation of the simple
structure of the superparticle cohomologies.

The $D=11$ superparticle is interesting on its own, as the simplest of the M-theory
superbranes, the M0-brane, and because its quantization produces, as noticed above, the
linearized $D$=11 supergravity multiplet. Nevertheless, our main motivation is to look
for the origin and geometric meaning of the `pure spinor' formalism by Berkovits
\cite{NB-pure}. Recently, a breakthrough in the covariant description of quantum
superstring theory has been reached in this pure spinor framework: a technique for loop
calculations was developed \cite{NBloops} and the first results were given in
\cite{NBloops,NBloopC,NBloopsR4}. In particular, two new multiloop theorems useful in a
resent investigations of the possible finiteness of $N$=8 $D$=4 supergravity
\cite{N8SG=fin} were proved in \cite{NBloopsR4}. On the other hand, the pure spinor
superstring was introduced -and still remains- as  a set of prescriptions for quantum
superstring calculations, rather than as a quantization of the Green-Schwarz
superstring. Despite a certain progress in relating the pure spinor superstring
\cite{NB-pure} to the original Green--Schwarz formulation \cite{pure-GS}, and also
\cite{Dima+Mario+02} to the superembedding approach
\cite{bpstv,hs1,bst,Dima}\footnote{Notice also the recent progress \cite{Skenderis07}
in derivation  of the pure spinor ghost measure for loop calculations, which was
originally proposed in \cite{NBloops} on the ground of a series of very elegant but
indirect arguments involving the picture changing operator characteristic of the RNS
(Ramond--Neveu--Schwarz) string model \cite{RNS,GSO}. This was reached, however, by
starting from the pure spinor superstring by Berkovits, covariantizing it with respect
to the worldsheet reparametrizations by introducing two dimensional gravity and
quantizing this sector {\it a la} Batalin-Vilkovisky \cite{BV83}. Thus, although the
subject of \cite{Skenderis07} was the quantization of Berkovits pure spinor model
rather than the original Green-Schwarz superstring, a deeper understanding of the loop
calculation technique has been reached already at this stage. The approach similar to
\cite{Skenderis07} was also developed in earlier \cite{GrassiPolocastro04}. }, the
origin and geometrical meaning of the pure spinor formalism is far from being clear.
Possible modifications  of the pure spinor approach are also being considered (see {\it
e.g.} \cite{nonmNB}).

In this context, the Lorentz harmonic approach
\cite{Sok,NPS,Lharm,B90,Ghsds,BZ-str,BZ-strH,BZ-p,GHT93}, in the frame of which a
significant progress in solving the problem of covariant superstring quantization had
already  been made in late eighties \cite{NPS,Lharm}, looks particularly interesting.
Although no counterpart of the recent progress in loop calculations
\cite{NBloops,NBloopC} has been reached (yet) in the Lorentz harmonics framework, its
relation with the superembedding approach \cite{bpstv,hs1,bst,Dima}, transparent
geometrical meaning \cite{Sok,Ghsds,BZ-str,BZ-strH,BZ-p} and twistor-likeness
\cite{BZ-str,BZ-strH,BZ-p} justifies the hope that its further development (in the
pragmatic spirit characteristic for the pure spinor approach of
\cite{NB-pure,NBloops,NBloopC}) may be helpful to understand  the origin and the
geometrical meaning of the pure spinor formalism \cite{NB-pure} as well as its
nonminimal modifications \cite{nonmNB} and even that it might provide a basis for an
alternative, convenient and algorithmic, technique for the superstring loop
calculations. A natural first stage in such a program is to study the covariant
quantization of superparticle, and in particular, of the $D$=11 massless superparticle
or M$0$--brane\footnote{See \cite{GrassiAnguelovaVanhove04} for the loop calculations
with the use of the $D=11$ pure spinor formalism. }, less studied as well in comparison
with the $D$=10 and $D$=4 superparticle models.

\subsection{Summary of the main results}

The BRST charge proposed by Berkovits \cite{NB-pure} has the form
\begin{eqnarray}\label{QbrstB}
\mathbb{Q}^{B}= {\Lambda}^\alpha \; d_\alpha  \; ,
 \qquad
\end{eqnarray}
where $d_\alpha$ are the fermionic constraints of the (here $D$=11) superparticle
model, which obey the algebra
\begin{eqnarray}\label{dd=P}
 \{ d_\alpha\, , \,
d_\beta \}= 2iP\!\!\!/_{\alpha\beta}\equiv 2i \Gamma^m_{\alpha\beta}P_m \;
 \qquad (\hbox{here}\; \alpha=1,\ldots , 32\; , \; m= 0,1,\ldots , 9, \#)\; , \quad
\end{eqnarray}
 where $P_m$  is the superparticle momentum, and ${\Lambda}^\alpha$ is the complex {\it pure spinor}
which obeys
\begin{eqnarray}\label{NB-pureSp}
    {\Lambda}\Gamma_a{\Lambda}=0 \; , \qquad
    {\Lambda}^\alpha\not= ({\Lambda}^\alpha)^*\; .
 \qquad
\end{eqnarray}
This constraint guarantees the nilpotency $(\mathbb{Q}^{B})^2 =0$ of the Berkovits BRST
charge (\ref{QbrstB}).

The generic null spinor $\Lambda_\alpha$ contains $23$ complex or
$46$ real parameters \cite{NB-pure}\footnote{The direct counting
shows  32 - 11 = 21 complex or 42 real parameters, but one can show,
passing to the $SO(1,9)$ covariant representation of the (originally
$SO(1,10)$ covariant) $D$=11 pure spinor condition \cite{NB-pure},
that two of the $11$ complex conditions are satisfied automatically,
so that there are only nine independent complex conditions.}. A $39$
parametric solution $\widetilde{\Lambda}_\alpha$ of this constraint
is provided by
\begin{eqnarray}\label{pureSp=}
    \widetilde{\Lambda}_\alpha = \tilde{\lambda}^+_p v_{\alpha
    p}^{\;-}\; , \qquad \tilde{\lambda}^+_p\tilde{\lambda}^+_p=0\; ,
    \qquad \{v_{\alpha p}^{\;-}\} = {Spin(1,10) \over
[Spin (1,1)\otimes Spin(9)] \, \subset\!\!\!\!\!\times \mathbb{K}_9 } = \mathbb{S}^{9}
\; , \qquad
\end{eqnarray}
where $\tilde{\lambda}^+_p$ is a complex 16 component $SO(9)$ spinor
with zero norm, $\tilde{\lambda}^+_p\tilde{\lambda}^+_p=0$, carrying
$32-2=30$ degrees of freedom and $v_{\alpha p}^{\;-}$ are spinorial
Lorentz harmonics \cite{BL98'} (see also \cite{B90,Ghsds},
\cite{BZ-str,BZ-strH,BZ-p,BL98'}), a set of $16$ constrained $D$=11
bosonic spinors which, once the constraints are taken into account,
provide the homogeneous coordinates for the 11 dimensional celestial
sphere $S^9$  and thus carry $9$ degrees of freedom (see below). The
existence of such a solution already suggests  a relation among the
pure spinor and the Lorentz harmonics approaches.

 Notice that in $D=10$ dimensional case such a
relation is much more close. The solution of Eq. (\ref{pureSp=})
carries {\bf 16+8-2=22 } degrees of freedom, the same number as the
generic pure spinor, so that it is the general solution. This may be
important for the study of superstring covariant quantization on the
line similar to what we present here for the case of superparticle.

\bigskip

 Here we first  construct the Hamiltonian mechanics of the twistor-like Lorentz
harmonics  formulation of the $D=11$ superparticle and, with the help of the spinorial
Lorentz harmonics, separate {\it covariantly} the first and the second class
constraints (see \cite{BZ-strH} for an analogous result for the Green-Schwarz
superstring). Then we take into account the second class constraints by introducing
Dirac brackets \cite{Dirac}, and calculate the Dirac brackets algebra of the first
class constraints which happens to be a nonlinear algebra. Further, following the
pragmatic spirit of the Berkovit's approach \cite{NB-pure}, \cite{NBloops}, we take
care of the part of constraints separately and   left with a set of $16$ fermionic and
$1$ bosonic first class constraints, the generators of the fermionic $\kappa$--symmetry
(see \cite{A+L82,S83}) and its bosonic $b$--symmetry superpartner, the Dirac brackets
of which represent  the $d=1$ , $n=16$ (worldline) supersymmetry algebra. This set of
constraints is described by the BRST charge
\begin{eqnarray}\label{Qsusy-Int}
\mathbb{Q}^{susy}= \lambda^+_q  D_q^{-} + i c^{++} \partial_{++} -
 \lambda^+_q\lambda^+_q {\partial\over \partial c^{++}}\; ,
 \qquad \{ D_p^{-} , D_q^{-}  \} = 2i \delta_{qp} \partial_{++} \; ,
 \qquad
\end{eqnarray}
including $16$ real  bosonic ghosts $\lambda^+_q$ and
\footnote{The sign superscripts in $\lambda^+_q$ and $D^-_q$
denote the spinorial Majorana-Weyl (MW) representations of
$SO(1,1)$; double sign superscript $--$, $++$ or subscript, like
in $\partial_{++}$, would correspond to the $SO(1,1)$ vector.
Since the MW spinorial representation of $SO(1,1)$ is one
dimensional, the subscript $+$ is equivalent to superscript $-$
and {\it vice versa}, so that $\{ D_p^{-} , D_q^{-}  \} = 2i
\delta_{qp}
\partial_{++} $ in (\ref{Qsusy-Int}) is $SO(1,1)$ invariant.
This notation corresponds to the light--cone basis in two
dimensional space (or in two dimensional subspace of the
D--dimensional spacetime) with the (flat space) metric of the form
$g_{++\; --}={1\over 2}$, $g_{++\; ++}=0=g_{--\; --}$, so that,
{\it e.g.} $\partial_{++}= {1\over 2}\partial^{--}$, where the
coefficients ${1\over 2}$ (which then appear in Eq.
(\ref{UUT=eta})) are introduced to avoid the appearance of
 $\sqrt{2}$ coefficients in many equations. } one real
fermionic ghost $c^{++}$.

An analysis of  the cohomology of this BRST operator shows that it
is trivial if the norm $ \lambda^+_q\lambda^+_q$ of bosonic ghost $
\lambda^+_q$ is nonvanishing. In other words, the nontrivial
cohomology of $\mathbb{Q}^{susy}$ has support on
$\lambda^+_q\lambda^+_q=0$. For a real spinor
$\lambda^+_q\lambda^+_q=0$ implies $\lambda^+_q=0$. This produces a
technical problem which is sorted out by means of a regularization
which consists in allowing $\lambda^+_q$ to be {\it complex},
$\lambda^+_q \mapsto \tilde{\lambda}^+_q\not=
(\tilde{\lambda}^+_q)^*$ so that
$\tilde{\lambda}^+_q\tilde{\lambda}^+_q=0$ allows for nonvanishing
complex solutions. Furthermore, this implies the reduction of the
cohomology problem for the regularized BRST operator
$\mathbb{Q}^{susy}$ to the search for cohomology at vanishing
bosonic ghost, $\tilde{\lambda}^+_q=0$, for the following complex
BRST charge
\begin{eqnarray}\label{tQsusy-Int}
\tilde{\mathbb{Q}}^{susy}= \tilde{\lambda}^+_q  \; D_q^{-} + i
c^{++}
\partial_{++} \; , \qquad
 \tilde{\lambda}^+_q\tilde{\lambda}^+_q=0 \; ,
 \qquad \{ D_p^{-} , D_q^{-}  \} = 2i \delta_{qp} \partial_{++} \; .
 \qquad
\end{eqnarray}
We discuss the relation of the above non-hermitean
$\tilde{\mathbb{Q}}^{susy}$ operators with the (always complex)
Berkovits BRST charge and find that this comparison shows the
possible origin of the intrinsic complexity of the Berkovits
formalism.

The above results were briefly reported in \cite{IB07}; here we give details on their
derivation. The possible results of stringy generalizations are discussed in the
concluding sec. 6 of the present paper.

Let us stress that of all the cohomologies of the Berkovits--like BRST charge
$\tilde{\mathbb{Q}}^{susy}$ (\ref{tQsusy-Int}) only the ones calculated (and remaining
nontrivial) at $\tilde{\lambda}_q=0$ describe the cohomology of the superparticle BRST
operator ${\mathbb{Q}}^{susy}$. The full cohomology of $\tilde{\mathbb{Q}}^{susy}$ is
clearly reacher and is related with spinorial cohomologies of \cite{SpinCohom02}. As
far as the $\tilde{\mathbb{Q}}^{susy}$ cohomology for vanishing bosonic ghost
describing the M0--brane spectrum is concerned, this can be described by the function
of variables which are inert under $\kappa$-- and $b$--symmetry. This relatively simple
structure finds its explanation in the properties of the superparticle action in the
so--called covariantized light-cone basis (see \cite{Sok,B90,GHT93}).

The change of variable corresponding to  this basis in the
superparticle spinor moving frame action  results in an  automatical
gauge fixing of the $\kappa$--symmetry and $b$--symmetry. Thus, in
this basis, the set of superparticle first class constraints
contains only the generators of the Borel subgroup $[SO(1,1)\times
SO(9)]\subset\!\!\!\!\!\!\times K_9$ of the Lorentz group
$SO(1,10)$. We present here the BRST charge describing this set of
the first class constraints.   Then,  following Dirac \cite{Dirac},
we impose this set of first class constriants as conditions on the
wave function and discuss the quantization of physical degrees of
freedom (a covariantized light cone basis prototype of the
supertwistor quantization in \cite{BdAS2006}) which shows the hints
of hidden $SO(16)$ symmetry and suggests some speculations on
possible $E_8$ symmetry of $D=11$ supergravity.

\subsection{Structure of the paper}

This paper is organized as follows. Sec. \ref{LHSsp} reviews the
spinor moving frame (twistor like Lorentz harmonics) formulation of
the $D$=11 massless superparticle or M0-brane and shows its
classical equivalence with the standard Brink--Schwarz formulation.
In Sec. III we develop the Hamiltonian formalism for this
formulation of M0--brane, discuss its classical BRST charge and the
reduced BRST operator ${\mathbb{Q}}^{susy}$ corresponding to a
subset of the M0--brane first class constraints. Particularly, the
primary constraints are obtained in sec. \ref{Primary}. In sec.
\ref{DBsec} the Dirac brackets that allow us to treat harmonic
variables as coordinates on the Lorenz group manifold are defined.
These are related with the group-theoretical structure of Lorentz
harmonics in sec. \ref{CartanF}, where the
$SO(1,10)/[[SO(1,1)\otimes SO(9)]\subset\!\!\!\!\!\!\times K_9]$
Cartan forms are introduced. These are used in sec. \ref{PB-DB} to
define canonical Hamiltonian of the M0--brane model. The second
class constraints are found and the Dirac brackets allowing to treat
them in the strong sense are presented in sec. \ref{secIIclass}. The
Dirac bracket algebra of all the first class constraints is
presented in sec. \ref{secIclass}. The BRST charge
$\mathbb{Q}^{\prime}$ for the nonlinear (sub)superalgebra of the
first class constraints is obtained in sec. \ref{Qprime}. Finally,
the BRST charge ${\mathbb{Q}}^{susy}$  is obtained by reduction of
$\mathbb{Q}^{\prime}$ in sec. \ref{secQsusy}.

The cohomology of ${\mathbb{Q}}^{susy}$ is studied in Sec.
\ref{QsusyCoH}. In particular, the complex charge (\ref{tQsusy-Int})
is introduced in sec. \ref{CohQ-2} and its relation with the
Berkovits BRST charge is discussed in sec. \ref{CohQ-3}.  To explain
the relatively simple structure of the  ${\mathbb{Q}}^{susy}$
cohomology,  in Sec. \ref{SecAnalB} we study the superparticle
spinor moving frame action in the covariantized light-cone basis
(see \cite{Sok,B90,GHT93}). The automatical gauge fixing of the
$\kappa$--symmetry and $b$--symmetry which occurs in the action when
changing variables to this basis is discussed in sec.
\ref{SecAnalB}.1.1. The BRST charge describing the set of the first
class constraints of the superparticle action in this basis is
presented in sec. \ref{SecAnalB}.1.2. The quantization of the
physical degrees of freedom of the superparticle using the
covariantized light cone basis is discussed in sec.
\ref{SecAnalB}.2. There we also discuss the hints of possible hidden
symmetries of the D=11 supergravity which appears on the way of such
a covariant quantization.

In Sec. \ref{Concl} we present our conclusions (sec. \ref{Concl}.1) and an outlook
(secs. \ref{Concl}.2, \ref{Concl}.3), including the discussion on possible results of
the generalization of our study of M0--brane to the case of type IIB superstring (sec.
\ref{Concl}.2). Some technical details on harmonics  are presented in the Appendix.

\section{The M0-brane in the spinor moving frame
formulation. Twistor--like action and its gauge  symmetries. }\label{LHSsp}
\setcounter{equation}0

\subsection{Towards the spinor moving frame action for the D=11 massless superparticle}

The Brink-Schwarz massless superparticle action, $S_{BS}  =
\int_{W^1} {1\over 2e}{\Pi}_{\tau m} {\Pi}_\tau^{m}$,  can be
written in the following first order form
\begin{eqnarray}\label{11DSSP-1st}
S_{BS}^{1}  & = &  \int_{W^1} \left(P_{{m}} {\Pi}^{m} -  {1\over 2}d\tau \; e\; P_{{m}}
P^{{m}} \right)\; ,   \qquad
\end{eqnarray}
where $P_m(\tau)$ is the auxiliary momentum variable, $e(\tau)$ is
the worldline einbein and $\Pi^m = d\tau \hat{\Pi}^{m}_\tau$ is the
pull-back of the bosonic supervielbein of the tangent superspace to
the superparticle worldline. In flat $D=11$ superspace this reads
\begin{eqnarray}
\label{11DPi} && \Pi^m := dx^m - id\theta \Gamma^m \theta = d\tau
\hat{\Pi}^{m}_\tau \; , \qquad \hat{\Pi}^{m}_\tau:=
\partial_\tau\hat{x}^m(\tau ) - i\partial_\tau \hat{\theta}(\tau)
\Gamma^m \hat{\theta}(\tau)\;
\end{eqnarray}
The action (\ref{11DSSP-1st}) is valid in any dimension; the $D$=11
massless superparticle action \cite{B+T=D-branes} corresponds to
$m=0\, , 1,\ldots 9, \# $, a $32$ component Majorana spinor
$\theta^\alpha $ and  $32\times 32$ eleven--dimensional gamma
matrices $\Gamma^m_{\alpha\beta}:= \Gamma^m{}_\alpha{}^\gamma
C_{\gamma\beta}= \Gamma^m_{\beta\alpha}$.

The einbein $e(\tau)$ plays the r\^ole of Lagrange multiplier and
produces the mass shell constraint
\begin{eqnarray}
\label{PmPm=0}  P_mP^m=0 \; .
\end{eqnarray}
Since Eq. (\ref{PmPm=0}) is algebraic, it may be substituted into the action
(\ref{11DSSP-1st}), which gives
\begin{eqnarray}\label{11DSSPwhen}
S^{\prime}_{M0}  & = &  \int_{W^1} \; P_{{m}} \hat{\Pi}^{m} \; , \qquad  P_{{m}}
P^{{m}}=0 \; .   \qquad
\end{eqnarray}
Thus, if the general solution of (\ref{PmPm=0}) is known, one may substitute it for
$P_m$ in (\ref{11DSSPwhen}) and obtain a classically equivalent formulation of the $D$-
(here 11-) dimensional Brink-Schwarz superparticle. The moving frame or twistor-like
Lorentz harmonics formulation of \cite{BL98',BdAS2006} (see \cite{B90} for $D$=4 and
\cite{IB+AN96} for $D=10$) can be obtained just in this way.

It is easy to solve the constraint (\ref{PmPm=0}) in a non-covariant
manner: in a special Lorentz frame a solution
 with positive energy, $P^{\!\!\!^{0}}_{(a)}$, reads {\it e.g.}
\begin{eqnarray}
\label{PmPm=00} & P^{\!\!\!^{0}}_{(a)} = {\rho\over 2} \; (1,\ldots , -1) = {\rho\over
2}  \; (\delta_{(a)}^0 -\delta_{(a)}^{\#}) \quad .
\end{eqnarray}
The solution in an arbitrary frame follows from (\ref{PmPm=00}) by making a Lorentz
transformation, $P_{m}= U_m{}^{(a)}P^{\!\!\!^{0}}_{(a)}$ with $U_m{}^{(a)} \in
SO(1,10)$,
\begin{eqnarray}
\label{PmPm=01} P_m := U_m{}^{(a)} P^{\!\!\!^{0}}_{(a)} = {\rho\over 2} \; (u_{m}^{\;\;
0} - u_{m}^{\; \#}) \; , \qquad U_m^{\, (a)}:= (u_{m}^{\;\; 0} , u_{m}^{\;\; i} ,
u_{m}^{\;\#}) \in SO(1,D-1) \; .
\end{eqnarray}
Note  that, since  $P_{m}=P_{m}(\tau)$ is dynamical variable in the action
(\ref{11DSSPwhen}),  the same is true for the Lorentz group matrix $U$ when it is used
to express $P_{m}$ through Eq. (\ref{PmPm=01}), $U_m{}^{(a)}=U_m{}^{(a)}(\tau)=
(u_{m}^{\;\; 0}(\tau) , u_{m}^{\;\; i}(\tau) , u_{m}^{\;\#}(\tau))$. Such {\it moving
frame variables} \cite{BZ-str} are called {\it Lorentz harmonics} \cite{B90,Ghsds} (see
\cite{GIKOS}, also light--cone harmonics in \cite{Sok}).

Substituting (\ref{PmPm=01}) for $P_{m}$ in (\ref{11DSSPwhen}) or, equivalently, in
(\ref{11DSSP-1st}), one arrives at the following action
\begin{eqnarray}\label{11DSSP(LH)}
S  =  \int_{W^1} \; {1\over 2}  \rho^{++} u^{--}_{{m}} \hat{\Pi}^{m} \; , \qquad
u^{--}_{{m}} u^{--{m}}=0 \quad & (\; \Leftarrow  \quad U:= \{ {u_m^{++} + u_m^{--}\over
2} , u_m^{\; i} , {u_m^{++} - u_m^{--}\over 2}  \} \in SO(1,10)\;)   \qquad
\end{eqnarray}
where the light--likeness of the vector $u^{--}_m=u^0_m - u^{\#}_m$  (see also
(\ref{harmUdef}) below) follows from the orthogonality and normalization of the
timelike $u_m^{0}$ and spacelike $u_m^{\# }$ vectors which, in their turn, follow  from
 $U\in SO(1,10)$
in Eq. (\ref{PmPm=01}) (as it is noticed in the brackets in (\ref{11DSSP(LH)})).

At this stage it might seem obscure what is the advantage of the action of Eq.
(\ref{11DSSP(LH)}) with respect to (\ref{11DSSPwhen}) or (\ref{11DSSP-1st}). However,
as we discussed below, the action (\ref{11DSSP(LH)}) {\it hides the twistor--like
action}, a higher dimensional ($D$=11 here) generalization of the D=4
Ferber--Schirafuji action \cite{Ferber}. The twistor like variables called {\it
spinorial harmonics} appears as `square roots' of the vector harmonics (see below);
they can be used to separate covariantly the first and the second class constraints and
to provide the {\it irreducible} form of the $\kappa$--symmetry \cite{A+L82,S83}
(infinitely reducible in the standard formulation of massless superparticle
\cite{S83}\footnote{Notice that in the case of massless N=2 superparticle, which
presently are identified with D0--branes, the covariant gauge fixing of the
$\kappa$--symmetry is possible already in the standard formulation \cite{A+L82}.}).
This also explains why the formulation based on the action (\ref{11DSSP(LH)}) is called
the {\it spinor moving frame} formulation.

\subsection{Twistor--like spinor moving frame action of M0--brane and its  gauge symmetries.  }

The spinor moving frame action for the $D=11$ massless superparticle can be written in
the following equivalent forms \cite{BL98'} (see \cite{IB+AN96} for D=10 and \cite{B90}
for D=4)
\begin{eqnarray}\label{11DSSP}
S:= \int d\tau L  &=&   \int_{W^1} {1\over 2}\rho^{++}\, u_{m}^{--} \, \Pi^m =
\int_{W^1} {1\over 32}\rho^{++}\, v_{\alpha q}^{\; -} v_{\beta q}^{\; -} \, \Pi^m
\tilde{\Gamma}_m^{\alpha\beta}\; ,  \qquad
\\ \nonumber
&& {} \alpha= 1,2, \ldots , 32 \, \quad (n\; in \; general ) \; , \quad q=1, \ldots ,
16 \, \quad (n/2\; in \; general ) \; , \qquad \\ \nonumber && {} \qquad m=0,\ldots ,
9, \# \quad  ( (D-1)\; in \; general) \;
\end{eqnarray}
where we use the symbol $\#$ to denote the tenth spatial direction ($X^{\#}:= X^{10}$)
and  the notation $\Gamma^m\equiv \Gamma^m{}_{\alpha\beta}:=
\Gamma^m{}_{\alpha}{}^{\gamma}C_{\gamma\beta}\,,\,
 {\tilde \Gamma}^m\equiv   {\tilde \Gamma}^m{\,}^{\alpha\beta}:=C^{\alpha\gamma}
 \Gamma^m{}_{\gamma}{}^{\beta}$ for the $D=11$ gamma--matrices contracted with
 $C_{\alpha\beta}$ and $C^{\alpha\beta}$. The first from of the
action (\ref{11DSSP}) coincides with (\ref{11DSSP(LH)}); the second  form is
twistor--like, {\it i.e.} it resembles the Ferber--Schirafuji action \cite{Ferber} for
the massless $D=4$ superparticle. Instead of two--component Weyl spinor of the Ferber
supertwistor, the action of Eq. (\ref{11DSSP}) includes the set of $16$ bosonic
$32$--component Majorana spinors $v_{\alpha}{}^-_{q}$ which satisfy the following
kinematical constraints (see \cite{BZ-str,BZ-strH,BL98'}),
\begin{eqnarray}\label{vv=uG}
\left\{ \matrix{2 v_\alpha{}_{q}^{-} v_\beta{}_{q}^{-} &=&
 u_m^{--}{\Gamma}^m_{\alpha\beta}\;   & \qquad (a)\;
 ,  \cr  v_{q}^{-}\tilde{\Gamma}_m v_{p}^{-} &=&  \delta_{qp} \; u_m^{--}  \; & \qquad (b)\; ,
  \cr
v_\alpha{}_{q}^{-}C^{\alpha\beta}v_\beta{}_{p}^{-}&=& 0 \;  \qquad & \qquad (c)\; , }
\right. \;
   \quad u_m^{--}u^{m --}=0 \qquad (d)\;\; . \qquad
\end{eqnarray}

Although, in principle, one can study the dynamical system using just the kinematical
constraints (\ref{vv=uG}) (see \cite{gs92,BCSV}), it is more convenient to treat the
light--like vector $u_m^{--}$ as an element of the $SO(1,10)$-valued matrix describing
{\it vector moving frame} and the set of $16$ $SO(1,10)$- spinors $v_\alpha{}_{q}^{-}$
as part of the corresponding $Spin (1,10)$--valued matrix describing the {\it spinor
moving frame}. These moving frame variables are also called ({\it vector} and {\it
spinor}) {\it Lorentz harmonics} and will be discussed in Sec. 2.4 below.

Let us conclude this section by noticing that the action (\ref{11DSSP}) possesses a set
of gauge symmetries which includes \\
i) the {\it irreducible} $\kappa$--symmetry
\begin{eqnarray}\label{kappa-irr}
\delta_\kappa x^m = i \delta_\kappa \theta^\alpha
\Gamma^m_{\alpha\beta}\theta^\beta\; , \qquad \delta_\kappa
\theta^\alpha = \kappa^{+q} v_q^{-\alpha} \; , \qquad \delta_\kappa
v_\alpha{}^-_q =0 = \delta_\kappa u_m^{--}\; ; \qquad
\end{eqnarray}
the possibility to reformulate the $\kappa$--symmetry in the
irreducible form is due to  the presence of the constrained bosonic
spinor variables $v_\alpha{}_{q}^{-}$ (see \cite{BZ-str,IB+AN96} and
the discussion below);
\\
ii) its superpartner, the tangent space copy of the worldvolume
reparametrization symmetry, which we, following the pioneer paper
\cite{A+L82}, call $b$--symmetry,
\begin{eqnarray}\label{b-sym}
\delta_b x^m = b^{++} u^{--m} \; , \qquad \delta_b \theta^\alpha = 0 \; , \qquad
\delta_b v_\alpha{}^-_q =0 = \delta_b u_m^{--}\; ; \qquad
\end{eqnarray}
iii) a scaling $GL(1,\mathbb{R})$ symmetry
\begin{eqnarray}\label{SO(1,1)}
\rho^{++} \mapsto e^{2\alpha} \rho^{++}\; , \qquad u_m^{--} \mapsto
e^{-2\alpha} u_m^{--}\; , \qquad v_{\alpha q}{}^- \mapsto e^{-
\alpha} v_{\alpha q}{}^- \; ,  \qquad
\end{eqnarray}
with the wait determined by the sign indices $^{++}$, $^{--}$ and
$^{-}$. In the light of Lorentz harmonic treatment of $v_{\alpha
q}{}^{\!\! -}$ and $u_m^{--}$, which will be presented below, we
prefer to identify this scaling symmetry as $SO(1,1)$ group
transformations.
\\ iv) The action (\ref{11DSSP}) is also invariant under the  $Spin(9)$ symmetry acting
on the $q=1, \ldots 16$ index of the constrained bosonic spinor
variable  $v_{\alpha q}{}^-$,
\begin{eqnarray}\label{SO(9)}
v_{\alpha q}{}^{\!\!\! -} \mapsto  v_{\alpha p}{}^{\!\!\! -}
S_{pq}\; , \qquad S_{pq} \in Spin(9)\; \quad \Leftrightarrow  \quad
\cases{ S^TS=\mathbb{I}_{16\times 16} \; , \cr S\gamma^I S^T =
\gamma^J U^{JI} \; , \quad U^TU= \mathbb{I}_{9\times 9} }\; ,
\qquad
\end{eqnarray}
Notice that the nine dimensional charge conjugation matrix is
symmetric and can be identified with the Kroneker delta symbol, $
\delta_{qp}\; $, so that the contraction $v_{\alpha q}{}^{\!\!\!
-}v_{\beta q}{}^{\!\!\! -}$, entering the action, is $Spin(9)$
invariant.

This $Spin(9)$ symmetry is used as an identification relation when
the spinorial Lorentz harmonics are defined as homogeneous
coordinates of the coset $SO(1,10)\over [SO(1,1)\otimes
SO(9)]\subset\!\!\!\!\times K_9=\mathbb{S}^9$ (see below) given by a
$Spin(1,10)$ valued matrix $V_\alpha{}^{(\beta)}=(v_{\alpha
q}{}^{\!\! -} , v_{\alpha q}{}^{\!\!+})\in Spin(1,10)$, one of the
two $32\times 16$ blocks of which is identified with our $v_{\alpha
q}{}^-$.

However, when the action (\ref{11DSSP}) with the variable $v_{\alpha
q}{}^-$ subject only to the constraints (\ref{vv=uG}) is considered,
one immediately finds that neither constraints nor the action
involve the $d=9$ gamma matrices;  all the contractions are made
with $16\times 16$ Kroneker symbol $\delta_{qp}$, and the same
matrix only is used in the constraints.

\subsection{On O(16) gauge symmetry}
\label{O(16)} Thus we have observed that {\it the action}
(\ref{11DSSP}), when considered as constructed from spinorial
variables restricted by the constraints (\ref{vv=uG}),
\begin{eqnarray}\label{11DSSP-16}
S &=& \int_{W^1} {1\over 32}\rho^{++}\, \tilde{v}_{\alpha q}^{\; -}
\tilde{v}_{\beta q}^{\; -} \, \Pi^m \tilde{\Gamma}_m^{\alpha\beta}\;
, \qquad  \cases{ 2 \tilde{v}_\alpha{}_{q}^{-}
\tilde{v}_\beta{}_{q}^{-} =
 {1\over 16} \tilde{v}_{p^\prime}^{-}\tilde{\Gamma}_m \tilde{v}_{p^\prime}^{-} {\Gamma}^m_{\alpha\beta}\;
 , \quad (a)\; \cr  \tilde{v}_{q}^{-}\tilde{\Gamma}_m \tilde{v}_{p}^{-} =   \delta_{qp} \;
{1\over 16} \tilde{v}_{p^\prime}^{-}\tilde{\Gamma}_m \tilde{v}_{p^\prime}^{-} \; ,
\quad (b)
 \cr
\tilde{v}_\alpha{}_{q}^{-}C^{\alpha\beta}\tilde{v}_\beta{}_{q}^{-}=0\; ,
   \qquad {} \qquad {}\quad  (c) \; } \;   \qquad
\\ \nonumber
&& {} \qquad \alpha= 1,2, \ldots , 32 \, \qquad \; , \quad q=1, \ldots , 16 \, \quad
 \; ,
\end{eqnarray}
actually possesses the local $SO(16)$ symmetry acting on the
$q=1,\ldots , 16$ indices of $\tilde{v}_{\alpha q}^{\; -}$
variables,
\begin{eqnarray}\label{SO(16)}
\tilde{v}_{\alpha q}{}^- \mapsto  \tilde{v}_{\alpha p}{}^- O_{pq}\; , \qquad O_{pq} \in
O(16)\quad \Leftrightarrow  \quad  O^TO=\mathbb{I}_{16\times 16} \;  . \qquad
\end{eqnarray}
One can conclude that the relation between spinorial harmonic
${v}_{\alpha q}{}^- $, which transforms under $Spin(9)$ symmetry,
and the above $\tilde{v}_{\alpha p}{}^- $, carrying the $SO(16)$
index $p$ is given by
\begin{eqnarray}\label{tv-=v-L}
\tilde{v}_{\alpha p}{}^- =  {v}_{\alpha q}{}^- L_{qp}\; , \qquad L_{qp} \in O(16)\quad
\Leftrightarrow  \quad  L^TL=\mathbb{I}_{16\times 16} \;  , \qquad
\end{eqnarray}
where $L_{qp}$ is an arbitrary orthogonal $16\times 16$ matrix.
Clearly, $\tilde{v}_{\alpha p}{}^-$ of Eq.  (\ref{tv-=v-L}) solves
the constraints (\ref{vv=uG}a-d) if these are solved by ${v}_{\alpha
q}{}^-$. But if ${v}_{\alpha q}{}^-$ is the spinorial harmonic, this
is to say a $32\times 16$ block of the $Spin(1,10)$ valued matrix
$V_\alpha{}^{(\beta)}=(v_{\alpha q}{}^- , v_{\alpha q}{}^+)\in
Spin(1,10)$, then $\tilde{v}_{\alpha p}{}^-$ cannot be such a block
if the $O(16)$ matrix $L_{pq}$ does not belong to the $Spin(9)$
subgroup of $SO(16)$. However,
  $\tilde{v}_{\alpha p}{}^- \tilde{v}_{\beta p}{}^- =
{v}_{\alpha q}{}^- {v}_{\beta q}{}^-$ so that substituting (\ref{tv-=v-L}) for
$\tilde{v}_{\alpha p}{}^-$ in (\ref{11DSSP-16}), one observes the cancelation  of
 the contributions of the matrix $L_{qp}$.

On one hand this is tantamount to the statement of the $O(16)$ of the action
(\ref{11DSSP-16}), with variable restricted only by the constraints presented
explicitly. On the other hand, this can be used to treat the variables ${v}_{\alpha
q}{}^- $ in the action (\ref{11DSSP}) as spinorial harmonics (allowing only the
$Spin(9)$ transformations (\ref{SO(9)}) on $q$ index). In the next section we accept
this latter point of view as it is technically more convenient for the Hamiltonian
analysis. The reason is that the constraints (\ref{vv=uG}) are reducible\footnote{This
is seen already from the fact that their number, $2122$, exceed the number $512$ of the
components of $32\times 16$ matrix. The above number of constraints is composed as
$2122$=$528-11+1496-11+120$, where $-11$ come from the facts of coincidence of the
gamma--trace parts of constraints (a) and (b) and of that $u_m^{--}$ can be defined by
means of one of these parts; the light--likeness of $u_m^{--}$, Eq. (\ref{vv=uG}d),
follows from the fact that the rank of the matrix in the {\it l.h.s.} of the constraint
(\ref{vv=uG}a) is $16$ or less and, thus, is not counted. } and even to calculate the
number of degrees of freedom becomes a nontrivial problem. This can be solved passing
through the identification of ${v}_{\alpha q}{}^- $ with spinorial harmonics: also one
introduces additional variables ${v}_{\alpha q}{}^+ $, one gains a clear group
theoretical and geometrical meaning which helps to deal with the reducible constraints.

To conclude this section, let us note that the (seemingly
fictitious) $SO(16)$ symmetry of the M0--brane, which we have
observed studying different versions of its twistor-like
formulation, reappears inevitably in the  quantization of physical
degrees of freedom which we will consider in Sec. \ref{SecAnalB}
(see also \cite{BdAS2006}).

\subsection{Vector and spinor Lorentz harmonics: moving frame  and spinor  moving frame}

The {\it vector} Lorentz harmonics variables $u_m^{\pm\pm}$, $u_m{}^i$ \cite{Sok}  are
defined as elements of the $SO(1,10)$ Lorentz group matrix, Eq. (\ref{PmPm=01}). In the
lightlike basis they are given by
\begin{eqnarray}
\label{harmUin} && U_m^{(a)}= (u_m^{--}, u_m^{++}, u_m^{i})\;  \in \; SO(1,10) \; ,
\qquad
\\ \nonumber && {m= 0,1,\dots,9,\# \; , } \qquad (a)=++,--, i \; , \qquad i=1,\dots,9 \; , \qquad
\end{eqnarray}
where $u^{\pm\pm}_m=u^0_m \pm u^{\#}_m$. The three-blocks splitting (\ref{harmUin}) is
invariant under  $SO(1,1)\otimes SO(9)$; $SO(1,1)$ rotates $u^0_m$ and $u^{\#}_m$ among
themselves and, hence,  transforms their sum and differences, $u^{\pm\pm}_m=u^0_m \pm
u^{\#}_m$, by inverse scaling factors,  see Eq. (\ref{SO(1,1)}). The fact that $U\in
SO(1,10)$ implies the following set of  constraints
\begin{eqnarray}
\label{harmUdef} U^T\eta U = \eta  \quad \Leftrightarrow \cases{ u_m^{--}u^{m--}=0 \; ,
\quad u_m^{++}u^{m++}=0 \; , \quad u_m^{\pm\pm}u^{m\, i}=0 \; , \cr u_m^{--}u^{m++}=2
\; , \qquad u_m^{i}u^{m\, j}=- \delta^{ij} }
\end{eqnarray}
or, equivalently, the unity decomposition
\begin{eqnarray}\label{UUT=eta}
\delta_m^n= {1\over 2}u_m^{++}u^{n--} + {1\over 2}u_m^{--}u^{n++} - u_m^{i}u^{n
i}\qquad  \Leftrightarrow \qquad U\eta U^T=\eta\; .
\end{eqnarray}

The {\it spinor}  harmonics \cite{Ghsds} or spinor moving frame variables
\cite{BZ-str,BZ-strH,BZ-p} $v^{\;\;\,\pm}_{\alpha q}$ are elements of the $32\times32$
$Spin(1,10)$--valued  matrix
\begin{eqnarray}
\label{harmVin} V_\alpha^{(\beta)}= (v_\alpha{}_q^{-}\; , v_\alpha{}_{q}^{+})\; \in \;
Spin(1,10) \qquad (\alpha=1,\dots 32\; , \; q=1,\dots,16) \; .
\end{eqnarray}
They are `square roots' of the associated vector harmonics in the sense that
\begin{eqnarray}
\label{harmVdef} V \Gamma^{(a)} V^T = \Gamma^m U_m ^{(a)} \qquad (a) \; , \qquad V^T
\tilde{\Gamma}_m V = U_m^{(a)} \tilde{\Gamma}_{(a)} \qquad (b) \;   ,
\end{eqnarray}
which express the $Spin(1,10)$ invariance of the Dirac matrices.

Equation in (\ref{vv=uG}a) is just the $(a)=(--)\equiv (0)-(\# )$ component of Eq.
(\ref{harmVdef}a) in the Dirac matrices realization in which $\Gamma^0$ and $\Gamma^{\#
}$ are diagonal; the nine remaining $\Gamma^I$ are off-diagonal. Eq. (\ref{vv=uG}b)
comes from the upper diagonal block of Eq. (\ref{harmVdef}b). To complete the set of
constraints defining the spinorial harmonics, we have to add the conditions expressing
the invariance of the charge conjugation matrix $C$,
\begin{eqnarray}
\label{harmVdefC}
 VCV^T=C \quad, \quad V^TC^{-1}V=C^{-1}\; ,
\end{eqnarray}
which give rise to the constraint (\ref{vv=uG}c).

In a theory with the local $SO(1,1)\otimes SO(9)$ symmetry (\ref{SO(1,1)}),
(\ref{SO(9)}), containing only one of the two sets of $16$ constrained spinors
(\ref{harmVin}), say $v_{\alpha p}^{\;-}\,$, these can be treated as homogeneous
coordinates of the $SO(1,10)$ coset giving the celestial sphere $S^9$; specifically
(see \cite{Ghsds})
\begin{eqnarray}
\label{v-inS11} {} \{v_{\alpha q}^{\;-}\} = {Spin(1,10) \over [Spin (1,1)\otimes
Spin(9)] \, \subset \!\!\!\!\!\!\times {\mathbb{K}_9} } = \mathbb{S}^{9}  \quad ,
\end{eqnarray}
where $\mathbb{K}_9$ is the abelian subgroup of $SO(1,10)$ defined by\footnote{The
$\mathbb{K}_9$ symmetry (\ref{K9-def}) is tantamount to stating that the model contains
only one, $v_{\alpha p}^{\;-}\,$, of the two sets of $16$ constrained spinors
$(v_\alpha{}_q^{-}\; , v_\alpha{}_{q}^{+})$ in  (\ref{harmVin}).}
 \begin{eqnarray}
\label{K9-def} \delta v_{\alpha q}^{\; -}=0\; , \qquad \delta v_{\alpha q}^{\; +}=
 k^{++ i} \gamma^i{}_{qp}\,v_{\alpha p}^{\; -}\; , \qquad i=1,\ldots , 9 \; . \qquad
 \end{eqnarray}
 Our superparticle model contains just $v_{\alpha q}^{\; -}$ and is invariant
under $SO(1,1)\otimes Spin(9)$ transformations. Hence the harmonics sector of its
configuration space parametrize $S^9$ sphere.

\subsubsection{On harmonics and explicit parametrization of  $SO(1,D-1)/H$ cosets}

The vector harmonic variables, when constrained only by Eqs. (\ref{harmUdef}),
parametrize the eleven dimensional Lorentz group $SO(1,10)$, Eq. (\ref{harmUin}). This,
in principle, can be solved by expressing the harmonics in terms of $55$ parameters
$l^{(a)(b)}=- l^{(b)(a)}$, $U_m{}^{(a)}=U_m{}^{(a)}(l^{(b)(c)})$,
\begin{eqnarray}
\label{harmU=} U_m^{(a)}&=& \left(u_m^{--}, u_m^{++}, u_m^{i}\right)= \delta_m^{(a)} +
\eta_{m(b)} l^{(b)(a)} + {\cal O}(l^2)\; , \qquad \nonumber {{}\over {}}
\\  && u_m^{\pm\pm}=
\delta_m^{\pm\pm} - \eta_{m(b)} l^{\pm\pm \, (b)} + {\cal O}(l^2)\; , \quad u_m^{i}=
\delta_m^{i} + \eta_{m(b)} l^{ (b) i} + {\cal O}(l^2)\; , \qquad
\\ \label{l-param} && \delta_m^{\pm\pm}:=  \delta_m^0 \pm \delta_m^{\#}\; ,  \qquad
l^{(a)(b)}=- l^{(b)(a)}= \left(\matrix{  0 & - 4 l^{(0)} & l^{++j} \cr
  \,  4 \, l^{(0)} &  0 & l^{--j} \cr
 - l^{++i} & - \; l^{--i} & l^{ij}\;}\right)
 \; , \qquad
\end{eqnarray}
where we used the `light-like' splitting  $(a)=++,--, i$, $i=1,\ldots , 9$,  so that
\begin{eqnarray}
\label{g(a)(b)} \eta_{(a)(b)}:= \left(\matrix{  0 & {1\over 2} &  0 \cr
 {1\over 2} &  0 & 0 \cr
 0 & 0 & - \delta_{ij}\;
}\right) \quad \; , \qquad \eta^{(a)(b)}:= \left(\matrix{  0 & 2 &  0 \cr
 {2} &  0 & 0 \cr
 0 & 0 & - \delta_{ij}\;
}\right) \quad \; , \qquad
\end{eqnarray}

The same can be said about spinorial harmonics. Eqs. (\ref{harmVdef}),
(\ref{harmVdefC}) imply that spinorial harmonics parametrize the $Spin(1,10)$-valued
matrix providing the double covering  of the $SO(1,10)$ group element (\ref{harmUin})
and, hence, that they can be expressed (up to the sign) through the same $l^{(a)(b)}=-
l^{(b)(a)}$ parameters, $V_\alpha^{(\beta)}= \pm  V_\alpha^{(\beta)}(l)$,
\begin{eqnarray}
\label{harmV=} V_\alpha^{(\beta)}(l)= (v_\alpha{}_q^{-}(l)\; , v_\alpha{}_{q}^{+}(l))\;
= \left(\delta_\alpha^{(\beta)} + {1\over 4} l^{(a)(b)}\Gamma_{(a)(b)}{}
_\alpha^{(\beta)}  + {\cal O}(l^2) \right)\; . \qquad
\end{eqnarray}

The identification of the harmonics with the coordinates of $SO(1,10)/H$ corresponds to
setting to zero the $H$ coordinates in the explicit expressions (\ref{harmU=}),
(\ref{harmV=}). In our case with $H=[SO(1,1)\otimes SO(9)]\otimes \mathbb{K}_9$ this
implies $l^{(0)}=l^{ij}=l^{++j}=0$ so that the $SO(1,10)$ and $Spin(1,10)$ matrices are
constructed with the use of $9$ parameters $l^{--j}$,
$U_m{}^{(a)}=U_m{}^{(a)}(l^{--j})$, $V_\alpha^{(\beta)}= V_\alpha^{(\beta)}(l^{--j})$.
These expressions are not so complicated and read
\begin{eqnarray}
\label{U=l--} u_a^{--}= \delta_a^{--} + \delta_a{}^i l^{--i} + {1\over 2}\delta_a^{++}
(l^{--j}l^{--j})\; , \qquad  u_a^{++}= \delta_a^{++}\; , \qquad u_a{}^{i}=
\delta_a{}^{i} + {1\over 2}\delta_a^{++} {l}^{--i} \; . \qquad
\end{eqnarray}
for the vector harmonics. The expressions for spinor harmonics are even simpler,
\begin{eqnarray}
\label{V=l--} v_{\alpha}{}^-_q = \delta_{\alpha}^{-q}  + {1\over 2}\,
l^{--i}\gamma^i_{qp} \delta_{\alpha}^{+q} \; , \qquad v_{\alpha}{}^+_q =
\delta_{\alpha}^{+q} \; . \qquad
\end{eqnarray}
The disadvantage of the above  equations  Eqs. (\ref{U=l--}), (\ref{V=l--}) with
respect to the general Eqs. (\ref{harmU=}), (\ref{harmV=}), is that they are not
Lorentz covariant; this follows from that they are the gauge fixed version of
(\ref{harmU=}), (\ref{harmV=}) obtained with the use of $[SO(1,1)\otimes
SO(9)]\subset\!\!\!\!\!\!\times K_9$ symmetry.

Although the use of the  explicit expressions (\ref{harmU=}), (\ref{harmV=}) is not
practical (so that we even would not like to present them beyond the linear
approximation; these explicit expressions can be found in \cite{GomisWest06}), it is
convenient to keep in mind the  mere  fact of there existence. To make calculations we
rather use the `admissible variation' technique \cite{BZ-strH,BZ-p} and/or, when the
Hamiltonian mechanics is considered, the Dirac brackets for the constraints
(\ref{harmUin}), (\ref{harmVin}) on the Harmonic variables and their conjugate. These
Dirac brackets for $U$, $V$ and their momenta, which can be identified as Poisson
brackets for $l^{(a)(b)}$ in (\ref{harmUin}), (\ref{harmVin}) and its conjugate
momentum, are discussed in the next section.

\section{ Hamiltonian mechanics of the D=11 superparticle in the spinor moving frame formulation and the BRST charge
$\mathbb{Q}^{susy}$} \setcounter{equation}0

In \cite{BdAS2006} we presented the supertwistor quantization of
M0--branes. Albeit heuristic, it has the advantage of being simple,
 formulated in terms of physical variables (like the light-cone
gauge quantization in \cite{Green+99}), and of being covariant (in contrast with
\cite{Green+99}). Here we perform the complete Hamiltonian analysis of the dynamical
system and consider its BRST quantization.

\subsection{Primary constraints of the D=11 superparticle model
(M0--brane)}\label{Primary}

 The primary constraints of the M0-brane in the spinor moving
 frame formulation (\ref{11DSSP}) include the defining relations of the
harmonic variables, Eqs. (\ref{vv=uG}), plus other relations in
(\ref{harmVdef}), as well as
\begin{eqnarray}\label{P-rvv}
\Phi_a &:= &  P_a - {1\over 2} \rho^{++} u_a^{--}\approx 0 \qquad
\Leftrightarrow \qquad \Phi\!\!\!/_{\alpha\beta}:= \Phi_a
\Gamma^a_{\alpha\beta}= P\!\!\!/_{\alpha\beta} - \rho^{++}
v_\alpha{}_{q}^{-} v_\beta{}_{q}^{-} \approx 0 \qquad \; , \\
\label{df=} d_\alpha &:= & \pi_\alpha + i
P\!\!\!/_{\alpha\beta}\theta^{\beta}\approx 0 \; , \qquad
\pi_\alpha:= {\partial L \over
\partial \dot{\theta}^{\alpha}  } \; , \quad  P_m:= {\partial L \over
\partial \dot{x}^{m} } \\
\label{Pr=0} P_{++}^{(\rho )} &:= &  {\partial L \over
\partial \dot{\rho}^{++}  } \approx 0  \; , \qquad
\end{eqnarray}
and
\begin{eqnarray}
\label{Pharm=0} P^{[u]}{}_{(a)}{}^{m}&:= &  {\partial L \over
\partial \dot{u}_m^{(a)}  } \approx 0 \;  \qquad or \qquad P^{[v]}{}_{(\alpha)}{}^{\beta}:=
  {\partial L \over
\partial \dot{V}{}_{\beta}^{(\alpha)}  } \approx 0 \; .
 \end{eqnarray}

The definition of the momenta
\begin{eqnarray}
\label{Pdef} P_{_{{\cal N}}}=  {\partial L \over
\partial \dot{{\cal Z}}^{{\cal N}}}  := \left(P_a \, , \, \pi_\alpha \, ,
\,  P^{(\rho)}_{++}\, , \, P^{[u]}{}_{(a)}{}^{m} \; or \;
P^{[v]}{}_{(\alpha)}{}^{\beta}\right)
\end{eqnarray}
for the configuration space coordinates
\begin{eqnarray}
\label{cZdef} {\cal Z}^{{\cal N}}  := \left( x^a \, , \, \theta^\alpha \, , \,
\rho^{++}\, , \, u_m^{(a)} \; or \; {V}{}_{\beta}^{(\alpha)} \right)
\end{eqnarray}
determines the form of the (equal--proper--time) Poisson brackets $ [ \ldots \; ,  \;
\ldots \}_{_{PB}}$ ($:= (  [ \ldots \; ,  \; \ldots ]_{_{PB}}\,$ , \,  $\{ \ldots \; ,
\; \ldots \}_{_{PB}})$)
\begin{eqnarray}
\label{PBdef} {}[ {\cal Z}^{{\cal N}} \; , \; P_{_{{\cal N}^\prime}}
\}_{_{PB}} := (-)^{{{\cal N}}}  \delta_{{\cal N}^\prime}{}^{{\cal
N}} \; , \qquad [ \ldots \; ,  \; \ldots \}_{_{PB}} :=   {\partial
... \over
\partial {\cal Z}^{{\cal N}} } (-)^{{{\cal N}}}   {\partial ... \over
\partial P_{_{{\cal N}}}} - {\partial ... \over
\partial P_{_{{\cal N}}}}  {\partial ... \over
\partial {\cal Z}^{{\cal N}} }\; .
\end{eqnarray}
The canonical Hamiltonian $H_0$ is defined by
\begin{eqnarray}
\label{H0:=} d\tau H_0 := d{\cal Z}^{{\cal N}}\; P_{_{{\cal N}}}  - d\tau \, L \; .
\end{eqnarray}
Since  the canonical Hamiltonian of the massless superparticle is zero in the weak
sense ({\it i.e.}, when constraints are taken into account \cite{Dirac}), its
Hamiltonian analysis reduces to the analysis of the constraints. Following Dirac
\cite{Dirac}, we shall split the whole set of the constraints into first and second
class ones and will deal with the second class constraints by using Dirac brackets.

To make the analysis more transparent it is convenient deal first with the second class
constraints imposed on the harmonic variables.

\subsection{Dirac brackets in Hamiltonian mechanics on the $SO(1,D-1)$ group manifold}
\label{DBsec}

Eqs. (\ref{harmU=}), (\ref{harmV=}) make manifest that the vector and the spinor
Lorentz harmonics can be expressed through the same parameter $l^{(a)(b)}$. Hence one
can, in principle, use the local $l^{(a)(b)}=-l^{(b)(a)}$ coordinate in the
configurational space (${\cal Z}^{{\cal N}} = ( x^a \, , \, \theta^\alpha \, , \,
\rho^{++}\, , l^{(a)(b)})$ in our case of massless superparticle) and develop the
Hamiltonian mechanics using this variable and its conjugate momentum. This way is,
however, technically involved.

Much more practical is to work with the whole set of Harmonic variables $U$ and/or $V$
and to take Eqs. (\ref{harmUin}), (\ref{harmVin}) into account by passing to the
associated Dirac brackets. (This may be treated as an implicit use of Eqs.
(\ref{harmU=}), (\ref{harmV=}) which, in terms of \cite{Dirac}, would correspond to
explicit solution of the corresponding second class constraints). It is more convenient
to work in terms of vector harmonics; the correpsonding Dirac brackets (as they
actually coincide with the Poisson brackets for $l$) can be then applied to the spinor
harmonics as well.

When the harmonics enter as auxiliary variables, the primary constraints include the
statement of vanishing of all the momentum conjugate to the vector harmonics,
$P_{(a)}{}^m =0$ (Eq. (\ref{Pharm=0})). This set of constraints can be easily split in
a set of 55 constraints $\mathbf{d}_{(a)(b)}:= P_{(a)}{}^m U_{m(b)} - P_{(b)}{}^m
U_{m(a)}$ and the $66$ constraints $\mathbf{K}_{(a)(b)}:= P_{(a)}{}^m U_{m(b)} +
P_{(b)}{}^m U_{m(a)}$. These latter are manifestly second class ones as far as they are
conjugate to the (also second class) $66$ kinematical constraints (\ref{harmUdef}),
\begin{eqnarray}\label{IIccH}
\mathbf{\Xi}^{(a)(b)} := U_m^{(a)}U^{m(b)} - \eta^{(a)(b)}\approx 0\; , \qquad
\mathbf{K}_{(a)(b)}:=
P_{(a)}{}^m U_{m(b)} + P_{(b)}{}^m U_{m(a)} \approx 0 \; , \qquad \\
\label{PBIIccH} {}[\; \mathbf{\Xi}^{(a)(b)} \; , \; \mathbf{K}_{(a^\prime)(b^\prime)}\;
]_{_{PB}} = 4 \delta^{((a) }{}_{(a^\prime)} \delta^{(b))} {}_{(b^\prime)}  +4
\delta^{((a) }{}_{((a^\prime)} \mathbf{\Xi}^{(b))} {}_{(b^\prime))} \approx 4
\delta^{((a) }{}_{(a^\prime)} \delta^{(b))} {}_{(b^\prime)}  \; , \qquad
\end{eqnarray}
while the 55 constraints $\mathbf{d}_{(a)(b)}:= P_{(a)}{}^m U_{m(b)} - P_{(b)}{}^m
U_{m(a)}$ commute with the kinematical constraints $\Xi^{(a)(b)}$,
\begin{eqnarray}\label{[d,Xi]=0}
\mathbf{d}_{(a)(b)}:= P_{(a)}{}^m U_{m(b)} - P_{(b)}{}^m U_{m(a)}\approx 0\; , \qquad
 {}[\; \mathbf{\Xi}^{(a)(b)} \; , \; \mathbf{d}_{(a^\prime)(b^\prime)}\; ]_{_{PB}} = 0\; .  \qquad
\end{eqnarray}
The brackets of these constraints represent the Lorentz group algebra while their
brackets with $\mathbf{K}_{(a)(b)}$ show that these are transformed as symmetric second
rank tensor under the Lorentz group,\footnote{Furthermore, on can see that the Poisson
brackets of two $\mathbf{K}$'s close on $\mathbf{d}_{(a)(b)}$, so that the complete set
of brackets of $\mathbf{K}$ and $\mathbf{d}_{(a)(b)}$ constraints represent
$gl(D,\mathbb{R})$; the $\mathbf{K}_{(a)(b)}$ constraints correspond to the
${GL(D,\mathbb{R})\over SO(1,D-1)}$ coset generators. }
\begin{eqnarray}\label{dd,kk}
 {}[\mathbf{d}_{(a)(b)} \; , \; \mathbf{d}^{(c)(d)} \; ]_{_{PB}} = - 4\delta_{[(a)
}{}^{[(c)} \mathbf{d}_{(b)]}{}^{(d)]}\; ,  \qquad {}[\mathbf{d}_{(a)(b)} \; , \;
\mathbf{K}^{(c)(d)} \; ]_{_{PB}}= - 4\delta_{[(a) }{}^{((c)} \mathbf{K}_{(b)]}{}^{(d))}
 \; .  \qquad
\end{eqnarray}

Hence in the Lorentz harmonics sector of phase space one can define Dirac brackets
\begin{eqnarray}\label{DB-harm}
 {}[\; \ldots \; , \; \ldots \; ]_{_{DB_{harm}}}=[\; \ldots \; , \; \ldots \; ]_{_{PB}} & - {1\over 4}
[\; \ldots \; , \; \mathbf{K}_{(a)(b)} \; ]_{_{PB}} [\; \mathbf{\Xi}^{(a)(b)}\; , \;
\ldots \; ]_{_{PB}} \qquad \nonumber \\ & + {1\over 4} [\; \ldots \; , \;
\mathbf{\Xi}^{(a)(b)} \; ]_{_{PB}} [\; \mathbf{K}_{(a)(b)} \; , \; \ldots \; ]_{_{PB}}
\qquad
\end{eqnarray}
allowing us to use (\ref{harmUdef}) and, moreover, all the $122$
constraints (\ref{IIccH}) in the strong sense,
\begin{eqnarray}\label{IIccH=0}
\mathbf{\Xi}^{(a)(b)} := U_m^{(a)}U^{m(b)} - \eta^{(a)(b)}= 0\; , \qquad
\mathbf{K}_{(a)(b)}:= P_{(a)}{}^m U_{m(b)} + P_{(b)}{}^m U_{m(a)} = 0 \; . \qquad
\end{eqnarray}
Using (\ref{IIccH=0}) one sees that in the phase space sector that involves the
harmonics $U_{m(a)}$ and the `covariant momenta' $\mathbf{d}_{(a)(b)}:= P_{(a)}{}^m
U_{m(b)}  - P_{(b)}{}^m U_{m(a)}$, but not the canonical momenta $P_{(b)}{}^m$
themselves, the above defined Dirac brackets coincide with the Poisson brackets; in
particular (see (\ref{dd,kk}))
\begin{eqnarray}\label{dab-DB=PB}
{}[\; \mathbf{d}_{(a)(b)} \; , \; \ldots \; ]_{_{DB_{harm}}}=[\; \mathbf{d}_{(a)(b)} \;
, \; \ldots \; ]_{_{PB}}\; . \qquad
\end{eqnarray}
This reflects the fact that $\mathbf{d}_{(a)(b)}$ provide a representation of the
Lorentz group generators {\it i.e.} generate a parallel transport ('translations')
along the Lorentz group manifold: $[\mathbf{d}_{(a)(b)} \, , f(U) ]_{_{PB}} =
({\partial\over
\partial l^{(a)(b)}} + \ldots )  f (U(l))$ in terms of explicit parametrization in
(\ref{harmU=}) (and (\ref{harmV=}) for spinorial harmonics, $[\; \mathbf{d}_{(a)(b)} \;
, \; f(V) \; ]_{_{PB}} = (\partial/\partial l^{(a)(b)} +  \ldots ) \;  f (V(l))\;$).
The above described Dirac brackets give a convenient way to represent the Poisson
brackets on the Lorentz $SO(1,D-1)$ group manifold (which can also be formulated in
terms of $l^{(a)(b)} = - l^{(b)(a)}$ and its conjugate momentum).

This gives a reason for not distinguishing notationally these Dirac
brackets ${}[ ... , ...]_{_{DB_{harm}}}$ from the original Poisson
brackets (\ref{PBdef}), denoting them also by ${}[ ... ,
...]_{_{PB}}$ or ${}\{ ... , ...\}_{_{PB}}$ for the case of two
fermionic constraints, and reserve the notation ${}[ ... ,
...]_{_{DB}}$, ${}\{ ... , ...\}_{_{PB}}$ for the Dirac brackets
allowing to resolve {\it all} the second class constraints for the
M0-brane model.

\bigskip

\subsection{Cartan forms and Hamiltonian mechanics on the Lorentz group manifold}
\label{CartanF}

The above Dirac brackets can be also applied \cite{BZ-strH} to
calculations with the spinorial Lorentz harmonics. This is
particularly important because the simple constraints on these
variables, Eqs. (\ref{harmVin}), are reducible, and the irreducible
constraints are not so easy to extract and to deal with. However, a
relatively simple method to obtain  the definite expressions for the
above Dirac brackets and, more generally, to deal with the
derivatives and variations of harmonic variables can be formulated
using just the group--theoretical meaning of the harmonic variables
(see \cite{BZ-strH} and also Appendix for more detail on this {\it
admissible variation technique}).

Using the kinematic constraints (\ref{harmUdef}) (first of Eqs. (\ref{IIccH})) and
(\ref{harmVdef}), one can express the  derivatives of both the vector and the spinor
harmonics through the $55$ Cartan forms,
\begin{eqnarray}
\label{Omab} \Omega^{(a)(b)}:= U^{m(a)}dU_m^{(b)} = - \Omega^{(b)(a)} = \left(\matrix{
0 & - 4 \Omega^{(0)} & \Omega^{++j} \cr
  \,  4 \, \Omega^{(0)} &  0 & \Omega^{--j} \cr
 - \Omega^{++i} & - \; \Omega^{--i} & \Omega^{ij}\;
}\right) \quad \in\quad so(1,10)\;  .
\end{eqnarray}
Indeed, the equation
\begin{eqnarray}
\label{dU=UOm} dU_m^{(a)}= U_{m(b)}\Omega^{(b)(a)}
\end{eqnarray}
is just equivalent to the definition of the Cartan forms, Eq.
(\ref{Omab}), when (\ref{harmUdef}) (or equivalent (\ref{UUT=eta}))
is taken into account. As, according to (\ref{harmVdef}), the
spinorial harmonic matrix $V$ provides the spinoral representation
of  the $Spin(1,D-1)$ element $g$ which correspond to the Lorentz
rotation $U$, its derivative can be expressed through the same
Cartan form $g^{-1}dg= {1\over 2}
\Omega^{(a)(b)}\mathbb{T}_{(a)(b)}$, but with $\mathbb{T}_{(a)(b)}=
{1\over 2}\Gamma_{(a)(b)}$ instead of
$\mathbb{T}_{(a)(b)}{}_{(c)}{}^{(d)}= 2\eta_{(c)[(a)}
\delta_{(b)]}{}^{(d)}$  giving rise to  Eq.  (\ref{Omab}),
\begin{eqnarray}
\label{VdV=UdUG} V^{-1}dV =   {1\over 4} \; \Omega^{(a)(b)}\; \Gamma_{(a)(b)} \quad \in
\; spin(1,10) \; , \qquad \Omega^{(a)(b)}:= U^{m(a)}dU_m^{(b)} \quad  . \qquad
\end{eqnarray}
Eq. (\ref{VdV=UdUG}) can be equivalently written in the form of $dV =   {1\over 4} \;
\Omega^{(a)(b)}\; V\Gamma_{(a)(b)}$. This equation implies, in particular, the
following expression for the differential $dv_\alpha{}_q^{-}$ of the harmonics
$v_\alpha{}_q^{-}$ entering the action (\ref{11DSSP}):
\begin{eqnarray}
\label{dv-q} & dv_q^{-}=  - \Omega^{(0)} v_q^{-} - {1\over 4} \Omega^{ij}
v_p^{-}\gamma_{pq}^{ij} + {1\over 2} \Omega^{--i} \gamma_{qp}^{i}v_p^{+} \quad .
\end{eqnarray}
The particular ($(a)=--$) case of Eq. (\ref{dU=UOm}) gives
\begin{eqnarray}
\label{du--}  & du^{--}_m = - 2u^{--}_m \Omega^{(0)} + u^{i}_m \Omega^{--i} \;  \qquad
 \end{eqnarray}
for the derivative of the only vector harmonics that appear explicitly in the action
(\ref{11DSSP}). Notice that (\ref{dv-q}) and (\ref{du--}) do not contain the Cartan
form $\Omega^{++ i}$, corresponding to the abelian $\mathbb{K}_9$ subgroup (see Eq.
(\ref{K9-def})) of $SO(1,10)$ parametrized by the harmonics. This actually reflects the
$\mathbb{K}_9$ gauge invariance of the action (\ref{11DSSP}), allowing, together with
its manifest $SO(1,1)$ and $SO(9)$ invariance, to identify the relevant harmonics
$u_m^{--}$ and $v_{\alpha q}^{\;\;\, -}$ with the homogeneous coordinates of
$\mathbb{S}^{9}$,  Eq. (\ref{v-inS11}).

\bigskip

When the Hamiltonian formalism for a dynamical system involving harmonic variables is
considered, one can use, as above,  the standard way to define hamiltonian, $H_0=
\partial_\tau u P^{[u]} + ... - L $ or $H_0= \partial_\tau V P^{[v]} + ... - L $,  Eq.
(\ref{H0:=}), and introduce the Dirac brackets (\ref{DB-harm}). Alternatively one can
use Eqs. (\ref{dU=UOm}), (\ref{VdV=UdUG}) in the above expressions for $H_0$, or better
for $d\tau H_0$, and, in such a way, to arrive at the Hamiltonian of the form
 \begin{eqnarray}
\label{H0:=Om(gen)--} d\tau H_0= - {1\over 2} \Omega^{(a)(b)} \mathbf{d}_{(a)(b)}+
\ldots - d\tau L\; , \qquad \mathbf{d}_{(a)(b)}:= u_{(b)}{}^m P_{(a)m}- u_{(a)}{}^m
P_{(b)m} \;
\end{eqnarray}
containing the Cartan form (\ref{Omab}) and the `covariant momentum'
$\mathbf{d}_{(a)(b)}$ (see (\ref{[d,Xi]=0})) instead of $dU$ or $dV$
and its conjugate momentum.

Such a Hamiltonian can be thought of as the one with the kinematical constraints solved
in terms of the independent parameter $l$ ($U=U(l)$, $V=V(l)$, see Eqs. (\ref{harmU=}),
(\ref{harmV=})), but, as we see, one does not need using the explicit form of such a
solution. In particular, to find the Poisson bracket of the 'covariant momentum'
$\mathbf{d}_{(a)(b)}$ with harmonics one can just use the general form of the
Hamiltonian equations $\dot{U}:= [\;  U \; , H_0 \; ]_{_{PB}}$ or $\dot{V}:= [\;  V \;
, H_0 \; ]_{_{PB}}$, and the explicit expression for the Cartan form, (\ref{Omab}) and
(\ref{VdV=UdUG}) for the case of spinor Harmonics. Indeed, for the vector harmonic
$d{U}_m{}^{(a)}:= d\tau [\; U_m{}^{(a)} \; , H_0 \; ]_{_{PB}}= - {1\over 2}
\Omega^{(c)(d)} [\; U_m{}^{(a)} \; , \mathbf{d}_{(c)(d)}\; ]_{_{PB}} = - {1\over 2}
dU_n^{(d)} [\; U_m{}^{(a)} \; , \mathbf{d}_{(c)(d)}\; ]_{_{PB}} U^{n(c)} $ implies
\begin{eqnarray}
\label{[d,U]=} {}[\; \mathbf{d}_{(a)(b)}  \; , \; U_m{}^{(a^\prime)}\;  ]_{_{PB}} = 2
U_{m[(a)}\delta_{(b)]}{}^{(a^\prime)}\; .
\end{eqnarray}
Making the similar calculation with the spinor harmonics, one finds
\begin{eqnarray}
\label{[d,V]=} {}[\; \mathbf{d}_{(a)(b)}  \; , \; V_{\alpha}{}^{(\beta)}\; ]_{_{PB}} =
{1\over 2} V_{\alpha}{}^{(\gamma)}
\Gamma_{(a)(b)}{}_{(\gamma)}{}^{(\beta)}\delta_{(d)]}{}^{(a^\prime)}\; .
\end{eqnarray}
Then, calculating  the Poisson bracket of (\ref{[d,U]=}) and  $\mathbf{d}_{(a)(b)}$,
and using the Jacobi identities for the Poisson brackets we find the first of Eqs.
(\ref{dd,kk})
\begin{eqnarray}
\label{[d,d]=} {}[\mathbf{d}_{(a)(b)} \; , \; \mathbf{d}^{(c)(d)} \; ]_{_{PB}} = -
4\delta_{[(a) }{}^{[(c)} \mathbf{d}_{(b)]}{}^{(d)]}\; ,  \qquad
\end{eqnarray}
which implies that $\mathbf{d}_{(a)(b)}$ are the Lorentz group
generators.

Thus,  using the kinematical constraints (\ref{harmUdef}) and/or
(\ref{harmVdef})  in the strong sense we also can easily construct
the canonical Hamiltonian and the Poisson brackets directly on the
$SO(1,D-1)$ group manifold, thus overcoming the stage of introducing
the Dirac brackets (\ref{DB-harm}) and escaping the use of explicit
parametrization (\ref{harmU=}), (\ref{harmV=}).

\bigskip

\subsection{Canonical Hamiltonian and Poisson/Dirac brackets of the M0--brane model}
\label{PB-DB}

The discussion and  equations of the previous section hold for Hamiltonian mechanics on
any space including Lorentz group $SO(1,D-1)$ or its coset $SO(1,D-1)/H$ as a subspace.
The harmonics used in the twistor--like formulations of super--$p$--branes with $p\geq
1$ \cite{BZ-str,BZ-strH,BZ-p} are homogeneous coordinates of the coset with
$H=SO(1,p)\otimes SO(D-p-1)$. The case of  massless superparticle ($p=0$) is special.
Here the $H=[SO(1,1)\otimes SO(D-2)]\subset\!\!\!\!\!\!\times \; \mathbb{K}_{D-2}$ is
the Borel (maximal compact) subgroup of $SO(1,D-1)$. In this case (as well as in the
string case \cite{BZ-str,BZ-strH})  one uses the $H$--covariant splitting (\ref{Omab})
to arrive at
\begin{eqnarray}
\label{H0:=OmD-L} & d\tau H_0 :=  -  {1\over 2}\Omega^{--i}\mathbf{d}^{++i} - {1\over
2}\Omega^{++i}\mathbf{d}^{--i}-   \Omega^{(0)} \mathbf{d}^{(0)} +  {1\over 2}
\Omega^{ij} \mathbf{d}^{ij} +\qquad \nonumber \\ & + d x^a P_a + d \theta^\alpha
\pi_\alpha + d\rho^{++} P^{(\rho)}_{++}  - d\tau \, L \; . \quad
\end{eqnarray}
Then the  Poisson/Dirac brackets  can be defined by the following
set of non-zero relations (see (\ref{PBdef}))
\begin{eqnarray}
\label{PB=XP} {}[P_{a}\; ,  \; x^{b}]_{_{PB}} = - \delta_{a}{}^{b} \; , \qquad \{
\pi_{\alpha}\; ,  \; \theta^{\beta}\}_{_{PB}} = - \delta_{\alpha}{}^{\beta}\; , \qquad
[P^{(\rho)}_{++} \; ,  \; \rho^{++}]_{_{PB}} = - 1 \; ,  \qquad
\end{eqnarray}
as well as  Eqs. (\ref{[d,U]=}), (\ref{[d,V]=}) and the Lorentz group algebra
(\ref{[d,d]=}) which splits as
\begin{eqnarray}\label{PB=d'd} {}[\mathbf{d}^{++i}\; ,  \; \mathbf{d}^{--j}]_{_{PB}}
= 2\mathbf{d}^{ij} + \mathbf{d}^{(0)}\delta^{ij}\; , \qquad{}[\mathbf{d}^{(0)}\; ,  \;
\mathbf{d}^{\pm\pm i}]_{_{PB}} = \pm 2 \mathbf{d}^{\pm\pm i} \; , \qquad \nonumber
\\ {} [\mathbf{d}^{ij}\; , \; \mathbf{d}^{\pm\pm k}]_{_{PB}} =
2\mathbf{d}^{\pm\pm [i} \delta^{j]k}\; , \qquad  [\mathbf{d}^{ij}\; , \;
\mathbf{d}^{kl}]_{_{PB}} = 2\mathbf{d}^{k[i} \delta^{j]l} - 2\mathbf{d}^{l[i}
\delta^{j]k}\; . \qquad
\end{eqnarray}

 The splitting $\mathbf{d}_{(a)(b)}= (\mathbf{d}^{(0)}\, ,
\mathbf{d}^{\pm\pm j}\, , \mathbf{d}^{ij})$ of the $SO(1,10)$ generators (see
\ref{Omab}) is invariant under $SO(1,1)\otimes SO(9)$ symmetry the generators of which
are represented by $\mathbf{d}^{(0)}\, , \mathbf{d}^{ij}$. The set of remaining
generators $ \mathbf{d}^{++ j}$, $ \mathbf{d}^{-- j}$ can be conventionally split on
two Abelian subsets, one, say $ \mathbf{d}^{-- j}$, representing the $\mathbb{K}_9$
generator, and other, $ \mathbf{d}^{++ j}$, corresponding to the
$SO(1,10)/SO(1,1)\otimes SO(9)]\subset\!\!\!\!\!\!\times \; \mathbb{K}_9$ coset.

The split form of  Eqs. (\ref{[d,U]=}), (\ref{[d,V]=}) include
\begin{eqnarray}
\label{[d,u--]=}  & {}[\mathbf{d}^{(0)}\; , u^{--}_m \; ]_{_{PB}}  = - 2u^{--}_m\; ,
\quad {}[\mathbf{d}^{--i}\; , u^{--}_m \; ]_{_{PB}}  = 0\; , \quad
{}[\mathbf{d}^{++i}\; , u^{--}_m \; ]_{_{PB}}  =2 u^{i}_m \; , \quad \nonumber \\ &
{}[\mathbf{d}^{ij}\; , u^{--}_m \; ]_{_{PB}}  =0\; , \qquad
\\ \label{[d,v-q]=} & {}[\mathbf{d}^{(0)}\; , v_q^{-} \; ]_{_{PB}}  = - v_q^{-}\; , \quad
{}[\mathbf{d}^{--i}\; , v_q^{-} \; ]_{_{PB}}  =  0 \; , \quad {}[\mathbf{d}^{++i}\; ,
v_q^{-} \; ]_{_{PB}} =
 \gamma_{qp}^{i}v_p^{+}\; , \quad \nonumber \\ &  {}[\mathbf{d}^{ij}\; ,
v_q^{-} \; ]_{_{PB}} =
 {1\over 2} v_p^{-}\gamma_{pq}^{ij}\; .  \qquad
\\ \label{[d,v+q]=} & {}[\mathbf{d}^{(0)}\; , v_q^{+} \; ]_{_{PB}}  = \; v_q^{-}\; , \quad
{}[\mathbf{d}^{--i}\; , v_q^{+} \; ]_{_{PB}}  =  \gamma_{qp}^{i}v_p^{-} \; , \quad
{}[\mathbf{d}^{++i}\; , v_q^{+} \; ]_{_{PB}} =  0 \; , \quad \nonumber \\ &
{}[\mathbf{d}^{ij}\; , v_q^{+} \; ]_{_{PB}} =
 {1\over 2} v_p^{+}\gamma_{pq}^{ij}\; ,   \qquad
\end{eqnarray}
and  the relations for the brackets of $\mathbf{d}_{(a)(b)}$ with $u_m^{++}$ and
$u_m^{i}$ vectors, which are not needed in this paper. All these relations can be
collected in
\begin{eqnarray}
\label{PB-d=D} {}[\mathbf{d}^{(a)(b)} , U \}_{_{PB}} := \mathbb{D}^{(a)(b)} U \; ,
\qquad {}[\mathbf{d}^{(a)(b)} , V \}_{_{PB}} := \mathbb{D}^{(a)(b)} V \; , \qquad
\end{eqnarray}
where  $\mathbb{D}_{(a)(b)}=(\mathbb{D}^{\pm\pm i}, \mathbb{D}^{ij}, \mathbb{D}^{(0)})$
are the covariant harmonic derivatives which provide the differential operator
representation for the Lorentz group generators ($\mathbb{D}_{(a)(b)}=\partial/\partial
l^{(a)(b)} + \ldots$ in terms of explicit parametrization) which are defined by the
decomposition of the differential on the Cartan forms (\ref{Omab})\footnote{ The minus
signs in (\ref{H0:=OmD-L}) are chosen to provide the plus sign in (\ref{PB-d=D}).},
\begin{eqnarray}
\label{d=OmD/2} d:=  {1\over 2}  \Omega^{(a)(b)} \mathbb{D}_{(a)(b)}=: \Omega^{(0)}
\mathbb{D}^{(0)} + {1\over 2}\Omega^{++i}\mathbb{D}^{--i}+ {1\over
2}\Omega^{--i}\mathbb{D}^{++i} - {1\over 2}  \Omega^{ij} \mathbb{D}^{ij} \; .
\end{eqnarray}

\bigskip

\subsection{Second class constraints of the D=11 superparticle
model}\label{secIIclass}

\bigskip

With the Poisson/Dirac brackets (\ref{PB=XP})--(\ref{[d,v+q]=}),  the phase space
$(Z^{{\cal N}}, P_{{\cal N}})$ of our superparticle model includes, for the moment, the
$Spin(1,10)$ group manifold, parametrized by harmonics, and the corresponding momentum
space parametrized by the non--commutative generalized momenta $\mathbf{d}_{(a)(b)}$ of
Eqs. (\ref{H0:=Om(gen)--}), (\ref{[d,d]=}), (\ref{PB-d=D}). In all we have
\footnote{Here it is convenient to consider vector harmonics $U_m^{(a)} \in SO(1,10)$
as composites of the spinoral ones, ${V}{}_{\beta}^{(\alpha)} \in Spin(1,10)$, defined
by the gamma--trace parts of Eqs. (\ref{harmVdef}), $U_{m}^{(a)}= {1\over 32} \, tr
V\Gamma^{(a)}V^T\tilde{\Gamma}_m$. }
\begin{eqnarray}
\label{Z,Pdef2} P_{{{\cal N}}}= \left(P_a \, , \, \pi_\alpha \, , \,
P^{(\rho)}_{++}\, , \; \mathbf{d}_{(a)(b)} \right) \; , \qquad {\cal
Z}^{{\cal N}} := \left( x^a \, , \, \theta^\alpha \, , \,
\rho^{++}\, , \; {V}{}_{\beta}^{(\alpha)}\right) \in Spin(1,10)
\end{eqnarray}
This phase space (\ref{Z,Pdef2}) is restricted by the constraints
(\ref{P-rvv}), (\ref{df=}), (\ref{Pr=0}) and
\begin{eqnarray}
\label{d-harm-c}  \mathbf{d}_{(a)(b)} \approx 0 \;  \qquad \Leftrightarrow \qquad
\cases{ \mathbf{d}^{(0)}\approx 0 \; , \quad \mathbf{d}^{ij}\approx 0 \; , \quad
\mathbf{d}^{--i}\approx 0 \; ,  \cr \mathbf{d}^{++i}\approx 0 \;  } \;  \qquad
\end{eqnarray}
for the non--commutative momentum of the $Spin(1,10)$ group valued spinor moving frame
variables  $V\in Spin(1,10)$ [instead of the `original' (\ref{Pharm=0}) for an
apparently unrestricted  $V$ matrices]. The algebra of primary constraints
(\ref{P-rvv}), (\ref{df=}), (\ref{Pr=0}) and (\ref{d-harm-c}) is characterized by the
following nonvanishing brackets
\begin{eqnarray}
\label{(C,C)=} & {} [\Phi_a \; , \; P_{++}^{[\rho]} ]_{_{PB}}= -{1\over 2} u_a^{--} \;
, \quad  [\Phi_a \; , \; \mathbf{d}^{(0)} ]_{_{PB}}= -\rho^{++} u_a^{--} \; , \quad
[\Phi_a \; , \; \mathbf{d}^{++i} ]_{_{PB}}= -\rho^{++} u_a^{i} \; , \quad
\\ \label{(d,d)=}  & {} \{ d_\alpha \; , \; d_{\beta} \}_{_{PB}}= -
2i P\!\!\!/_{\alpha\beta} \; \equiv - 2i \Phi\!\!\!/_{\alpha\beta} -
2i \rho^{++} v_{\alpha}{}^-_q  v_{\beta}{}^-_q \; ,
\end{eqnarray}
and the Lorentz algebra relations (\ref{PB=d'd}). This allows us to
find the following {\it fermionic and bosonic second class
constraints}, the latter split in  mutually conjugate pairs
\begin{eqnarray}
\label{IIcl} & d^+_q:= v^{+\alpha}_q d_\alpha \approx 0 \; , \qquad & \qquad \{ d^+_q
\; , \; d^+_p \}_{_{PB}}= - 2i \rho^{++} \delta_{pq}
 \; , \nonumber \\ & u^{a++}\Phi_a \approx 0 \, , \quad   P_{++}^{[\rho]} \approx 0 \, ,
 \qquad & {}\qquad [u^{a++}\Phi_a \; , \; P_{++}^{[\rho]} \}_{_{PB}}= -1
 \; , \nonumber \\ & u^{a i}\Phi_a \approx 0 \, , \qquad \mathbf{d}^{++j} \approx 0 \, ,
 \qquad  & {}\qquad  [u^{ai}\Phi_a \; , \; \mathbf{d}^{++j}   \}_{_{PB}}=
 - \rho^{++}
 \; . \qquad
\end{eqnarray}
Here $v^{+\alpha}_q$ is an element of the inverse spinor moving
frame  matrix $V^{-1}{}_{(\beta)}^{\; \alpha}= ( v^{+\alpha}_q \; ,
\; v^{-\alpha}_q) \in Spin(1,10)$
which obeys
$v^{+\alpha}_qv_{\alpha q}^{\; +}= 0$ and $ v^{+\alpha}_qv_{\alpha
q}^{\; -}= \delta_{qp}$.  In $D$=11 (as in the other cases when the
charge conjugation matrix exists) this is expressed through the
original spinor harmonics with the help of Eqs. (\ref{harmVdefC}),
 \begin{eqnarray}
\label{V-1=CV} D=11\; : \qquad  v^{\pm \alpha}_q = \pm i C^{\alpha\beta}v_{\beta q}^{\;
\pm}\;
 \end{eqnarray}
(notice that the $D=11$ charge conjugation matrix is imaginary in our `mostly minus'
signature).

Introducing the Dirac brackets
\begin{eqnarray}
\label{DB}
   {} [\ldots \; , \; \ldots ]_{_{DB}}  &=&  [\ldots \; , \; \ldots ]_{_{PB}}
 + [\ldots \; , \; P_{++}^{[\rho]} ]_{_{PB}} \cdot [ (u^{++}P-\rho^{++})
\; , \; \ldots ]_{_{PB}}
 - \qquad \nonumber \\ &&  \qquad - \; [\ldots \; , \; (u^{++}P-\rho^{++}) ]_{_{PB}} \cdot [ P_{++}^{[\rho]} \; , \; \ldots
]_{_{PB}} - \qquad \nonumber \\ &&
 - [\ldots \; , \; u^{j}P ]_{_{PB}} {1\over \rho^{++}} [ \mathbf{d}^{++j}\; , \; \ldots
]_{_{PB}}
 + [\ldots\; , \; \mathbf{d}^{++j} ]_{_{PB}} {1\over \rho^{++}} [ u^{j}P\; , \; \ldots
]_{_{PB}} - \qquad \nonumber \\ &&
 - [\ldots \; , \; d^+_q ]_{_{PB}} {i\over 2\rho^{++}} [ d^+_q\; , \; \ldots
]_{_{PB}} \; ,
\end{eqnarray}
one can treat the second class constraints as the strong relations
\begin{eqnarray}
\label{strongIIcl} & d^+_q:= v^{+\alpha}_q d_\alpha = 0 \; ; \qquad
\rho^{++}= u^{a++}P_a  \, , \quad   P_{++}^{[\rho]} = 0 \, ;
 \qquad  u^{a i}P_a = 0 \, , \quad \mathbf{d}^{++j} = 0
 \; . \qquad
\end{eqnarray}

\subsection{First class constraints and their (nonlinear) algebra} \label{secIclass}

The remaining constraints are
\begin{eqnarray}
\label{pre-Icl} & d^-_q:= v^{-\alpha}_q d_\alpha \approx 0 \; ,
\qquad  u^{a--}\Phi_a = u^{a--}P_a =: P^{--} \approx 0 \, ,  \\
\label{pre-IclH} & \mathbf{d}^{ij} \approx 0 \, ,
 \qquad \mathbf{d}^{(0)} \approx 0  \, ,
 \qquad \mathbf{d}^{--i} \approx 0 \; . \qquad
\end{eqnarray}
They give rise to the first class constraints. Namely, the Dirac
bracket algebra of the constraints (\ref{pre-Icl}), (\ref{pre-IclH})
is closed and contains the following nonvanishing brackets
\begin{eqnarray}
\label{(IH,IH)=DB} {} & [\mathbf{d}^{ij} , \; \mathbf{d}^{kl}]_{_{DB}} =
4\mathbf{d}^{[k|[i} \delta^{j]|l]} \; ,  & \;
 [\mathbf{d}^{ij} , \; \mathbf{d}^{--k}]_{_{DB}} =
2\mathbf{d}^{-- [i} \delta^{j]k}\; , \qquad  [\mathbf{d}^{(0)} , \;
\mathbf{d}^{\pm\pm i}\}_{_{DB}} = \pm 2 \mathbf{d}^{\pm\pm i}  , \qquad  \\
\label{d--d--DB} && {} \fbox{$[\mathbf{d}^{--i} \; , \; \mathbf{d}^{--j} ]_{_{DB}}  =
  {i\over 2P^{\!^{++}}} \;  d^-_q \gamma^{ij}_{qp}  d^-_p $} \; ,  \qquad \\
\label{(IH,I)=DB} {} & [\mathbf{d}^{ij} \; , \; d^-_p]_{_{DB}} = - {1\over 2}
\gamma^{ij}_{pq} d_q^-  \; , & \; {} [\mathbf{d}^{(0)} \; , \; d^-_p]_{_{DB}} = -
 d_q^-  \; ,  \qquad [\mathbf{d}^{(0)} \; , \; P^{--}]_{_{DB}} = -2
 P^{--}
  \; ,  \qquad
\\
\label{(I,I)=DB} && {} \fbox{$ \; \{ d_q^- \; , \; d^-_{p} \} _{_{DB}}= - 2i
\delta_{qp} P^{--} \;$} \;  .  \qquad {}
\end{eqnarray}
 Notice that the right hand side of Eq.
(\ref{d--d--DB}) includes the product of the two fermionic first class constraint and,
hence, implies moving outside the Lie algebra (to the enveloping algebra) \footnote{
One may also think of this as an analogy of the very well known phenomenon of the
non--commutativity of the bosonic spacetime coordinates of the superparticle which
appears in standard formulation \cite{Casalbuoni}  after transition to the Dirac
brackets for the second class constraints; see also the second reference in
\cite{A+L82}. There the Dirac brackets of two bosonic coordinates are proportional to
the product of two Grassmann coordinates \cite{Casalbuoni,A+L82}.  In four dimensions
such a noncommutativity is overcame by passing to the so called chiral basis of $D=4$
superspace the imaginary part of the bosonic coordinate of which is given by the
Grassmann coordinates bilinear. The use of the Gupta-Bleuler technique
\cite{Casalbuoni,JdA+L88} also helps. The appearance of a nonlinear algebra of
constraints was also observed for the twistor--like formulation of $D$=4 null
superstring and null--supermembranes in \cite{BZnull}. Notice finally that among the
`nonlinear algebras', the most popular are the W--algebra intensively studied some
years ago (see {\it e.g.} \cite{SKrASo} and reference therein). }. If this term were
absent, one would state that the first class constraints (\ref{pre-IclH}) generated $H=
SO(1,1)\otimes SO(9)]\subset\!\!\!\!\!\!\times K_9$ group symmetry, and the whole gauge
symmetry would be described by its semidirect product (see (\ref{(IH,I)=DB}))  $H
\subset\!\!\!\!\!\!\times \Sigma^{(1|16)}$ with the $d=1, N=16$ supersymmetry group
$\Sigma^{(1|16)}$ of the $\kappa$--symmetry and $b$--symmetry, Eqs. (\ref{pre-Icl}),
(\ref{(I,I)=DB}). Then the actual algebra of Eqs. (\ref{(IH,IH)=DB}), (\ref{d--d--DB}),
(\ref{(IH,I)=DB}), (\ref{(I,I)=DB}) is a {\it `generalized W--deformation'}  of the Lie
superalgebra of this semidirect product $[ SO(1,1)\otimes
SO(9)]\subset\!\!\!\!\!\!\times K_9]\subset\!\!\!\!\!\!\times \Sigma^{(1|16)}$. The
role of the constant parameter for the standard deformation here is taken by the
function ${1\over P^{++}}$ (hence the name {\it generalized} for this `W-deformation').
However,  although  momentum $P^{++}= u^{a ++}P_a$  is a dynamical variable, it has
vanishing Dirac brackets with all the first class constraints.

One may guess that the complete BRST charge $\mathbb{Q}$  for the algebra of the first
class constraints (\ref{(I,I)=DB}) is quite complicated and its use is not too
practical. Following the pragmatic spirit of the pure spinor approach
\cite{NB-pure,nonmNB}, it is tempting to take care of the constraints corresponding to
the (deformed) $[SO(1,1)\otimes SO(9)]\subset\!\!\!\!\!\!\times K_9$ part of the gauge
symmetries in a different manner, by imposing them as conditions on the wavefunctions
in quantum theory, and to leave with a short and fine BRST charge corresponding to the
supersymmetry algebra (\ref{(I,I)=DB}) of the $\kappa$--symmetry and the $b$--symmetry
generators.

However, the appearance of the deformation given by the product of the fermionic first
class constraints in the  {\it r.h.s.} of Eq. (\ref{d--d--DB}) might produce doubts on
the  consistency of such a prescription. Indeed, imposing, for instance, the deformed
(now non--Abelian) $K_9$ constraint $\mathbf{d}^{--i}$ as a condition on the wave
function in quantum theory, $\widehat{\mathbf{d}}{}^{--i}\Phi=0$, one should also
impose by consistency the condition, $[\widehat{\mathbf{d}}{}^{--i},
\widehat{\mathbf{d}}{}^{--j}]\Phi=0$ which implies  $\gamma^{ij}_{qp}\, \widehat{d}^-_q
\widehat{d}^-_p\Phi=0$.

To clarify the situation with the BRST quantization of the nonlinear
algebra (\ref{(IH,IH)=DB})--(\ref{(I,I)=DB}) and its possible
simplification we begin with studying the BRST charge
$\mathbb{Q}^\prime$ corresponding to the subalgebra of $\kappa$--,
$b$-- and $K_9$--symmetry generators, $d_q^-$, $P^{--}$ and
$\mathbf{d}^{--i}$.

\subsection{BRST charge for a nonlinear sub(super)algebra of the first class constraints} \label{BRST-min}
\label{Qprime}

The sub--superalgebra of the  $\kappa$--, $b$--  and the deformed
$K_9$--symmetry generators, $d_q^-$, $P^{--}$ and $\mathbf{d}^{--i}$
is described by Eqs. (\ref{d--d--DB}) and (\ref{(I,I)=DB}) plus
vanishing brackets for the rest,
\begin{eqnarray}
\label{didj,dqdq} && {} [\mathbf{d}^{--i} \; , \; \mathbf{d}^{--j}
]_{_{DB}}  =
  {i\over 2P^{\!^{++}}} \;  d^-_q \gamma^{ij}_{qp}  d^-_p  \quad (a) \; ,  \qquad
\qquad   {} \{ d_q^- \; , \; d^-_{p} \} _{_{DB}}= - 2i \delta_{qp}
P^{--} \quad (b)
 \; . \qquad
\end{eqnarray}
It is obtained  from (\ref{(IH,IH)=DB})-- (\ref{(I,I)=DB}) by
setting the generators of $SO(9)\otimes SO(1,1)$ equal to zero,
$\mathbf{d}^{ij}=0$ and $\mathbf{d}^{(0)}=0$. Notice that, when
acting on the space of $SO(9)\otimes SO(1,1)$ invariant functions,
the full BRST charge $\mathbb{Q}$ of our $D=11$ superparticle
reduces to the BRST charge of the algebra (\ref{didj,dqdq}). In the
quantum theory such an algebra reduction can be realized by imposing
$\mathbf{d}^{ij}$ and $\mathbf{d}^{(0)}$ as conditions on the state
vectors $\widehat{\mathbf{d}}^{ij}\; \Phi =0$ and
$\widehat{\mathbf{d}}^{(0)}\;\Phi=0$. This specifies the
wavefunction dependence on the harmonics making it  a function on
the non--compact coset $SO(1,10)/[SO(9)\times SO(1,1)]$ (dependence
on $l^{\pm\pm i}$ parameters only in the case of explicit
parametrization (\ref{harmU=}), (\ref{harmV=})).

We denote the  BRST charge corresponding to the non--linear
superalgebra (\ref{didj,dqdq}) by $\mathbb{Q}^{\prime}$ which
reflects the fact that it gives only a part of the full BRST charge
describing the complete gauge symmetry algebra (\ref{(IH,IH)=DB})--
(\ref{(I,I)=DB}) of the M0--brane in spinor moving frame
formulation. The master equation
\begin{eqnarray}
   \label{QmQm=0}
    {}  \{ \; \mathbb{Q}^{\prime}\; ,  \; \mathbb{Q}^{\prime}\;\}_{_{DB}} =0  \;
\end{eqnarray}
has the solution
\begin{eqnarray}
   \label{Qmin}
 \mathbb{Q}^{\prime}&=& {\lambda}^+_q d_q^{-} + c^{++} P^{--}  +
 c^{++j}\mathbf{d}^{--j}     - \; i
 \lambda^+_q\lambda^+_q \pi^{[c]}_{++}\; + {i\over 2P^{\!^{++}}} c^{++i}c^{++j}
 d_q^{-}\gamma^{ij}_{qp} P^{-[\lambda]}_p + \qquad \nonumber \\  &&   +
 {1\over P^{\!^{++}}} c^{++i}c^{++j} \lambda_q^{+}\gamma^{ij}_{qp} P^{-[\lambda]}_p
\pi^{[c]}_{++}
 - {i\over 4(P^{\!^{++}})^2} c^{++i}c^{++j}c^{++k}c^{++l}P^{-[\lambda]}_q\gamma^{ijkl}_{qp} P^{-[\lambda]}_p
  \pi^{[c]}_{++} \; .
   \qquad
\end{eqnarray}
Here ${\lambda}^+_q$ is the bosonic ghost for the fermionic
$\kappa$--symmetry gauge transformations, $c^{++}$ and $c^{++j}$ are
the fermionic ghosts for the bosonic $b$--symmetry and deformed
$\mathbb{K}_9$ symmetry transformations, and $P^{-[\lambda]}_q$,
$\pi^{[c]}_{++}$ are the (bosonic and fermionic) ghost momenta
conjugate to ${\lambda}^+_q$ and $c^{++}$,
\begin{eqnarray}
\label{ghostDB} && {} [{\lambda}^+_q  \; , \; P^{-[\lambda]}_p
]_{_{DB}} = \delta_{qp}\; , \qquad {} \{ c^{++} \; , \;
\pi^{[c]}_{++} \} _{_{DB}}= -1\; , \qquad {} \{ c^{++i} \; , \;
\pi^{[c]}_{++ j} \} _{_{DB}}= - \delta^i_{j}
 \; . \qquad
\end{eqnarray}
Notice that the fermionic ghost momentum $\pi^{[c]}_{++ j}$
conjugate to $c^{++ j}$ does not enter $\mathbb{Q}^\prime$
(\ref{Qmin}).

The $\mathbb{Q}^\prime$ of Eq. (\ref{Qmin}) is the third rank BRST charge in the sense
that  the series stops on the third degree in the ghost momenta $P^{-[\lambda]}_p$,
$\pi^{[c]}_{++}$. Technically, the decomposition stops due to nilpotency of
$\pi^{[c]}_{++}$. The nilpotency of the BRST charge (\ref{Qmin}) is preserved in the
quantum theory, $(\mathbb{Q}^{\prime})^2=0$, as far as no products of noncommuting
operators (like {\it e.g.} $\lambda^+_q P^{-[\lambda]}_q$) appear in the calculation of
$(\mathbb{Q}^{\prime})^2$.

\subsection{The further reduced BRST charge $\mathbb{Q}^{susy}$}
 \label{secQsusy}

The (already restricted) BRST charge  (\ref{Qmin}) is (still) too much complicated to
discuss it as a counterpart of (or as an alternative to) the Berkovits pure spinor BRST
charge. A (further) reduction looks necessary. To this end let us notice that
$\mathbb{Q}^{\prime}$ of Eq.  (\ref{Qmin}) can be presented as a sum
\begin{eqnarray}\label{Q'=}
\mathbb{Q}^{\prime}&=& \mathbb{Q}^{susy} +
c^{++j}\widetilde{\mathbf{d}}{}^{--j}\; ,
\end{eqnarray}
of the much simpler operator
\begin{eqnarray}\label{Qbrst1}
\fbox{$\; \mathbb{Q}^{susy}= \lambda^+_q  \; d_q^{-} + c^{++} \;
P^{--} \; - \; i
 \lambda^+_q\lambda^+_q \pi^{[c]}_{++}\;$}\;  , \qquad
{} \{ \mathbb{Q}^{susy} \; , \; \mathbb{Q}^{susy} \}_{_{DB}} = 0 \;
,
 \qquad
\end{eqnarray}
and the term containing the $c^{++j}$ ghost fields. The operator  (\ref{Qbrst1}) can be
identified as BRST charge corresponding to the $d=1$, $N=16$ supersymmetry algebra
\begin{eqnarray}\label{16+1al} & {} \{ d_q^{-} \; , \; d_p^-
\}_{_{DB}} = -2i P^{--} \; , \qquad [ P^{--} \; , \; d_p^- ]_{_{DB}}
= 0 \; , \qquad [ P^{--} \; , \; P^{--} ]_{_{DB}} \equiv 0 \; .
 \qquad
\end{eqnarray}
of  the $\kappa$-- and $b$--symmetry generators (\ref{16+1al}). The
second term in (\ref{Q'=}), $c^{++j}\widetilde{\mathbf{d}}{}^{--j}$,
contains the deformed $K_9$ generator modified by additional ghost
contributions,
\begin{eqnarray}\label{td--j:=}
\widetilde{\mathbf{d}}{}^{--i} & = {\mathbf{d}}{}^{--i}
 + {i\over 2P^{\!^{++}}} c^{++j}
 d_q^{-}\gamma^{ij}_{qp} P^{-[\lambda]}_p   +
 {1\over P^{\!^{++}}} c^{++j} \lambda_q^{+}\gamma^{ij}_{qp} P^{-[\lambda]}_p
\pi^{[c]}_{++} - \qquad \nonumber \\ &
 - {i\over 4(P^{\!^{++}})^2} c^{++j}c^{++k}c^{++l}P^{-[\lambda]}_q\gamma^{ijkl}_{qp} P^{-[\lambda]}_p
  \pi^{[c]}_{++} \; .
   \qquad
\end{eqnarray}
The `nilpotency' of the $\mathbb{Q}^{susy}$(\ref{Qbrst1}) (${} \{ \mathbb{Q}^{susy} \;
, \; \mathbb{Q}^{susy} \}_{_{DB}} = 0$) guaranties the consistency of the reduction of
the $\mathbb{Q}^{\prime}$--cohomology problem to the $\mathbb{Q}^{susy}$--cohomology.
For the classical BRST charge such a reduction can be reached  just by setting the
$K_9$ ghost equal to zero, $c^{++j}=0$. In classical mechanics one can consider this
reduction as a result of the gauge fixing, {\it e.g.} in the explicit parametrization
(\ref{harmU=}), (\ref{harmV=}) by setting $l^{++i}=0$ and (as $l^{ij}=l^{(0)}=0$ can be
fixed by $SO(1,1)\otimes SO(9)$ transformations) expressing all the harmonics in terms
of nine parameters $l^{--i}$ (related to the projective coordinates of the $S^9$
sphere) as in Eqs. (\ref{U=l--}), (\ref{V=l--}).

Although technical, the question of how to realize a counterpart of such a classical
gauge fixing in quantum description looks quite interesting. The problem is whether in
this way one arrives just at scalar functions on $S^9= SO(1,10)/[[SO(1,1)\otimes
SO(9)]\subset\!\!\!\!\!\!\times K_9]$, or the interplay of the $v_q^+$ (or $u_m^{++}$,
$u_m^i$) harmonics and the $K_9$ ghost $c^{++j}$  may result in wavefunctions
transforming nontrivially under $SO(1,1)\otimes SO(9)$ (a counterpart of the effect of
the $D$=4 helicity appearance in the quantization of $D$=4 superparticle, see
\cite{B90} and refs. herein). Such an interplay could appear, {\it e.g.} when one
imposes the quantum counterpart  of the deformed $K_9$ constraints modified by ghost
contribution (\ref{td--j:=}) on the wavefunctions. However, this interesting problem is
out of the score of the present paper devoted to a search for the origin and geometric
meaning of the Berkovits approach in the frame of spinor moving frame formulation of
(presently) M0--brane.

Thus, let us  accept, following the pragmatic spirit of the pure spinor approach
\cite{NB-pure}, the simple prescription of the reduction of the first class constraint
Dirac brackets algebra down to the $d=1$ $N=16$  supersymmetry algebra of
$\kappa$--symmetry and $b$--symmetry, Eq. (\ref{(I,I)=DB}) (taking care of other
constraints in a different manner), which implies the reduction of
$\mathbb{Q}^{\prime}$ to the much simpler $\mathbb{Q}^{susy}$, and let us turn to the
study of the  cohomology problem for   the BRST charge $\mathbb{Q}^{susy}$
(\ref{Qbrst1}).

\section{BRST quantization of the D=11 superparticle. Cohomology of $\mathbb{Q}^{susy}$
and the origin of the complexity of the Berkovits approach }\label{QsusyCoH}
 \setcounter{equation}0

\subsection{Quantum BRST charge $\mathbb{Q}^{susy}$}

It is practical, omitting the overall  $\pm i$ factor, to write
the quantum BRST charge obtained from (\ref{Qbrst1}) as
\begin{eqnarray}\label{Qsusy}
\mathbb{Q}^{susy}= \lambda^+_q  \; D_q^{-} + i c^{++} \partial_{++} -
 \lambda^+_q\lambda^+_q {\partial\over \partial c^{++}}\; , \qquad
{} \{ \mathbb{Q}^{susy} \; , \; \mathbb{Q}^{susy} \} = 0 \; , \qquad
 \qquad
\end{eqnarray}
where the quantum operators $D^-_q$ and $\partial_{++}$, associated with $d^-_q$ and
$P_{++}$, obey the $d=1, n=16$ supersymmetry algebra ({\it cf.} (\ref{(I,I)=DB}))
\begin{eqnarray}\label{DD=d--}
 {} \{ D_p^{-} , D_q^{-}  \} = 2i \delta_{qp} \partial_{++} \; , \qquad [
 \partial_{++}\, , D_p^{-}]=0 \; .
\end{eqnarray}

The quantum BRST operator $\mathbb{Q}^{susy}$ (\ref{Qsusy}) should act on the space of
wavefunctions that depend on the physical (gauge invariant) variables and on a number
of variables which transform nontrivially under the action of generators
$\partial_{++}$, $D_q^{-}$ (in general case, the  variables of a model cannot be split
covariantly on gauge invariant and pure gauge ones, but for our model this is actually
possible, see Sec. 5). It is convenient to use a realization of $\partial_{++}$,
$D_q^{-}$ as differential operators on the $1+16$ dimensional superspace $W^{(1|16)}$
of coordinates $(x^{++},\theta^+_q)$,
\begin{eqnarray}\label{D-q=}
 D_q^{-} = \partial_{+q}  + i \theta^+_q  \partial_{++}\; , \qquad \partial_{++} :=
 {\partial \over \partial x^{++}}, \qquad \partial_{+q} :=
 {\partial \over \partial \theta^+_q} \; .
\end{eqnarray}
These variables have straightforward counterparts in the covariant light--cone basis,
$\theta^+_q= \theta^\alpha v_{\alpha}{}^+_q$ and $x^{++}=x^{m} u_{m}^{++}$ (see
\cite{Sok,GHT93} and Sec. 5). The other `physical' variables, on which the
wavefunctions should also depend, can be related to other coordinates of this basis,
including $x^{--}=x^{m} u_{m}^{--}$ and $\theta^-_q= \theta^\alpha v_{\alpha}{}^-_q$
and the harmonics $v_{\alpha}{}^-_q$ parametrizing $S^9$ (and carrying $9$ of $10$
degrees of freedom of the light--like momentum). However, to study the cohomology of
the BRST operator (\ref{Qsusy}) the dependence on these latter coordinates is
inessential and, in this section, we will use the notation $\Phi = \Phi ( \lambda^+_q
\, , c^{++} \, ; \, x^{++} , \theta^{+}_q \; , ...)$  or $\Phi (c^{++} , \lambda^+_q\,
... )$ to emphasize the essential dependence of our wavefunctions.

The Grassmann odd $c^{++}\;$ variable, $\;c^{++}c^{++}=0$, and the bosonic variables
$\lambda^+_q$ in (\ref{Qsusy}) are  ghosts for the bosonic and 16 fermionic first class
constraints represented by the differential operators $\partial_{++}$ and $D^-_q$.
Their ghost numbers are $1$, and this also fixes  the ghost number of the BRST charge
to be one,
\begin{eqnarray}\label{ghN}
gh_\# (\lambda^+_q)=1 \; , \qquad gh_\# (c^{++})=1 \; , \qquad  gh_\# (
\mathbb{Q}^{susy})=1\; .
\end{eqnarray}
The cohomology problem has to be solved for functions with
definite ghost number $g := gh_\# (\Phi)$. Let us begin, however,
with some general observations for which the ghost number fixing
is not relevant.

\subsubsection{The nontrivial cohomology of $\mathbb{Q}^{susy}$
is located at $\lambda^+_q\lambda^+_q=0$}

BRST cohomology is determined by wavefunctions $\Phi$ which are BRST-closed,
$\mathbb{Q}^{susy}\Phi=0\;$, but not BRST-exact. They are defined modulo the BRST
transformations {\it i.e.} modulo BRST-exact wavefunctions $\mathbb{Q}^{susy}\chi$,
where $\chi$ is an arbitrary function of the same configuration space variables and of
ghost number $gh_\# (\chi)= gh_\#(\Phi)-1$,
\begin{eqnarray}
\label{Qcoh=def}
\mathbb{Q}^{susy}\Phi=0 \; , \qquad  \Phi \sim \Phi^\prime = \Phi +
\mathbb{Q}^{susy}\chi \; , \qquad gh_\# (\chi)= gh_\# (\Phi) - 1\; . \qquad
\end{eqnarray}
Decomposing the wave function $\Phi = \Phi (c^{++} , \lambda^+_q \,
; \, x^{++} , \theta^{+}_q \; , ...)$ in power series of the
Grassmann odd ghost $c^{++}$,
\begin{eqnarray}\label{Phi=Phi(c)}
 \Phi &=& \Phi_0 + c^{++} \Phi_{++}  \qquad  \\ \nonumber \qquad &=&
\Phi_0(\lambda^+_q \, ; \, x^{++} , \theta^{+}_q \, ;\ldots ) + c^{++}
\Phi_{++}(\lambda^+_q \, ; \, x^{++} , \theta^{+}_q \, ;\ldots ) \; ,
\end{eqnarray}
 one finds that
 $\mathbb{Q}^{susy} \Phi =0$ for the superfield (\ref{Phi=Phi(c)})
 implies for its components
\begin{eqnarray}\label{QPhi=0}
 \lambda^+_q D^-_q \Phi_0 = \lambda^+_q\lambda^+_q \Psi_{++}\quad (a) \; ,
  \qquad \lambda^+_q D^-_q \Psi_{++} = i \partial_{++}\Phi_0 \quad (b)\; .
  \qquad
\end{eqnarray}
Using a similar decomposition for the  arbitrary superfield in
(\ref{Qcoh=def}), $\chi= \chi_0 + c^{++} K_{++}$, one finds for the
BRST transformations,
\begin{eqnarray}\label{Phi=Phi+Qchi}
   \Phi \mapsto \Phi^\prime = \Phi + \mathbb{Q}^{susy} \chi \quad \Rightarrow \quad
 \cases{\Phi_0  \mapsto  \Phi_0^\prime = \Phi_0 +  \lambda^+_q D^-_q \chi_0 -
 \lambda^+_q \lambda^+_q  K_{++} \;\;\; (a) \; ,
 \qquad \cr \Psi_{++} \mapsto \Psi_{++}^\prime = \Psi_{++} + i
 \partial_{++}\chi_0 + \lambda^+_q D^-_q K_{++} \;\;\; (b) \;{}
  } \; . \qquad
\end{eqnarray}

If one assumes that the spinorial bosonic ghost $\lambda^+_q$ is non-zero, or,
equivalently,  that the square $\lambda^+_q\lambda^+_q\not= 0$, then one can use Eq.
(\ref{QPhi=0}a) to express the fermionic component of the wave function in terms of the
bosonic one, $ \Psi_{++}= \lambda^+_q D^-_q\Phi_0/ \lambda^+_p \lambda^+_p$. Then one
can also chose the second bosonic component $K_{++}$ of the parameter superfield
$\chi=\chi_0 + c^{++} K_{++}$ to be  $K_{++}= {1\over \lambda^+_p\lambda^+_p} (\Phi_0 +
\lambda^+_q D^-_q\chi_0)$ and arrive at $\Phi_0^\prime =0$ in (\ref{Phi=Phi+Qchi}a).
Thus, if the ghost variables $\lambda^+_q$ parametrize $\mathbb{R}^{16}- \{ 0\}$,
 $\lambda^+_q\lambda^+_q\not= 0$ and
the BRST cohomology of $\mathbb{Q}^{susy}$ is necessarily trivial:
all the BRST--closed states are BRST-exact.

Hence,  if $\mathbb{Q}^{susy}$ has to admit non-trivial closed states, they must have a
representation by wavefunctions with support on $\lambda^+_q\lambda^+_q\not= 0$. In
other words, the closed non-exact wavefunctions representing non-trivial cohomology
must be of the form $\Phi \propto \delta (\lambda^+_q\lambda^+_q)$ plus a possible BRST
trivial contribution.

\bigskip

\subsection{Cohomologies at vanishing bosonic ghost }
\label{CohQ-2}

Thus wavefunctions  describing the non-trivial cohomology of
$\mathbb{Q}^{susy}$, if exists, must have representation by closed
non-exact wavefunctions of the form $\Phi= \delta
(\lambda^+_q\lambda^+_q)\; \Phi^{++}$, where $\Phi^{++}= \Phi^{++}+
c^{++}\Psi^{0}$ has ghost number two units more than $\Phi$, $\;
gh_{\#}(\Phi^{++})\, = \, gh_{\#}(\Phi^{0}) + 2\;$. But there is a
difficulty with studying these states: since the bosonic ghosts
$\lambda^+_q$ are real, $\lambda^+_q\lambda^+_q=0$ implies
$\lambda^+_q=0$. Thus, since ${Q}^{susy}$ includes $\lambda^+_q$ in
an essential manner, it is necessary to make a `regularization'
allowing us to consider, at the intermediate stages, a nonvanishing
$\lambda^+_q$ which nevertheless satisfies
$\lambda^+_q\lambda^+_q=0$.

This is possible if we {\it allow $\lambda^+_q$ to be complex} ({\it cf.} with the pure
spinors by Berkovits \cite{NB-pure}),
\begin{eqnarray}\label{ll-cl}
\lambda^+_q\mapsto \tilde{\lambda}^+_q\; : \quad \fbox{$\;
\tilde{\lambda}^+_q\tilde{\lambda}^+_q=0 \; , \quad (\tilde{\lambda}^+_q)^*\not=
\tilde{\lambda}_q \qquad \Rightarrow \qquad  \tilde{\lambda}^+_q\not= 0\; $ is possible
}\; . \qquad
\end{eqnarray}
A suggestive form of the general solution of
$\tilde{\lambda}^+_q\tilde{\lambda}^+_q=0$ is
\begin{eqnarray}\label{l-null}
\tilde{\lambda}^+_q= \epsilon^+ \; (n_q +i m_q)\; ,  \qquad
\vec{n}^2:=n_q n_q=1\; , \quad \vec{m}^2:=m_qm_q=1 \; , \qquad
\vec{n}\vec{m}= n_qm_q= 0 \; ,
\end{eqnarray}
where $n_q$ and $m_q$ are two real mutually orthogonal unit $SO(16)$
vectors ($SO(9)$ spinors) and $\epsilon^+$ is a real number. The
only real representative of the family of complex $SO(9)$ spinors
$\tilde{\lambda}^+_q$ in (\ref{l-null}) is $\tilde{\lambda}^+_q=0$;
this corresponds to setting the `regularization parameter'
$\epsilon^+$ equal to zero.

The `regularized' BRST charge is thus complex. It contains the
complex ghost $\tilde{\lambda}^+_q$ rather than the real
${\lambda}^+_q$ in (\ref{Qsusy}), but does not contain
$(\tilde{\lambda}^+_q)^*$. It acts on the space of wavefunctions
depending, among other configuration space variables, on the
complex $\tilde{\lambda}^+_q$. Since the discussion of the
previous section is not affected by above complexification
${\lambda}^+_q\mapsto \tilde{\lambda}^+_q$, we conclude that the
non-trivial cohomology states of the complexified BRST charge are
wavefunctions of the form
\begin{eqnarray}\label{Phi(reg)}
 \Phi= \delta
(\tilde{\lambda}^+_q\tilde{\lambda}^+_q)\;
\Phi^{++}(\tilde{\lambda}^+_q \, , \, c^{++}\,  ; \; x^{++} ,
\theta^{+}_q \, , \; \ldots   )\; .
\end{eqnarray}
As the BRST charge $\mathbb{Q}^{susy}$ does not contain any
derivative with respect to the bosonic ghost ${\lambda}^+_q$, its
regularization acts on the $\Phi^{++}$ part of the function $ \Phi$
in  (\ref{Phi(reg)}) only. Namely, one finds
\begin{eqnarray}\label{QsusyPSI=}
\mathbb{Q}^{susy}\vert_{_{{\lambda}^+_p\mapsto \tilde{\lambda}^+_p}}
\;  \delta (\tilde{\lambda}^+_q\tilde{\lambda}^+_q)\;
\Phi^{++}(\tilde{\lambda}^+_q\; \, , \, c^{++}\, ; \ldots ) =
 \delta
(\tilde{\lambda}^+_q\tilde{\lambda}^+_q)\; \tilde{Q}^{susy}
\Phi^{++}(\tilde{\lambda}^+_q\; \, , \, c^{++}\,  ; \ldots ) \; ,
\qquad
\end{eqnarray}
where we introduced the non-Hermitian  BRST charge ({\it cf.}
(\ref{Qsusy}))
\begin{eqnarray}\label{tQsusy}
\fbox{$\; \tilde{Q}^{susy}= \tilde{\lambda}^+_q  \; D_q^{-} + i c^{++}
\partial_{++} \qquad \; , \qquad
 \tilde{\lambda}^+_q \tilde{\lambda}^+_q  = 0 \; $} \; ,  \qquad
\tilde{Q}^{susy}=\mathbb{Q}^{susy}\vert_{{\lambda}^+_q\mapsto
\tilde{\lambda}^+_q\; : \;
\tilde{\lambda}^+_q\tilde{\lambda}^+_q=0}\;   , \qquad
\end{eqnarray}
which can be used to reformulate the regularized cohomology problem. Note that, once we
have  concluded that the nontrivial cohomology of $\mathbb{Q}^{susy}$ is determined by
wavefunctions of the form (\ref{Phi(reg)}), we can reduce the nontrivial cohomology
search to the set of such functions, restricting as well the arbitrary superfields
$\chi$ of the BRST transformations (\ref{Phi=Phi+Qchi}) to have the form $\chi = \delta
(\tilde{\lambda}^+_q \tilde{\lambda}^+_q) \chi^{++}$.

Then,   the regularized cohomology problem for the complexified BRST operator
($\mathbb{Q}^{susy}$ of (\ref{Qsusy}) now depending on the complexified bosonic ghost
$\tilde{\lambda}^+_q$), reduces to the search for {\it $\tilde{\lambda}^+_q=0$ `value'}
of the functions describing non-trivial cohomologies of the $\tilde{Q}^{susy}$ operator
 in Eq. (\ref{tQsusy}),
\begin{eqnarray}\label{CHtQsusy}
\tilde{Q}^{susy} \Phi^{++} =0 \; , \qquad \Phi^{++} \sim \Phi^{++\,\prime }= \Phi^{++}
+ \tilde{Q}^{susy}\chi^{++}\;  . \qquad
\end{eqnarray}

This problem (\ref{CHtQsusy}) can be reformulated in terms of
components $\Phi_0^{++}$ and $\Psi^{(0)}$ of the wavefunction
superfield  $\Phi^{++} = \Phi_0^{++} + c^{++} \Psi^{(0)}$ giving
rise to the following equations
\begin{eqnarray}\label{EqCHtQ}
 & \tilde{\lambda}^+_q D^-_q \Phi^{++}_0 = 0\; , \qquad
  \qquad & \tilde{\lambda}^+_q D^-_q \Psi^{(0)} = i \partial_{++}\Phi^{++}_0\; .
  \qquad
\\ \label{Phi=Phi+tQchi}
  & \Phi^{++}_0  \sim  \Phi_0^{++}{}^\prime = \Phi^{++}_0 +
 \tilde{\lambda}^+_q D^-_q \chi^{++}_0 \; ,
 \qquad & \Psi^{(0)}   \sim \Psi^{(0)\prime} = \Psi^{(0)} + i
 \partial_{++}\chi^{++}_0 + \tilde{\lambda}^+_q D^-_q K^{(0)} \; . \qquad
\end{eqnarray}
To obtain the cohomology of $\mathbb{Q}^{susy}$, we have to set
$\tilde{\lambda}^+_q=0$   at the end to remove the regularization;
thus we are really interested in the wavefunctions for
$\tilde{\lambda}^+_q=0$: $\; \Phi_0^{++}(0):=
\Phi_0^{++}\vert_{\tilde{\lambda}^+_q=0}= \Phi_0^{++}(0 \; , \;
x^{++} , \theta^{+}_q \, ;\; \ldots  )$, $\Psi_0^{(0)}(0):=
\Psi_0^{(0)}\vert_{\tilde{\lambda}^+_q=0}= \Psi_0^{(0)}(0 \; , \;
x^{++} , \theta^{+}_q \, ;\; \ldots  )$.

Eqs. (\ref{EqCHtQ}), (\ref{Phi=Phi+tQchi}) show that the
`superfield' cohomology problem of Eq. (\ref{CHtQsusy}) includes a
 (pure-spinor like) cohomology problem for the
leading component $\Phi_0^{++}$  of the  $\Phi^{++}$ superfield,
\begin{eqnarray}\label{CHtQ0}
 \tilde{\lambda}^+_q D^-_q \Phi^{++}_0 = 0\; ,
  \qquad \Phi^{++}_0 & \mapsto  \Phi_0^{++}{}^\prime = \Phi^{++}_0 +
 \tilde{\lambda}^+_q D^-_q \chi^{++}_0 \; . \qquad
\end{eqnarray}
Let us recall that we are interested in the cohomology problems for
fixed ghost number
 \begin{eqnarray}\label{g0hN}
 g =gh_{\#}(\Phi )= g_0 -2 \; , \qquad g_0:= gh_{\#}(\Phi_0^{++} )\; . \qquad
 \end{eqnarray}
As far as the remaining part of the cohomology problem
(\ref{CHtQsusy}) (or (\ref{EqCHtQ}), (\ref{Phi=Phi+tQchi})) is
concerned,
\begin{eqnarray}\label{CHtQ00}
 \tilde{\lambda}^+_q D^-_q \Psi^{(0)} = i \partial_{++}\Phi^{++}_0\; ,
  \qquad
 \Psi^{(0)}  & \mapsto \Psi^{(0)\prime} = \Psi^{(0)} + i
 \partial_{++}\chi^{++}_0 + \tilde{\lambda}^+_q D^-_q K^{(0)} \; , \qquad
\end{eqnarray}
the presence of the $i
 \partial_{++}\chi^{++}_0$ term in the BRST transformations suggests its triviality
 (which is indeed the case, see below).

Thus we have reduced our cohomology problem for the Lorentz harmonics BRST charge
(\ref{Qsusy}) to the auxiliary cohomology problem (\ref{CHtQ0}) for the charge
(\ref{tQsusy}). Before turning to  it, we would like to comment on the relation of our
BRST charge (\ref{Qsusy}) involving a complex $SO(9)$ spinor $\tilde{\lambda}^+_q$,
satisfying $\tilde{\lambda}^+_q\tilde{\lambda}^+_q=0$, with the Berkovits  BRST charge
constructed with the $D$=11 pure spinors \cite{NB-pure}.

\bigskip

\subsection{Relation with the Berkovits's pure spinors}
 \label{CohQ-3}

 The $D=11$ pure spinors of Berkovits obey \cite{NB-pure}
${\Lambda}\Gamma_a{\Lambda}=0$ (\ref{NB-pureSp}) and, in general, carry $46$ ($23$
complex) degrees of freedom. A specific $39$ parametric solution $\tilde{\Lambda}$ can
be found using the spinor moving frame approach (see \cite{BZ-str,BL98'}). It is given
by \footnote{Indeed, using the constraint (\ref{vv=uG}) one finds that
$\tilde{\Lambda}\tilde{\Gamma}_a\tilde{\Lambda}= \tilde{\lambda}^+_q
v^-_q\tilde{\Gamma}_av^-_p \, \tilde{\lambda}^+_p = u_a^{--} \ \tilde{\lambda}^+_q
\tilde{\lambda}^+_q = 0$ since $\tilde{\lambda}^+_q \tilde{\lambda}^+_q = 0$.}
\begin{eqnarray}\label{Lpure=lv}
\fbox{$\; \tilde{\Lambda}_\alpha = \tilde{\lambda}^+_q
v_{\alpha}{}^-_q\;$} \;  , \qquad  \{ v_{\alpha}{}^-_q\}=
{SO(1,10)\over SO(1,1)\otimes SO(9)\otimes K_9}= S^9 \; , \qquad
\tilde{\lambda}^+_q\tilde{\lambda}^+_q=0 \quad \Rightarrow \quad
\tilde{\Lambda}\Gamma_a\tilde{\Lambda}=0 \; .
 \qquad
\end{eqnarray}
Thus, the complex $16$ component $SO(9)$ spinor $\tilde{\lambda}^+_q
\tilde{\lambda}^+_q = 0$ with  $\tilde{\lambda}^+_q \tilde{\lambda}^+_q = 0$, carries
$30$ of the $39$ degrees of freedom of the (Berkovits-type) pure spinor
(\ref{Lpure=lv}). The remaining $9$ degrees of freedom in this pure spinor correspond
to the $S^9$ sphere of the light--like eleven--dimensional momentum modulo its energy.

Furthermore, as far as the $\kappa$--symmetry generator $D^-_q$ is basically
$v^{-\alpha}_q \mathbf{d}_\alpha$, one finds that Berkovits BRST charge in Eq.
(\ref{QbrstB}) can be obtained from our (\ref{tQsusy}) by replacing the composite  pure
spinor $\tilde{\lambda}^+_qv^{-\alpha}_q$  (\ref{Lpure=lv}) by a generic pure spinor
$\tilde{\Lambda}^\alpha$ and by ignoring the second quite simple $c^{++}$ term in
(\ref{tQsusy}). In other words,
\begin{eqnarray}\label{tQsusy=QB}
 \tilde{Q}^{susy} =
 \mathbb{Q}^{B}\vert_{\Lambda^\alpha = \tilde{\lambda}^+_q v^{-\alpha}_q }
 +
i c^{++}
\partial_{++}\; , \qquad
\end{eqnarray}

Of course, the generic Berkovits's pure spinor \cite{NB-pure}  in
$D$=11 carries $46$ real degrees of freedom, while the composite
pure spinor (\ref{Lpure=lv}) only carries $39$. However, it is not
obvious that all degrees of freedom in a pure spinor are equally
important for the description of superparticle in the Berkovits
approach. Notice in particular that only the pure spinor cohomology
at vanishing bosonic ghost describe the superparticle, while the
complete pure spinor cohomology is much reacher and correspond to
the spinorial cohomologies of \cite{SpinCohom02}.

As far as the generalization for the case of superstring is concerned,  it is important
to note that {\it in $D=10$ dimensional case}, which corresponds to the Green--Schwarz
superstring,   {\it Eq. (\ref{Lpure=lv}) does provide the {\bf general solution} of the
pure spinor constraint (\ref{NB-pureSp})}. Indeed, in $D=10$ this solution  carries
{\bf 16+8-2=22 } degrees of freedom, the same number as the generic pure spinor. Thus
one may expect that the substitution of the solution (\ref{Lpure=lv}) for pure spinor
used to describe superstring in \cite{NB-pure} ({\it i.e.} replacing the pure spinor
approach by a pragmatically designed Lorentz harmonic approach) should not produce any
additional anomaly.

Coming back to our M0--brane case, we conclude that a counterpart (\ref{tQsusy}) of the
Berkovits BRST charge (\ref{QbrstB}) appears on the way of regularization
($\lambda^-_q\mapsto \tilde{\lambda}^-_q\not= (\tilde{\lambda}^-_q)^*$) from the
directly obtained BRST charge (\ref{Qsusy}) when the $D=11$ superparticle is quantized
in its twistor--like Lorentz harmonics formulation (\ref{11DSSP}).

\bigskip

\subsection{Cohomology of $\tilde{\lambda}^+_q D^-_q$ }
 \label{CohQ-4}

The physical spectrum of the model is found by solving the BRST
cohomology problem in a sector of Hilbert space with a fixed ghost
number. When dealing with the $\Phi_0^{++}$ part of the wavefunction
$\Phi^{++}$, $\;\Phi^{++}= \Phi_0^{++}+ c^{++}\Psi_0$, the only
remaining carrier of the ghost number is the bosonic ghost
$\tilde{\lambda}^+_q$. Thus the ghost number $g_0:=g-2$   of the
wavefunction $\Phi_0^{++}$ (see (\ref{g0hN})) coincides with its
homogeneity degree in $\tilde{\lambda}^+_q$,
\begin{eqnarray}\label{gh=homD}
  &
  \Phi_0^{++}( z \tilde{\lambda}^+_q \, , \, \ldots )  =
  z^{{g_{_0}}} \Phi_0^{++}(\tilde{\lambda}^+_q , \, \ldots )   \qquad   \Leftrightarrow \qquad gh_{\#} (\Phi_0^{++}) =g_0 \; .  \qquad
\end{eqnarray}
We are interested in  $ \Phi= \delta (\tilde{\lambda}^+_q\tilde{\lambda}^+_q)\;
\Phi^{++}(\tilde{\lambda}^+_q, \, \ldots )$, Eq. (\ref{Phi(reg)}) which, after removing
regularization, can be written as $ \Phi= \delta ({\lambda}^+_q{\lambda}^+_q)\;
(\Phi_0^{++}\vert_{{\lambda}^+_q=0} + c^{++} \Psi_0\vert_{{\lambda}^+_q=0})$. This
means that we are actually interested in the cohomologies of the operator
$\tilde{\lambda}^+_q D^-_q$  at vanishing bosonic ghost, $\tilde{\lambda}^+_q=0$.

As such, one immediately concludes that we cannot have nontrivial
cohomology with $\Phi_0^{++}$ of ghost number $g_0>0$ since, due to
(\ref{gh=homD}), $\Phi_{0\, g_0>0}^{++}({\lambda}^+_q=0)=0$.
Furthermore, the values of the ghost number $g_0<0$ are actually
prohibited for $\Phi^{++}=\Phi_0^{++}+ \ldots  $ in
(\ref{Phi(reg)}), because $ \Phi_{0\, g_0<
0}^{++}\longmapsto\!\!\!\!\!\!\!\!\!\!\!\!_{_{{\lambda}^+_q\mapsto
0} } \infty $ and the expression for $\Phi$ in (\ref{Phi(reg)})
diverges (as $\delta (\lambda^2) \cdot \infty$) and cannot describe
a physical state. Thus a non-trivial BRST cohomology for
(\ref{Qsusy}) may come from the $\tilde{\lambda}^+_q D^-_q$
cohomologies in the  Hilbert space sector of the ghost number
$g_0=0$ {\it only}. This corresponds to   $g:=g_0-2=-2$ for the
ghost number of the complexified $\mathbb{Q}^{susy}$--closed,
non-exact wavefunction $\Phi$ in Eq. (\ref{Phi(reg)}) (see Eq.
(\ref{g0hN})).

Assuming the wavefunctions $\Phi_0^{++}$ to be analytic in $\tilde{\lambda}^+_q$, one
finds that, being homogeneous of degree zero, the wavefunction is actually independent
of $\tilde{\lambda}^+_q$. Then $\tilde{\lambda}^+_q D^-_q \Phi_0^{++}=0$ actually
implies $D^-_q \Phi_0^{++}=0$. As far as the BRST transformations  $\Phi_0^{++}\mapsto
\Phi_0^{++}{}^\prime = \Phi^{++}_0 +
 \tilde{\lambda}^+_q D^-_q \chi^{++}_0$ of Eq. (\ref{CHtQ0}) are considered, the above
 assumptions requires $\chi^{++}_0$ to be an analytic function of
$\tilde{\lambda}^+_q$ with degree of homogeneity $-1$, and such a
nonvanishing function does not exist.

Hence the calculation of the reduced  BRST cohomology (\ref{CHtQ0})
($(\tilde{\lambda}^+_q D^-_q$)--cohomology) in the space of the analytic wavefunctions
$\Phi_0^{++}$ of ghost number zero is reduced to calculating the kernel of the
$\tilde{\lambda}^+_q D^-_q$ operator which, in the sector of ghost number zero,
coincides with the kernel, $D^-_q\Phi_0^{++}=0$, of the $\kappa$--symmetry generator
$D^-_q$,
\begin{eqnarray}\label{gh=0(coh)}
  & g_0:= gh_{\#} \Phi^{++}_0 =0 \; , \qquad   \tilde{\lambda}^+_q D^-_q \Phi^{++}_0 =0   \quad &
  \Rightarrow   \quad D^-_q \Phi^{++}_0 =0 \; .  \qquad
\end{eqnarray}
With the realization (\ref{D-q=}), this equation implies the
vanishing of all the coefficients in the decomposition of
$\Phi^{++}_0$ in the power series on $\theta^+_q$, and requires that
the leading ($\theta^+_q$ independent)  component does not depend on
$x^{++}$. In other words the general solution of this equation is a
function independent on both $\theta^+_q$ and $x^{++}$,
\begin{eqnarray}\label{(gh0coh)=}
   g_0:= gh_{\#} \Phi^{++}_0 =0 \; , \quad   \tilde{\lambda}^+_q D^-_q \Phi^{++}_0 =0   \;\;  \Rightarrow   \quad
& \Phi^{++}_0 \not= \Phi^{++}_0(x^{++}\, , \, \theta^+_q)   \; \qquad \\ \nonumber
 & \qquad \left({\partial\;\;\; \over \partial x^{++}}\Phi^{++}_0=0\; , \;\; {\partial\;\;\over
\partial \theta^+_q }\Phi^{++}_0=0\; \right)\;  .
\end{eqnarray}
The ghost number of the second component $\Psi_0$ of the wavefunction
$\Phi^{++}=\Phi_0^{++} + c^{++}\Psi_0$ is  $gh_\# (\Psi_0)= g_0 - 1$, so that when
$g_0=0$ and the nontrivial cohomologies can be carried by $\Phi_0^{++}$, $gh_\#
(\Psi_0)= -1$ which, according to the discussion above, requires $\Psi_0=0$. On the
other side, when $g_0=1$ and the wavefunction $\Phi_0^{++}$ cannot describe a
nontrivial cohomology of $Q_{susy}$, one can find a nonzero BRST closed $\Psi_0$
obeying the first equation in (\ref{CHtQ00}). However, the second equation in
(\ref{CHtQ00}) allows one to `gauge' $\Psi_0$ away by using the parameter $\chi^{++}_0$
so that the cohomology problem defined by Eqs. (\ref{CHtQ00}) has only the trivial
solution.

Thus the nontrivial {\it cohomology of the BRST charge $\mathbb{Q}^{susy}$}
(\ref{Qsusy}) is described by the cohomology of the complex $\tilde{Q}^{susy}$
(\ref{tQsusy}) in the sector of ghost number $g_0:=gh_{\#}(\Phi^{++})=0$ (which
corresponds to $g:=gh_{\#}(\Phi)=-2$ for $\Phi$ in (\ref{Phi(reg)})), which in turn
{\it is described by  wavefunctions that depend on the `physical variables' only}. This
actually reduces the covariant quantization problem to the quantization of the physical
degrees of freedom, {\it i.e.} to a counterpart of the twistor quantization presented
in \cite{BdAS2006}.

The fact that the cohomologies of the BRST operator are described by wavefunctions that
do not dependent on  variables on which the constraints $D^-_q$ and $\partial_{++}$ act
nontrivially ($x^{++}$ and $\theta^+_q$ in  (\ref{Phi(reg)})) is related to properties
that are specific for  the superparticle case, where there exists a coordinate basis in
which the action is written in terms of variables invariant under both
$\kappa$--symmetry (generated by $D^-_q$ above) and $b$--symmetry (generated by
$\partial_{++}$). The action in such a coordinate basis will be discussed in the next,
concluding Sec. \ref{Concl}. Let us note that the above effect does not happen in the
superstring case, and hence in the cohomology problem for the superstring counterpart
of the BRST charge (\ref{Qsusy}) such a simplification cannot occur.

We have to stress that of all the cohomologies of the complex Berkovits--like BRST
charge $\tilde{\mathbb{Q}}^{susy}$ only their values at vanishing bosonic ghost,
$\tilde{\lambda}^-_q=0$, describe the cohomologies of the M0--brane BRST charge
$\mathbb{Q}^{susy}$ and, hence, the superparticle spectrum. The
$\tilde{\mathbb{Q}}^{susy}$  cohomologies for $\tilde{\lambda}^-_q\not= 0$,
corresponding to the higher ghost numbers, are reacher and are related with the
spinorial cohomologies of \cite{SpinCohom02}.

\bigskip

\section{M0--brane and its quantization in the covariantized light--cone basis. }
\label{SecAnalB}

The simple structure of the cohomology of the M0--brane BRST charge
$\mathbb{Q}^{susy}$ can be explained by studying the spinor moving
frame action (\ref{11DSSP}) in different basis of canonical
variables, particularly in the  {\it covariantized light--cone
basis} \cite{Sok,NPS,GHT93}. The coordinates of this,
$(x^{\pm\pm},x^{i}, \theta^{\pm}_q )$,  are constructed from the
ones of the standard basis of superspace $Z^M=(x^m \; , \;
\theta^\alpha )$ and harmonics as (see \cite{GHT93}, {\it cf.}
\cite{Sok})
\begin{eqnarray}\label{analX-Th}
x^{\pm\pm}=x^m u_m^{\pm\pm}\; , \qquad x^{i}=x^m u_m^{\; i}\; , \qquad \theta^{\pm}_q
:= \theta^\alpha v_{\alpha q}^{\; \pm} \; .
\end{eqnarray}

The change of variables (\ref{analX-Th}) in the superparticle action (\ref{11DSSP})
gives
\begin{eqnarray}\label{11DSSPan}
S:= \int d\tau L  &=&  \int_{W^1} \left( {1\over 2}  \rho^{++} Dx^{--} -
 {1\over 2}  \rho^{++} \Omega^{--i} \tilde{x}^i
 - i  D\theta_q \; \theta_q \right) , \qquad
\end{eqnarray}
where
\begin{eqnarray}\label{tXi=}
& \tilde{x}^i = x^i + i \theta^-_p \gamma^i_{pq} \theta^+_q := x^i + i
\theta^{\alpha}\, v_{\alpha p}^{\; -} \gamma^i_{pq} v_{\beta q}^{\; +} \,
\theta^{\beta} \; , \qquad  & Dx^{--}:= dx^{--}+ 2\Omega^{(0)} x^{--}\; , \qquad
\\ \label{Th=}
 & \theta_q = \sqrt{\rho^{++}}\; \theta^-_q :=
\sqrt{\rho^{++}}\; \theta^{\alpha} v_{\alpha q}^{\; -}\; , \qquad & D\theta_q :=
d\theta_q + {1\over 4}\Omega^{ij}\theta_p \gamma^{ij}_{pq} \; , \qquad
\end{eqnarray}
and $\Omega^{(0)}$, $\Omega^{ij}$ are the $SO(1,1)$ and $SO(9)$ Cartan forms, see Eq.
(\ref{Omab}).

Notice that the action (\ref{11DSSPan}) is given in terms of $\kappa$-- and $b$--
invariant variables, so that no further gauge fixing is needed. Indeed, the irreducible
 $\kappa$--symmetry of the action (\ref{11DSSP}) is characterized by Eq.
 (\ref{kappa-irr}),
\begin{eqnarray}\label{kappa-sym}
\delta_\kappa x^m = i \delta_\kappa \theta^\alpha \Gamma^m_{\alpha\beta}\theta^\beta\;
, \qquad \delta_\kappa \theta^\alpha = \kappa^{+q} v_q^{-\alpha} \; , \qquad
\delta_\kappa v_\alpha{}^-_q =0 = \delta_\kappa u_m^{--}\; .
\end{eqnarray}
For the fermionic coordinate functions in the covariantized light cone basis one finds
that $\theta^+_q$ is transformed additively by the $16$--component $\kappa$--symmetry
parameter,  $\delta_\kappa \theta^+_q= \kappa^{+q}$, while $\delta_\kappa
\theta^-_q=0$. Furthermore, $\delta_\kappa x^{++} = 2i \kappa^{+q} \theta^+_q$, while
$\delta_\kappa x^{--}=0$ and, although $\delta_\kappa x^{i} = i \kappa^{+q}
\gamma^i_{qp}\theta^-_p$, $\tilde{x}^i $ of Eq. (\ref{tXi=}) is $\kappa$--invariant,
$\delta_\kappa \tilde{x}^i =0$. Thus all the variables entering the action
(\ref{11DSSPan}) are inert under $\kappa$--symmetry,
\begin{eqnarray}\label{kappa-inv}
\delta_\kappa x^{--}=0\; , \qquad \delta_\kappa \tilde{x}^i := \delta_\kappa x^{i} - i
\kappa^{+q} \gamma^i_{qp}\theta^-_p =0 \; , \qquad \delta_\kappa \theta^-_q=0\; ,
\qquad i_\kappa \Omega^{--i}=0 \qquad
\end{eqnarray}
This completes the proof of that  just the change of variable (\ref{analX-Th}) in the
spinor moving frame action (\ref{11DSSP}) results in the functional (\ref{11DSSPan})
which  involves $\kappa$--invariant variables only. This phenomenon of an automatic
gauge fixing, noticed already in \cite{Sok}, explains the mentioned simple structure of
the cohomology of the BRST operator constructed from just the $\kappa$-- and
$b$--symmetry generators $D^-_q$ and $\partial_{++}$.

The above `automatic' gauge fixing does not occur in the superstring case and, hence,
the cohomology of the corresponding Lorentz harmonics BRST operators are expected to be
richer.

\subsection{On BRST quantization of M0--brane  in the
covariantized light cone basis} \label{SecAnal}

Hence,  a difference between the original action of Eq.  (\ref{11DSSP}) and the action
in the covariantized light--cone basis (\ref{analX-Th}), Eq. (\ref{11DSSPan}), is that
the latter contains only variables invariant under the $\kappa$-- and $b$--symmetries.
Thus changing the basis to (\ref{analX-Th}) automatically provides the
$\kappa$--symmetry  and $b$--symmetry gauge fixed action (this effect was firstly
noticed in \cite{Sok}). Another difference between the two actions is that the
harmonics $v_\alpha{}^-_{q}$ enter in (\ref{11DSSPan})
 {\it only} through the Cartan forms $\Omega^{--j}$, $\Omega^{(0)}$, $\Omega^{ij}$
 defined by  Eqs. (\ref{dv-q}),
(\ref{Omab})) and entering the  canonical Liouville one form on the $SO(1,D-1)$ group
manifold as defined in Eqs. (\ref{H0:=Om(gen)--}), (\ref{H0:=OmD-L}),
\begin{eqnarray}
\label{H0:=OmD-L1} & {1\over 2}\Omega^{(a)(b)}\mathbf{d}_{(a)(b)} :=  -  {1\over
2}\Omega^{--i}\mathbf{d}^{++i} - {1\over 2}\Omega^{++i}\mathbf{d}^{--i}-   \Omega^{(0)}
\mathbf{d}^{(0)} +  {1\over 2} \Omega^{ij} \mathbf{d}^{ij}  \; . \qquad
\end{eqnarray}

\subsubsection{\small Hamiltonian mechanics in the covariantized light--cone basis}

Let us define the canonical momenta in the usual way and the covariant canonical
momenta by (\ref{H0:=OmD-L}) and remove the second class constraints on the harmonics
by using the Dirac brackets (\ref{DB-harm}) (see Sec. \ref{DBsec}). Doing the same for
the fermionic second class constraints, we identify the $16$ Grassmann variables with
their momenta,
\begin{eqnarray}\label{DB(ThTh)=}
{} \{ \theta_{q}\, , \,
 \theta_{p} \}_{_{DB}} = -{i\over 2}  \delta_{qp} \; .
\end{eqnarray}
Then the bosonic `primary' constraints implied by the action (\ref{11DSSPan}) read
\begin{eqnarray}
\label{IclAnn0}
 \mathbf{d}^{(0)}+ \rho^{++}x^{--}\approx 0\; , \qquad \mathbf{d}^{ij}+
{i\over 2}\theta \gamma^{ij}\theta
 \approx 0\; , \qquad \mathbf{d}^{--i}\approx 0\; ,
 \qquad \\
 \label{AnPrimDj} \mathbf{d}^{++i}- \rho^{++}\tilde{x}^{i}\approx 0\; , \qquad \tilde{P}_{j}\approx 0\; ,
 \qquad  \\
 \label{AnPrimCP}
  P_{--}-{1\over 2}\rho^{++}\approx 0\; ,
 \qquad  P^{(\rho )}_{++}\approx 0\; .  \qquad
\end{eqnarray}
Clearly, the last two constraints, Eqs. (\ref{AnPrimCP}), provide the resolved pair of
the second class constraints, which allows us to remove the $\rho^{++}$ variable by
replacing it by $2P_{--}$. The same is true about the pairs of  constraints in
(\ref{AnPrimDj}), which allows us to remove the orthogonal $\tilde{x}^{i}$ coordinates
(the non-covariant counterparts of which describe the physical degrees of freedom in
the standard light--cone gauge description of the Brink-Schwarz superparticle and
Green-Schwarz superstring) by expressing them through the covariant momenta
$\mathbf{d}^{++i}$ for the harmonic variables and the $P_{--}$ momentum
\begin{eqnarray}
\label{tXi=d/2P} \tilde{x}^{i} = \; {\; \mathbf{d}^{++i} \over 2 P_{--}} \; .
 \qquad
 \end{eqnarray}
The remaining constraints, Eqs. (\ref{IclAnn0}),
\begin{eqnarray}
\label{IclAnn}
 \widetilde{\mathbf{d}^{(0)}}:= \mathbf{d}^{(0)}+ 2x^{--}P_{--}\approx 0\; , \qquad
 \widetilde{\mathbf{d}^{ij}}:=\mathbf{d}^{ij}+
{i\over 2}\theta \gamma^{ij}\theta
 \approx 0\; , \qquad \widetilde{\mathbf{d}^{--i}}:= \mathbf{d}^{--i}\approx 0\; ,
 \qquad
\end{eqnarray}
are first class ones. Their Dirac brackets produce the $(so(1,1)\oplus so(9))
\subset\!\!\!\!\!\!+ K_9$ algebra, which can be obtained from the $so(1,10)$ of Eq.
(\ref{PB=d'd}) by omitting the relations  involving
 $\mathbf{d}^{++i}$,
 \begin{eqnarray}\label{DB(d,d)=Ann}
 &&  {}  [\widetilde{\mathbf{d}^{ij}}\; , \;
\widetilde{\mathbf{d}^{kl}}]_{_{DB}} = 2\widetilde{\mathbf{d}^{k[i} } \delta^{j]l} -
2\widetilde{\mathbf{d}^{l[i} } \delta^{j]k}\; , \qquad {}
[\widetilde{\mathbf{d}^{(0)}}\; ,  \; \widetilde{\mathbf{d}^{ij}}]_{_{DB}} = 0 \; ,
\qquad
 \nonumber \\ &&  {} [\widetilde{\mathbf{d}^{(0)}}\; ,  \; \widetilde{\mathbf{d}^{-- i}}]_{_{DB}} =
- 2 \widetilde{\mathbf{d}^{-- i}} \; , \qquad [\widetilde{\mathbf{d}^{ij}}\; , \;
\widetilde{\mathbf{d}^{-- k}}]_{_{DB}} = 2\widetilde{\mathbf{d}^{-- [i}} \delta^{j]k}\;
, \qquad   \nonumber \\ &&  {} [\widetilde{\mathbf{d}^{--i}}\; ,  \;
\widetilde{\mathbf{d}^{-- j}}]_{_{DB}} = 0 \; . \qquad
\end{eqnarray}
No `W-deformation' occurs here. Actually this is natural, as the
 {\it r.h.s.} in Eq. (\ref{d--d--DB}) was proportional to the square of the $
 \kappa$--symmetry generator absent in the covariantized light--cone basis.

\subsubsection{BRST charge for the first class constraints  in the covariantized light--cone basis}

In the covariantized light--cone basis, where the $\kappa$--symmetry and $b$--symmetry
are automatically gauge fixed, the  superparticle quantization might be based on the
BRST operator for the algebra (\ref{DB(d,d)=Ann}) of the $SO(1,1)\otimes
SO(9)\subset\!\!\!\!\!\!\times K_9$ symmetry, appearing here as the full BRST operator
for the gauge symmetries of the M0-brane model,
\begin{eqnarray}\label{Q(H)}
  {} \mathbf{Q}^{\!^{[SO(1,1)\otimes SO(9)]\subset\!\!\!\!\times
  K_9 }} &=& c^{++ i}D^{-- i} + {1\over 2} c^{ij} D^{ij} +
  c^{(0)}D^{(0)} - \nonumber \\ && - {1\over 2} c^{++ i}c^{ij} {\partial\;\; \over
  \partial  c^{++ j}} + 2  c^{(0)}c^{++j} {\partial\;\; \over
  \partial  c^{++ j}} +  c^{ik}c^{jk} {\partial\;\; \over
  \partial  c^{ij}}\; . \qquad
\end{eqnarray}
Here  $D^{(0)}$, $ D^{ij}$ and  $D^{-- i}$ are harmonic covariant derivatives
representing the $SO(1,1)$, $SO(9)$ and $K_9$ generators and, thus, obeying the Lie
algebra
\begin{eqnarray}\label{H(d,d)=Ann}
  {} [D^{(0)}\; ,  \; D^{-- i} ]=
- 2 D^{-- i} \; , \quad {} [D^{ij}\; , \; D^{-- k}] = 2D^{-- [i} \delta^{j]k}\; , \quad
\nonumber \\ {} [D^{ij}\; , \; D^{kl}] = 2D^{k[i} \delta^{j]l} - 2D^{l[i}
\delta^{j]k}\; , \quad  {} [D^{(0)}\; , \; D^{ij}] = 0 \; , \qquad {} [D^{--i}\; ,  \;
D^{-- j}] = 0 \; ,
\end{eqnarray}
 $c^{(0)}$, $ c^{ij}$ and  $c^{++ i}$ are the fermionic ghosts for these symmetries
and the derivative with respect to the tensorial ghost is defined by ${\partial
c^{i^\prime j^\prime} \over \partial \, c^{ij}}=
2\delta_{[i}^{i^\prime}\delta_{j]}^{j^\prime}$.

\subsection{Covariant quantization of the physical degrees of freedom and hints of hidden symmetries}

Although the quantization of the physical degrees of freedom in the covariantized light
cone basis ({\it cf.}  \cite{Sok}, where the vector harmonics were used for the first
time in quantization of such a type) is similar to the supertwistor quantization of
\cite{BdAS2006}, we briefly  discuss it here as it gives hints about possible hidden
symmetries of the 11D supergravity (see \cite{BdAS2006} for the discussion on
$SO(16)$).

As the first class constraints (\ref{IclAnn}) obey the Dirac bracket algebra
(\ref{DB(d,d)=Ann}) isomorphic to $[SO(1,1)\otimes SO(9)]\subset\!\!\!\!\!\!\times K_9$
(no deformation appear), we can, following Dirac  \cite{Dirac}, just impose their
quantum counterparts $D^{(0)}$, $ D^{ij}$ and  $D^{-- i}$  (\ref{H(d,d)=Ann}) as
differential operator conditions on the wavefunction $\Phi$,
\begin{eqnarray}\label{HPhi=0}
  {} D^{(0)}\Phi =0 \; , \qquad  D^{ij}\Phi =0 \; , \qquad  D^{--i}\Phi =0 \; . \qquad
\end{eqnarray}
In the purely bosonic limit the differential equations (\ref{HPhi=0}) are imposed on
the wavefunction which depends on the spinorial harmonics (which, due to the second
class constraints, parametrize the $Spin(1,10)$ group manifold, see Secs. 3.2-2.4) and
$\rho^{++}$. \footnote{Alternatively, one can consider a wavefunction dependent on
harmonics and  $x^{--}$, but for our line of arguments the use of wavefunctions
dependent on $\rho^{++}$ ($=2P_{--}$, see (\ref{AnPrimCP})) is more convenient.}
Imposing the conditions (\ref{HPhi=0}) is tantamount to requiring that, as a function
of harmonics, the wavefunction is now a function on the $S^9$ sphere which (in the
light of the primary constraint (\ref{P-rvv}) generalizing the Cartan--Penrose
representation for a light--like vector to D=11) can be identified with the space of
light--like momentum
 modulo its scale. This scale of the massless particle momentum, the energy, can be
 identified then (again in the light of the Cartan--Penrose
constraint (\ref{P-rvv})) with the Lagrange multiplier $\rho^{++}$.

Then, as the canonical Hamiltonian $H_0$ corresponding to the action (\ref{11DSSPan})
is zero, $H_0\approx 0$, one concludes that, in the purely bosonic limit, the
wavefunction is just an arbitrary function of the above listed physical bosonic
variables, namely
\begin{eqnarray}\label{Phi=S9R} \Phi \vert_{\theta_q =0} = \Phi_0(\mathbb{R}_+
\otimes \mathbb{S}^9)\; , \qquad \{ (v_{\alpha q}^-\, , \; \rho^{++}) \} = \mathbb{R}_+
\otimes \mathbb{S}^9 = \{ (p_{\underline{m}}\; : \; p^2:=
p_{\underline{m}}p^{\underline{m}}=0 )\}\;  . \qquad
\end{eqnarray}
This result coincides with one  obtained in  \cite{BdAS2006} in the framework of
supertwistor quantization of the M0--brane model.

The complete M0--brane action (\ref{11DSSPan}) includes also the fermionic contribution
$D\theta_q \; \theta_q= d\theta_q \; \theta_q + \Omega_{pq}\theta_{[p} \; \theta_{q]}$,
where $\Omega_{pq}= - \Omega_{qp}:= {1\over 4} \Omega^{ij}\gamma^{ij}_{pq}$ is the
$Spin(9)$ connection. Their presence modifies the $SO(9)$ generator by the term
bilinear in fermions (see Eq. (\ref{IclAnn})), but this does not change the conclusion
about the wavefunction dependence on the bosonic configurational space coordinates
(which, from the spacetime point of view, happen to parametrize the light-like
momentum).

  Then, the fermionic variables $\theta_q $ obey the
second class constraints stating that they are momenta for
themselves, which can be treated in the strong sense after passing
to the Dirac brackets (\ref{DB(ThTh)=}). In quantum theory the Dirac
brackets relation (\ref{DB(ThTh)=}) give rise to the
anti--commutational relation stating that the Grassmann coordinate
function of the M0--brane becomes a Clifford algebra valued,
\begin{eqnarray}\label{hThhTh=}
{} \{ \hat{\theta}_{q}\, , \,
 \hat{\theta}_{p} \} = {1\over 2}  \delta_{qp} \; , \qquad q=1,2,\ldots , 16 . \qquad
\end{eqnarray}
This $O(16)$ covariant Clifford algebra $\mathrm{C}\ell^{16}$ has a finite dimensional
representation by $256\otimes 256$ sixteen dimensional gamma matrices
\begin{eqnarray}\label{th=Gamma16}
{} & \hat{\theta}_{q} = \, {1\over 2 }\, ({\Gamma}_{q})_{{\cal A}}{}^{{\cal B}} \; ,
\qquad {\cal A}\, , \, {{\cal B}} = 1, \ldots , 256 \; , \qquad q=1,2,\ldots , 16 \; .
\qquad
\end{eqnarray}

Notice that the $O(16)$ symmetry of the  Clifford algebra $\mathrm{C}\ell^{16}$ is the
same $O(16)$ which we have met in the classical analysis of the spinor moving frame
action, sec. \ref{O(16)}. Indeed, it acts in the same way and on the same indices, as
far as $\theta_q = \sqrt{\rho^{++}}\; \theta^{\alpha} v_{\alpha q}^{\; -}$, Eqs.
(\ref{analX-Th}), (\ref{Th=}). Thus our spinor moving frame formulation (\ref{11DSSP})
makes manifest, already at the classical level, the $SO(16)$ symmetry playing, as we
will see in a moment,  an important role in the M0--brane quantization.

But before, let us make the following comments.

Firstly, substituting for $\theta_q$ its contraction with an $SO(16)$ matrix, $\theta_q
\mapsto \theta_p S_{pq}$ would produce the covariant derivative with the SO(16)
connection $\Omega_{pq} \mapsto (dS\, S^T)_{pq}+  {1\over 4}
\Omega^{ij}(S^T\gamma^{ij}S)_{pq}$,
\begin{eqnarray}\label{DthSO16}
&& D(\theta S)_q \; (\theta S)_q =  \tilde{D} \theta_q \; \theta_q= d\theta_q \;
\theta_q + \tilde{\Omega}_{pq} \theta_{[p} \; \theta_{q]} \; , \qquad S\, S^T =I \; ,
\qquad \nonumber
\\ && \qquad \tilde{\Omega}_{pq} = (dS\, S^T)_{pq}+ {1\over 4} \Omega^{ij}(S^T\gamma^{ij}S)_{pq}\equiv (dS\, S^T)_{pq}+
{1\over 4} \Omega^{ij}((S^T\gamma^{[i}S)(S^T\gamma^{j]}S))_{pq} \; . \qquad
\end{eqnarray}
It is not evident that such transformation leave the  model invariant. To be convinced
that they do (when supplemented by the corresponding transformations of the bosonic
variables), one can recall that $\theta_q= \sqrt{\rho^{++}} \theta^\alpha v_{\alpha
q}^-$ (Eq. (\ref{analX-Th})), that the action (\ref{11DSSPan}) is equivalent to
(\ref{11DSSP}) (obtained from it just by moving derivatives) and that the change $
v_{\alpha q}^-\mapsto  v_{\alpha p}^-S_{pq}$ leaves the action (\ref{11DSSP}) unchanged
as far as $S\, S^T= I$ ({\it i.e.} $S\in O(16)$).

Secondly, taking into account the results of quantization in the bosonic case, in which
the state vector is described by the wavefunction of the the light--like momentum,
$\Phi_0= \Phi_0(p_{\underline{m}}\vert_{p^2=0} )$, one might think that the state
vector of the supersymmetric particle is described by the {\it Clifford superfield}
\cite{Dima88}, {\it i.e.} by the  wavefunction dependent on such a light--like momentum
$p_{\underline{m}}$ {\it and} on the Clifford algebra valued $\hat{\theta}_{q}$
variable,
\begin{eqnarray}\label{Phi(Cl)}
 \Phi (p_{\underline{m}}\vert_{p^2=0}\, , \, \hat{\theta}_q )=
\Phi_0(p_{\underline{m}}\vert_{p^2=0} ) + 2\hat{\theta}_q \Psi_q
(p_{\underline{m}}\vert_{p^2=0} ) + \ldots + {2^n\over n! } \hat{\theta}_{q_1} \ldots
\hat{\theta}_{q_{n}}  \Phi_{q_{n}\ldots q_{1}}(p_{\underline{m}}\vert_{p^2=0} ) +
\qquad \nonumber \\ + \ldots + {2^{16}\over 16! } \hat{\theta}_{q_1} \ldots
\hat{\theta}_{q_{16}} \Phi_{q_{16}\ldots q_{1}}(p_{\underline{m}}\vert_{p^2=0} ) \; ,
\qquad \hat{\theta}_{q}\hat{\theta}_{p}+ \hat{\theta}_{q}\hat{\theta}_{q}= {1\over 2
}\delta_{qp}\hat{\mathbb{I}} \; , \qquad
\end{eqnarray}
where the coefficients are antisymmetric on their indices, $\Phi_{q_{n}\ldots
q_{1}}(p_{\underline{m}}\vert_{p^2=0} )=\Phi_{[q_{n}\ldots
q_{1}]}(p_{\underline{m}}\vert_{p^2=0} )$.

However, {\it such a representation of the $SO(16)$ symmetry is reducible}. It is
reducible also as a represenation of the Clifford algebra  $\mathrm{C}\ell^{16}$. To
see this, one can use the matrix representation (\ref{th=Gamma16}) substituting the
sixteen dimensional gamma--matrices for $2\hat{\theta}_q$. Then the (\ref{Phi(Cl)})
becomes represented by the $256\times 256$ matrix wavefunction, $\Phi
(p_{\underline{m}}\vert_{p^2=0}\, , \, \hat{\theta}_q ) \qquad \mapsto \qquad
\Phi_{{\cal A}}{}^{{\cal B}} (p_{\underline{m}}\vert_{p^2=0})$,
\begin{eqnarray}\label{Phi(Cl)G}
\Phi_{{\cal A}}{}^{{\cal B}}  (p_{\underline{m}})\, := \Phi_0(p_{\underline{m}} )
\delta_{{\cal A}}{}^{{\cal B}} + \Psi_q (p_{\underline{m}})\Gamma_q{}_{{\cal
A}}{}^{{\cal B}}  + \ldots + {1\over n! } \Phi_{q_{n}\ldots q_{1}} (p_{\underline{m}})
 \Gamma_{q_1\ldots q_{n}}{}_{{\cal A}}{}^{{\cal B}}  + \qquad \nonumber \\  + {} \ldots
+ {1\over 16!}
 \Phi_{q_{16}\ldots
q_{1}}(p_{\underline{m}} ){\Gamma}_{q_1\ldots q_{16}}{}_{{\cal A}}{}^{{\cal B}}  \; ,
\qquad  {p^2=0} \; .  \qquad
\end{eqnarray}
This is a general $SO(16)$ {\it bi}--spinor carrying the $\mathbf{256 \times 256}$
representation which is {\it reducible} both as the representation of the $SO(16)$
symmetry  and of the Clifford algebra $\mathrm{C}\ell^{16}$.

The appearance of a reducible representation contradicts to the very spirit of the
quantization procedure. The result of quantization of a particle mechanics is assumed
to be an elementary particle, the definition of which (see {\it e.g.}
\cite{Novozhilov}) was formulated involving the requirement to be irreducible
representation of Poincar\'e  and other physical   symmetry groups. This makes
accessible the procedure of projecting out a part of quantum state spectrum in
quantization of spinning particle \cite{spinQuant} and the famous GSO
(Gliozzi--Scherk--Olive) projection in quantization of the RNS string model \cite{GSO}.

Hence, the prescription of an unrestricted Clifford superfield does not work, at least
in our D=11 massless superparticle case. A simplest irreducible representation of
$\mathrm{C}\ell^{16}$ is the $SO(16)$ Majorana spinor, $\mathbf{256}$, and the choice
of the wavefunction $\Phi_{{\cal A}} (p_{\underline{m}}\vert_{p^2=0})$ gives the
linearized supergravity supermultiplet (see \cite{Green+99,BdAS2006}).

The physical degrees of freedom of the linearized $D=11$ supergravity multiplet are
described by symmetric traceless $SO(9)$ tensor $h_{IJ}=h_{(IJ)}$,  an antisymmetric
third rank $SO(9)$ tensor $A_{J_1J_2J_3} =A_{[J_1J_2J_3]}$ and a $\gamma$-traceless
fermionic $SO(9)$ vector-spinor $ \Psi_{Ip}$.  Indeed, the solution of the linearized
Einstein, three--form gauge field and the Rarita-Schwinger equations can be written in
terms of the above $h_{IJ}$, $A_{J_1J_2J_3}$, $ \Psi_{Ip}$ and Lorentz harmonics
 as (see \cite{GHT93,BdAS2006})
\begin{eqnarray}
\label{11DSGlin} & \cases{h_{mn}(p)= u^I_m u^J_n h_{(IJ)}(p) \; , \cr A_{m n p}(p)=
u^I_m u^J_n u^K_p A_{IJK}(p) \; , \cr \Psi_{m\,\alpha}(p) = \Psi_{I\, q}(p)u^I_m\;
v_{\alpha q}^{\;\; -}\sqrt{\rho^{++}}\; , } \qquad   \left. \matrix{ p_m
=\rho^{++}u_m^{--}\; , \qquad \cr u_m^{--} \Gamma^m_{\alpha\beta}= 2v_{\alpha
q}^{\;\;-}v_{\beta q}^{\;\;-}} \right\} \quad \Rightarrow \quad  p^2=0\; .
\end{eqnarray}

The action of $\hat{\theta}_q$ on these on-shell fields are defined by (see
\cite{Green+99} for the light--cone gauge quantization and \cite{BdAS2006} for the
supertwistor quantization)
\begin{eqnarray}
\label{repr-16b}  {2}\hat{\theta}_{q} h_{IJ} & =& \gamma^I_{qp} \Psi_{Jp}+
\gamma^J_{qp}
\Psi_{Ip} \; , \qquad \nonumber \\
 {2}\hat{\theta}_{q} A_{IJK} &= &  \gamma^{IJ}_{qp} \Psi_{Kp}+ \gamma^{KI}_{qp}
\Psi_{Jp}+ \gamma^{JK}_{qp} \Psi_{Ip} \; , \qquad
\\ \label{repr-16f}
{2}\hat{\theta}_{q} \Psi_{Ip} &= & \gamma^J_{qp} h_{IJ}+ {1\over 3!}
\left(\gamma^{IJ_1J_2J_3}_{qp} - 6 \delta^{I[J_1} \gamma^{J_2J_3]}_{qp}\right)
A_{J_1J_2J_3} \; , \qquad
\end{eqnarray}
To see that Eq. (\ref{repr-16b}) is nothing but an action of the $d=16$ gamma matrix
(see (\ref{th=Gamma16})) on one Majorana spinor of $SO(16)$, let us begin by splitting
 the Majorana spinorial representation of $SO(16)$ on two
Majorana--Weyl (MW) spinor representations,
$\mathbf{256}=\mathbf{128}+\widetilde{\mathbf{128}}$,
\begin{eqnarray}\label{Phi(256)=}
\Phi_{{\cal A}}  (p_{\underline{m}}\vert_{p^2=0})\, :=  \left( \matrix{\Phi_A
(p_{\underline{m}}\vert_{p^2=0}) \cr \Psi^{\tilde{A}} (p_{\underline{m}}\vert_{p^2=0})}
\right)\; . \qquad
\end{eqnarray}
The observation is that the balance of the bosonic and fermionic
degrees of freedom in $D=11$ supergrvaity multiplet is just
$128+128$ and that ({\it e.g.}) the first, $\mathbf{128}$, of the
above MW spinor representations can be used to describe the physical
degrees of freedom of the bosonic fields of the linearized
supergravity supermultiplet, while the second,
$\widetilde{\mathbf{128}}$,- to describe the physical degrees of
freedom of gravitino field
\begin{eqnarray}\label{Phi(128)}
\Phi_A &=& \left( \matrix{h_{IJ} \cr A_{IJK} }\right) \; , \qquad h_{IJ} = h_{(IJ)}\; ,
\quad h_{II}=0 \; , \qquad A_{IJK}=A_{[IJK]}\; , \quad \nonumber
\\ && \quad A=1,\ldots 128 \; , \quad I,J,K = 1,\ldots 9 \;   \quad \left({9\cdot 10\over 2}
- 1 + \left\{\matrix{ 3 \cr 9}\right\} = 44 + 84=128\;\right)\;  , \qquad
 \\
\label{Psi(t128)}  \Psi^{\tilde{A}} &=& \sqrt{2} \Psi_{Iq} \; ,
\qquad \Psi_{Iq}\gamma^I_{qp}=0 \; , \qquad    \nonumber \\ && \quad
\tilde{A}=1,\ldots 128 \; , \quad I = 1,\ldots 9 \; , \quad
q=1,\ldots , 16 \;  \qquad \left( \;9\cdot 16 - 16 =128\;\right)\;
. \qquad
\end{eqnarray}
To resume, the Majorana spinor of $SO(16)$, (\ref{Phi(256)=}), can be presented as
\begin{eqnarray}\label{Phi(256)=b+f}
\Phi_{{\cal A}} \, := \left( \matrix{\Phi_A \cr \Psi^{\tilde{A}} }
\right)\; = \left(\matrix{ \left( \matrix{ h_{IJ}  \cr A_{IJK}}
\right) \cr {} \cr \sqrt{2} \Psi_{Iq} } \right)\; , \qquad \cases{
h_{IJ} = h_{(IJ)}\; , \quad h_{II}=0 \; , \cr A_{IJK}=A_{[IJK]}\; ,
\cr {} \;
 \cr \Psi_{Iq}\gamma^I_{qp}=0 \; . } \qquad
\end{eqnarray}
Finally, assigning the Grassmann parity $0$ and $1$ to the first and second
Majorana--Weyl components, (\ref{Phi(128)}) and (\ref{Psi(t128)}),   of the (momentum
representation) wavefunction (\ref{Phi(256)=b+f}), one arrives at the linearized on
shell multiplet of $D=11$ supergravity.

\bigskip

 With the Weyl representation of the gamma-matrices
\begin{eqnarray}
\label{16dGamSig} &  ({\Gamma}_{q})_{{\cal A}}{}^{{\cal B}}= \left(\matrix{ 0 &
\sigma_{q\, A}{}^{\tilde{B}} \cr \tilde{\sigma}_q{}^{\tilde{B}}{}_{A}{} & 0 } \right)
\\
\label{16dPauliM} & (\sigma_q\tilde{\sigma}_p+ \sigma_p\tilde{\sigma}_q)=\delta_{qp}
\mathbb{I}_{128\times 128} \; , \qquad (\sigma_q\tilde{\sigma}_p)_{AB}=\delta_{AB}+
\sigma_{qp}{}_{AB} \; \qquad
\end{eqnarray}
Eqs. (\ref{repr-16b}) and (\ref{repr-16f}) can be formulated as an action of the d=16
Pauli matrices on  two Majorana--Weyl representations of the $SO(16)$,
\begin{eqnarray}
\label{11DSUSY} & 2\hat{\theta}_q \Phi_A = \sigma_q{}_A{}^{\tilde{B}} \Psi^{\tilde{B}}
\; , \qquad 2\hat{\theta}_q \Psi^{\tilde{A}}  = \tilde{\sigma}_q{}^{\tilde{A}} {}_B
\Phi_B \; . \qquad
\end{eqnarray}
This corresponds to the following representation of the $d=16$ Pauli matrices algebra
(\ref{16dPauliM}) in terms of $d=6$ Dirac matrices $\gamma^I_{qp}=\gamma^I_{(qp)}$:
\begin{eqnarray}
\label{16dPauli=} \sigma_{q\, A}{}^{\tilde{B}}&=& \left(\matrix{ \sqrt{2}
\gamma^{(I_1}{}_{qp} \,\delta^{I_2)J} - {\sqrt{2}\over 9} \delta^{I_1I_2}\,
\gamma_{qp}^{J} \qquad \cr \hline {} \cr {3\over \sqrt{2}}\gamma^{[I_1I_2}{}_{qp} \;
\delta^{I_3]J} - {1\over 3 \sqrt{2}}(\gamma^{I_1I_2I_3}\gamma^J){}_{qp} \; }
\right)\equiv  \left(\matrix{ \sqrt{2} \gamma^{(I_1}{}_{qp} \,\delta^{I_2)J} -
{\sqrt{2}\over 9} \delta^{I_1I_2}\, \gamma_{qp}^{J} \qquad \cr \hline {} \cr
\sqrt{2}\gamma^{[I_1I_2}{}_{qp} \; \delta^{I_3]J} - {1\over 3
\sqrt{2}}(\gamma^{I_1I_2I_3J}){}_{qp} \; } \right) \; , \qquad
\nonumber \\ {} \nonumber \\
\tilde{\sigma}_q{}^{\tilde{B}}{}_{A}&=& \left(\matrix{
\sqrt{2}\delta^{J(I_1} \gamma^{I_2)}{}_{qp} - {\sqrt{2}\over 9}
\delta^{I_1I_2} \gamma^{J}_{qp} \qquad \vert \qquad {1\over
3\sqrt{2}}(\gamma^J\gamma^{I_1I_2I_3})_{qp} - {3\over \sqrt{2}}
\delta^{J[I_1}\gamma^{I_2I_3]}{}_{qp}
 } \right)
\; . \qquad
\end{eqnarray}

\bigskip

Actually, the above results can be used {\it to speculate about possible $E_8$ symmetry
of the 11D supergravity}. For the 11D supergravity dimensionally reduced down to d=3
this symmetry was  conjectured already in \cite{CremmerJulia78} and proved in
\cite{E8}. Recently the appearance of $E_8$ symmetry in D=11 supergravity was discussed
in \cite{LambertWest07}.

Our line is a bit different and refers on the physical degrees of freedom in the
supergravity fields, associated to the irreducible representation of $SO(D-2)=SO(9)$,
as described above, rather than on the compactification of D=11 supergravity to d=3.

The generators of $E_8$ can be split onto the set of the generators
of its maximal compact subgroup $SO(16)$, $J_{qp}$, $128$ generators
$Q^A$ collected in the Majorana--Weyl spinor of $SO(16)$, whose
commutation relations close on the $SO(16)$ generator,
\begin{eqnarray}
\label{E8:SO} E_8\; : \qquad && [J_{qp}\, , \, J_{q^\prime p^\prime
}\, ]= 2 \delta_{q^\prime[q}J_{p] p^\prime} - 2 \delta_{p^\prime
[q}J_{p]q^\prime}\, \; , \qquad
\\\label{E8:SOQ} &&
[J_{qp}\, , \, Q_A]= {1\over 2}\sigma_{pq}{}_{AB} Q_B \; , \qquad \\
\label{QQ=SO} && [ Q_A\, , \, Q_B]= \sigma_{pq}{}_{AB} J_{pq} \; , \qquad
\end{eqnarray}
The Jacobi identities are satisfied due to the sigma-matrix identity
 $\sigma^{pq}{}_{(AB}\sigma^{pq}{}_{C)D}=0$.

Then, in the superparticle quantization above the linearized supergravity multiplet
appears in such a way that all the bosonic fields-- or, more precisely, their physical
components-- are collected in the Majorana--Weyl spinor of $SO(16)$. This makes
tempting to speculate on the relation of the bosonic field of D=11 supergravity with
the $SO(16)$ spinorial generator $Q^A$ and, further, with the  $E_8/SO(16)$ coset.
Furthermore, this suggests the speculation about possible $E_8$ symmetry of the
uncompactified $D=11$ dimensional supergravity (i.e. without dimensional reduction to
$d=3$).

Clearly, the linear approximation, which is seen from the superparticle quantization,
do not feel the difference between $E_8$ and its contraction given by extension of
$SO(16)$ by the mutually commuting spinorial generators (which includes $[Q_A \, , \,
Q_B ]=0$ instead of $[Q_A \, , \, Q_B ]=\sigma^{qp}_{AB}J_{qp}$ in (\ref{QQ=SO})). So,
to establish the hypothetic $E_8$ symmetry of the uncompactified $D=11$ supergravity,
one should define the $E_8$ transformations on eleven dimensional vielbein $e_m{}^a(x)$
and gauge field $A_{mnk}$ \footnote{The inclusion of fermions is a separate problem;
usually, when the $E_n$ symmetries of the compactified (to $d=11-n$) supergravity are
considered, the fermions are transformed as the field on nonlinear realization.} and to
show that (at least bosonic) supergravity equations are invariant under such
transformations. The experience of the description of the hidden $SO(16)$ symmetry
\cite{Nicolai87} suggests that this $E_8$ (if exists) might become manifest in a
formalism with broken Lorentz invariance. A new suggestion which brings our study is
that, a Lorentz symmetry breaking, which is appropriate to find the hidden $E_8$ (and
also $SO(16)$) symmetry might appear to be $SO(1,10)\mapsto SO(1,1) \otimes SO(9)$ (or
$SO(1,10)\mapsto [SO(1,1) \otimes SO(9)]\subset\!\!\!\!\!\!\times K_9$) rather than
$SO(1,10)\mapsto SO(1,2) \otimes SO(8)$ used in \cite{Nicolai87} to construct the
$SO(16)$ invariant formulation.

A check of whether the $D=11$ supergravity has indeed a hidden $E_8$ symmetry, even
without compactification, or the above described $SO(16)$ invariance of the linearized
supergravity and the coincidence of the number of physical polarizations of the bosonic
fields of the linearized supergravity multiplet with the dimension of the $E_8/SO(16)$
coset is purely occasional is an interesting subject for future study.

\bigskip

\section{Conclusions and outlook}\label{Concl}
\setcounter{equation}0

\subsection{Conclusions}

In this paper we have studied the BRST quantization of the M0-brane in the framework of
its spinor moving frame formulation \cite{BL98',BdAS2006} (see \cite{B90,IB+AN96} for
$D=4$ and $10$) where the action includes the spinorial Lorentz harmonics as
twistor--like auxiliary variables. Our main motivation was to search for the origin and
geometrical meaning  of the properties of the pure spinor approach to the quantum
superparticles and superstrings \cite{NB-pure}.

We have constructed here the Hamiltonian mechanics of the $D$=11 massless superparticle
in the spinor moving frame formulation separating covariantly the first and the second
class constraints (which has been possible due to the use of spinorial harmonics
\cite{BZ-str,BZ-strH}) and defining the Dirac brackets allowing to treat the second
class constraints as strong equalities.

We have shown that the set of the first class constraints of the M0--brane in the
spinor moving frame formulation can be separated into two groups. The first one
includes the $16$ fermionic generators of the $\kappa$--symmetry (which is irreducible
in the spinor moving frame formulation due to the presence of spinorial harmonics)  and
one bosonic generator of the $b$-symmetry. These generate the  $d=1, N=16$
supersymmetry gauge supergroup $\Sigma^{(1|16)}$. The remaining first class constraints
correspond to the generators of $H= [SO(1,1)\times SO(9)]\subset\!\!\!\!\!\! \times
K_9$ gauge symmetry. This eliminates the excess of variables in the harmonics used to
formulate the massless $D$=11 superparticle model making them the homogeneous
coordinates of $S^9$ which can be identified as D=11 celestial sphere. However, the
superalgebra of the Dirac brackets of the first class constraints is given by a {\it
`W--deformation'} of the one of the semidirect product $H\subset\!\!\!\!\!\! \times
\Sigma^{(1|16)}$, rather then by this semidirect product itself. This `W--deformation'
is produced by the appearance of the product of two $\kappa$--symmetry generators in
the Dirac brackets of two $K_9$ generators, so that $K_9$ is no longer  an abelian
subgroup and the Dirac brackets describes a generalized subalgebra of the enveloping
superalgebra rather than a Lie superalgebra.

The structure of the  complete BRST charge $\mathbb{Q}$ for all the first class
constraints of the M0--brane model is too complicated and its use is not practical.
This can be  seen already form the  BRST charge $\mathbb{Q}^\prime$ for the nonlinear
algebra of the $\kappa$-symmetry, $b$--symmetry and the deformed $K_9$ symmetry which
we have constructed in this paper (\ref{Q'=}). It already contains seven terms with up
to fourth power of the ghost fields. In the search for a counterpart of (or even an
alternative for) the Berkovits BRST charge we have accepted a further reduction of
$\mathbb{Q}^\prime$ down to the simple BRST charge $\mathbb{Q}^{susy}$ (\ref{Qbrst1})
associated to the $\kappa$- and $b$--symmetry gauge supergroup $\Sigma^{(1|16)}$.

We have shown that the non-trivial cohomologies of $\mathbb{Q}^{susy}$ can be described
by wavefunctions which have support on $\lambda_q^+\lambda_q^+=0$. This condition
requires the bosonic ghost $\lambda_q^+$, corresponding to the $\kappa$-symmetry, to be
zero. Since $\lambda_q^+$ defines essentially the BRST charge $\mathbb{Q}^{susy}$, this
makes a regularization necessary. Such a regularization is made by allowing the
$\kappa$-symmetry bosonic ghost to become complex, $\lambda_q^+\mapsto
\tilde{\lambda}_q^+\not= (\tilde{\lambda}_q^+)^*$, and by considering the non-Hermitian
BRST charge $\tilde{\mathbb{Q}}^{susy}$ resulting from it. The cohomology of the
original BRST charge $\mathbb{Q}^{susy}$ is then given by the cohomology of its
complexified and further reduced version  $\tilde{\mathbb{Q}}^{susy}$ (Eq.
(\ref{tQsusy})) at zero value of the bosonic ghost.

The need for a complex BRST charge at the regularization stage when computing the
non-trivial cohomology shows a reason for the intrinsic complexity of the Berkovits
pure spinor formalism for the superparticles and the superstring. This conclusion is
further supported by the observation that our $\tilde{\mathbb{Q}}^{susy}$ is
essentially a particular case of the Berkovits BRST charge for $D=11$ superparticle,
but with a composite pure spinor constructed from the $\kappa$--symmetry ghost and
Lorentz harmonics (Eq.(\ref{Lpure=lv}), see also below).

Computing the cohomology of the BRST charge $\mathbb{Q}^{susy}$ we have found that it
is nontrivial only in the sector with ghost number $-2$ (which corresponds to the ghost
number $g_0=0$ for the wavefunctions describing cohomologies of
$\tilde{\mathbb{Q}}^{susy}$) and are essentially  described by functions depending only
on the physical variables, which are inert under both the fermionic $\kappa$- and
bosonic $b$- gauge symmetries. The reason for such a simple structure is the existence
of a specific coordinate basis, the covariantized light-cone basis, the transition to
which results in the disappearance from the action of all the worldline fields that
transform nontrivially under the $\kappa$- and the $b$- gauge symmetries.

We have studied the covariant quantization of the physical degrees
of freedom in the covariant light--cone basis. This quantization,
quite close to the supertwistor one in \cite{BdAS2006}, shows the
hints of possible hidden symmetries of D=11 supergravity (or,
probably, of M-theory). These include the $SO(16)$ already mentioned
in \cite{BdAS2006} (and presumably related with the one of
\cite{Nicolai87}), but also some indication of possible $E_8$, which
brings us quite close to the $E_{10}$ and $E_{11}$ busyness of
\cite{Nicolai07} and \cite{WestE11}.

\subsection{Outlook 1: on BRST charge for superstring}

The main conclusion of our present study of the M0 case is that the twistor-like
Lorentz harmonic approach \cite{BZ-str,BdAS2006}, originated in \cite{Sok,NPS,Lharm},
is able to produce a simple and practical BRST charge. This suggests a similar
investigation of the $D=10$ Green--Schwarz superstring case. For instance, for the IIB
superstring the Berkovits BRST charge looks schematically like
\begin{eqnarray}\label{QIIB}
\mathbb{Q}^B_{IIB}=\int \Lambda^{\alpha 1}d_{\alpha} + \int \Lambda^{\alpha
2}d^2_{\alpha}\; , \qquad \Lambda^{\alpha 1}\sigma^a_{\alpha\beta}\Lambda^{\beta 1}= 0=
\Lambda^{\alpha 1}\sigma^a_{\alpha\beta}\Lambda^{\beta 1}
 \end{eqnarray}
with two complex pure spinors $\Lambda^{\alpha 1}$ and $\Lambda^{\alpha 2}$. By analogy
with our study of M0--brane (see (\ref{Lpure=lv})), one may expect that the BRST
quantization of the of the Green--Schwarz superstring  in its spinor moving frame
formulation \cite{BZ-str,BZ-strH} would result, after some reduction and on the way of
regularization of the `honest' ('true') hermitian BRST charge, in a complex charge of
the form (\ref{QIIB}), but with composite pure spinors
\begin{eqnarray}\label{pureSp12=}
    \widetilde{\Lambda}^{\alpha 1} = \tilde{\lambda}^+_p v^{-\alpha}_p\; , \qquad
 \widetilde{\Lambda}^{\alpha 2} = \tilde{\lambda}^-_p v^{+\alpha}_p\; ,
\qquad    \tilde{\lambda}^+_p\tilde{\lambda}^+_p=0=
\tilde{\lambda}^-_p\tilde{\lambda}^-_p\; . \qquad
\end{eqnarray}
Here, the $\tilde{\lambda}^{\pm}_p$ are two complex eight  component $SO(8)$ spinors
and the stringy harmonics $v^{\mp\alpha}_p$ are the homogeneous coordinates of the
non--compact $16$--dimensional coset
\begin{eqnarray}\label{harmV=IIB}
\{ V_{(\beta)}{}^{\alpha} \} = \{ ( v^{+\alpha}_p \; , \;
v^{-\alpha}_p )\}  = {Spin(1,9) \over SO(1,1)\otimes SO(8) } \; ,
\end{eqnarray}
characteristic for the spinor moving frame formulation of the (super)string
\cite{BZ-str,BZ-strH} and describing the spontaneous breaking of the spacetime Lorentz
symmetry by the string model.

It worth noticing  that, in contrast with the  M0--brane case, the $D=10$ solution
(\ref{pureSp12=}) of the pure spinor constraints in (\ref{QIIB}) {\it carries the same
number of degrees of freedom}, $44$($=2\times 8 + 2 \times 14$), that the pair of
Berkovits complex pure spinors $\Lambda^{\alpha 1}, \Lambda^{\alpha 2}$ ($22+22$).
Hence it provides {\it the general solution} of the $D=10$ pure spinor constraints in
terms of harmonics (\ref{harmV=IIB}) and two complex $SO(8)$ spinors of zero square so
that its substitution for the generic pure spinor of \cite{NB-pure} should not produce
any anomaly or other problem related to the counting of degrees of freedom.

\subsection{Outlook 2: $SO(16)$, $E_8$ and al that. }

Searching for the explanation of simple structure of the cohomologies of the M0--brane
BRST charge $\mathbb{Q}^{susy}$ we studied the M0--brane model in  different, the
so--called covariantized light cone basis \cite{GHT93}, the counterpart of which was
first considered in \cite{Sok}. The change of variables to this basis removes
automatically all the worldline fields which transformed nontrivially under the
$\kappa$--symmetry and $b$--symmetry. Such a phenomenon of automatical gauge fixing was
first  described in \cite{Sok}; one might observe it as well when passed to the pure
(super)twistor form of the action, as in  \cite{BdAS2006}.

Quantizing superparticle in this coordinate basis (as well as in the
supertwistor one \cite{BdAS2006}) one easily sees the $SO(16)$
symmetry of the model \footnote{In our spinor moving frame or
twistor--like Lorenz harmonics formulation
\cite{BL98',IB+AN96,BdAS2006,IB07} this symmetry can be seen also at
the classical level (see sec. 2.3); in the standard Brink--Schwarz
formulation it is hidden and appears after quantization in the
light--cone gauge.}. The reason is that, both in the covariantized
light cone basis and after fixing  the usual light--cone and the
(non--covariant) $\kappa$--symmetry gauge, the superparticle action
contains a set of   $16$ fermionic fields which, upon quantization,
become the $Cl^{16}$ Clifford algebra valued. The supergravity
multiplet appears in the superparticle spectrum when one choose the
wavefunction to be in ${\mathbf 256}$ Majorana spinor representation
of $Cl^{16}$. The bosonic and fermionic fields of the supermultiplet
appear as different ($\mathbf{128}$ and ${\mathbf{\overline{128}}}$)
Majorana Weyl parts of this Majorana spinor.

Furthermore, the observation of the well-known fact that $E_8$
exceptional group Lie algebra can be written in terms of the
generators of $SO(16)$ and $128$  bosonic generators carrying the
Majorana spinor ($\mathbf{128}$) representation of $SO(16)$ makes it
tempting to speculate on that the $E_8$ symmetry might be
characteristic of the $D=11$ supergravity itself rather than of its
reduction to $d=3$ only. In such a scenario the bosonic fields of
the D=11 supergravity multiplet appear to be associated to  the
generators of the $E_8/SO(16)$ coset. Notice that the assumption on
the Goldstone nature of graviton (physical degrees of freedom in our
case) is very much in spirit of the $E_{11}$ activity of
\cite{WestE11}, which develops in this respect the line of Borisov
and Ogievetsky \cite{BO74}. Also similarly to the case of $E_{10}$
and $E_{11}$ conjecture(s), the fermionic field (gravitino) appears
to be out of the consideration and have to be considered as a `field
of nonlinear realization' \cite{CCWZ}.

Surely, the superparticle quantization provides us only with the linearized fields
describing on-shell degrees of freedom. A check of whether the $D=11$ supergravity has
indeed a hidden $E_8$ symmetry, even without compactification, or the above described
$SO(16)$ invariance of the linearized supergravity and the coincidence of the number of
physical polarizations of the bosonic fields of the linearized supergravity multiplet
with the dimension of the $E_8/SO(16)$ coset is purely occasional, is an interesting
subject for future study.

Let us notice that $E_n/H_n$ cosets, which appeared as a manifold of scalar fields for
the $d=11-n$ compactifications of D=11 supergravity, were considered recently in
\cite{Hull07} in relation with the M--theoretic generalizations of the Hitchin's
generalized geometries \cite{Hitchin}. In particular, it was shown \cite{Hull07}, that
$E_7/SU(8)$ and $E_6/Sp(4)$ cosets can be described by $n$--dimensional components of
the bosonic fields of supergravity, $g_{ij}$, $A_3$ and $A_6$ (metric, three form gauge
field, and its 11--dimensional dual). The $E_n/H_n$ cosets with $n< 6$ can be described
by the $n$--dimensional components of $g_{ij}$ and $A_3$. The $128$--dimensional $n=8$
coset $E_8/SO(16)$ does not feet in this picture. Indeed, it is easy to see that the
number of the components of $8$--dimensional $g_{ij}$, $A_3$ and $A_6$, is
$36+56+28=120 < 128$\footnote{Of course, the simplest proposition to feet the coset
dimension ($128$) would be to add the eight--dimensional one form $A_1$, but this
field, in contrast with $g_{ij}$, $A_3$ and $A_6$,  does not have a straightforward
D=11 origin.} In the light of this the coincidence of the number of parameter of the
$E_8/SO(16)$ coset with the number of polarizations of the physical bosonic fields of
the supergravity multiplet, observed in sec. 5.2 and discussed above, looks even more
intriguing and worth  further thinking.

\bigskip

{\bf Acknowledgments. }

{The author thanks Jos\'e de Azc\'arraga, Paolo Pasti, Dmitri Sorokin, Mario Tonin for
useful discussions and Kelly Stelle for the conversation on $E_{10}$, $E_{11}$ and
$E_8$ issues. This work has been partially supported by research grants from the
Ministerio de Educaci\'on y Ciencia (FIS2005-02761) and EU FEDER funds, the Generalitat
Valenciana, the Ukrainian State Fund for Fundamental Research (N383), the INTAS
(2006-7928) and by the EU MRTN-CT-2004-005104 network {\it Constituents, Fundamental
Forces and Symmetries of the Universe} in which the author is associated to the
Valencia University. }

\bigskip

{\bf Notice added in proofs}. When the present work was finished, the author became
aware of the work \cite{Duff85} in which the possible hidden $E_8\times SO(16)$
symmetry of D=11 supergravity was conjectured for the first time.

\bigskip

\appendix{{\bf APPENDIX. Derivative of Harmonic variables and $SO(1,10)$ Cartan forms}}
\renewcommand{\theequation}{A.\arabic{equation}}
\renewcommand{\thesubsection}{A.\arabic{subsection}}
\setcounter{equation}{0}

\bigskip

 Vector ($u$) and spinor ($v$) Lorentz harmonics are elements of
$SO(1,10)$ and $Spin(1,10)$ matrices, $U\in SO(1,10)$, $V\in
Spin(1,10)$, Eqs. (\ref{harmUin}), (\ref{harmVin}). Their
interrelation is described by the constraints (\ref{harmVdef}) which
can be specified as
\begin{eqnarray}\label{vv=uG-all}
2v_\alpha{}^-_q v_\beta{}^-_q = \Gamma^{m}_{\alpha\beta} u_m^{--} \quad & (a) \; ,
\qquad
v^-_q\tilde{\Gamma}_mv^-_p = u_m^{--} \delta_{qp} \quad (d) & \; , \qquad \nonumber \\
2v_\alpha{}^+_q v_\beta{}^+_q = \Gamma^{m}_{\alpha\beta} u_m^{++} \quad & (b) \; ,
\qquad
v^+_q\tilde{\Gamma}_mv^+_p = u_m^{++} \delta_{qp} \quad (e) & \; , \qquad  \nonumber \\
2v_{(\alpha|}{}^+_q \gamma^i_{qp} v_{|\beta)}{}^+_q = \Gamma^{m}_{\alpha\beta}
u_m{}^{i} \quad & (c) \; , \qquad   v^-_q\tilde{\Gamma}_mv^+_p = u_m{}^{i }
\gamma^i_{qp} \quad (f) & \; , \qquad
\end{eqnarray}

The tangent space to the group can be associated to its Lie algebra. A basis of the
$55$ dimensional $so(1,10)$ algebra is provided by antisymmetric tensor generator
 $\mathbb{T}_{(a)(b)}=- \mathbb{T}_{(b)(a)}$, $(a)=(++,--,i)$. The
dual space is spanned by the $55$ left-invariant Cartan forms $\Omega^{(a)(b)}=-
\Omega^{(b)(a)}$ on the $SO(1,10)$ group manifold. This can be expressed in terms of
vector harmonics,
\begin{eqnarray}
\label{Omab-A} \Omega^{(a)(b)}:= U^{m(a)}dU_m^{(b)} = - \Omega^{(b)(a)} =
\left(\matrix{ 0 & - 4 \Omega^{(0)} & \Omega^{++j} \cr
  \,  4 \, \Omega^{(0)} &  0 & \Omega^{--j} \cr
 - \Omega^{++i} & - \; \Omega^{--i} & \Omega^{ij}\;
}\right) \quad \;  .
\end{eqnarray}
Eq. (\ref{Omab-A}) provides a vector representation for the $so(1,10)$ valued canonical
one-form $g^{-1}dg={1\over 2}\Omega^{(a)(b)}\mathbb{T}_{(a)(b)}$: with
$\mathbb{T}_{(a)(b)}{}_{(c)}{}^{(d)}= 2\eta_{(c)[(a)} \delta_{(b)]}{}^{(d)}$ and
 $g$  expressed in terms of the vector Lorentz harmonics (\ref{harmUin}), $g=U$, $g^{-1}dg=U^{-1}dU$.

The covariant harmonic derivatives $\mathbb{D}_{(a)(b)}$ provides a realization of the
generators $\mathbb{T}_{(a)(b)}$  in terms of the differential operators (vector
fields)  on the Lorentz group manifold. They can be obtained by decomposing the
exterior derivative $d$ in the basis of the Cartan forms (\ref{Omab-A}),
\begin{eqnarray}
\label{d=OmD} d:=  {1\over 2}  \Omega^{(a)(b)} \mathbb{D}_{(a)(b)}=: \Omega^{(0)}
\mathbb{D}^{(0)} + {1\over 2}\Omega^{++i}\mathbb{D}^{--i}+ {1\over
2}\Omega^{--i}\mathbb{D}^{++i} - {1\over 2}  \Omega^{ij} \mathbb{D}^{ij} \; .
\end{eqnarray}
For the light-like vector $u^{--}_m $, which is included in the action
(\ref{11DSSP(LH)}), Eqs. (\ref{d=OmD}) and (\ref{Omab-A}) imply
\begin{eqnarray}
\label{du--=}  & du^{--}_m = - 2u^{--}_m \Omega^{(0)} + u^{i}_m \Omega^{--i} \; .
\qquad
 \end{eqnarray}
 The presence
of $u^{i}_m$ in the {\it r.h.s.} of this equation shows the convenience of treating the
light-like vector $u^{--}_m$ in (\ref{11DSSP(LH)}) as an element of the moving frame:
its derivatives (or variations) are given in terms of variables already in the theory.

Notice that the Cartan forms $\Omega^{(0)}$ and $\Omega^{ij}$
transform as (composite) gauge fields under the local $SO(1,1)$ and
$SO(9)$ transformations, respectively. This allows to introduce the
$SO(1,1)\otimes SO(9)$ covariant differential $D$ and to write Eq.
(\ref{du--=}) and the expressions for the derivatives of other
vector harmonics in the form of
\begin{eqnarray}
\label{Du--}  &  Du^{--}_m :=du^{--}_m + 2u^{--}_m \Omega^{(0)} = u^{i}_m \Omega^{--i}
\; , \qquad \\ \label{Du++}  &  Du^{++}_m :=du^{++}_m - 2u^{++}_m \Omega^{(0)} =
u^{i}_m \Omega^{++i} \; , \qquad \\ \label{Dui}  & Du^{i}_m :=du^{i}_m + u^{j}_m
\Omega^{ji} = {1\over 2} u^{++}_m \Omega^{--i} + {1\over 2} u^{--}_m \Omega^{++i} \; ,
\qquad
\end{eqnarray}

When $g$ is realized in terms of the spinorial harmonics matrix $V$,
$g^{-1}dg=V^{-1}dV= {1\over 2} \Omega^{(a)(b)}\mathbb{T}_{(a)(b)}$, where now the
Lorentz algebra generators are in the spinorial representation, $\mathbb{T}_{(a)(b)}=
{1\over 2}\,\Gamma_{(a)(b)}\; \in \; spin(1,10)$ ,
\begin{eqnarray}
\label{VdV=UdUG=A} V^{-1}dV =   {1\over 4} \; \Omega^{(a)(b)}\; \Gamma_{(a)(b)} \quad
\in \; spin(1,10) \; , \qquad \Omega^{(a)(b)}:= U^{m(a)}dU_m^{(b)} \in \; so(1,10)
\quad  . \qquad
\end{eqnarray}
Eq. (\ref{VdV=UdUG}) can be equivalently written in the form of $dV =   {1\over 4} \;
\Omega^{(a)(b)}\; V\Gamma_{(a)(b)}$. This equation implies, in particular, the
following expression for the differential $dv_\alpha{}_q^{-}$ of the harmonics
$v_\alpha{}_q^{-}$:
\begin{eqnarray}
\label{dv-q=A} & dv_q^{-}=  - \Omega^{(0)} v_q^{-} - {1\over 4} \Omega^{ij}
v_p^{-}\gamma_{pq}^{ij} + {1\over 2} \Omega^{--i} \gamma_{qp}^{i}v_p^{+} \quad .
\end{eqnarray}
In terms of $SO(1,1)\otimes SO(9)$ covariant derivative this equation and its companion
read
\begin{eqnarray}
\label{Dv-q=} &  Dv_q^{-} := dv_q^{-} + \Omega^{(0)} v_q^{-} + {1\over 4} \Omega^{ij}
v_p^{-}\gamma_{pq}^{ij} = {1\over 2} \Omega^{--i} \gamma_{qp}^{i}v_p^{+} \; , \qquad
\\
 \label{Dv+q=} & Dv_q^{+} :=  dv_q^{+}  -
\Omega^{(0)} v_q^{+} + {1\over 4} \Omega^{ij} v_p^{+}\gamma_{pq}^{ij} =  {1\over 2}
\Omega^{++i} \gamma_{qp}^{i}v_p^{-} \; . \qquad
\end{eqnarray}
The covariant derivatives of the inverse harmonics, which in our $D=11$ case are
related to the original ones by
\begin{eqnarray} \label{V-1=CV-APP} D=11\; : \qquad v^{\pm \alpha}_q = \pm i
C^{\alpha\beta}v_{\beta q}^{\; \pm}
 \end{eqnarray}
have the related but not identical form
\begin{eqnarray}
\label{dv-1-q} &  Dv_q^{-\alpha} := dv_q^{-\alpha} + \Omega^{(0)} v_q^{-\alpha} +
{1\over 4} \Omega^{ij} v_p^{-\alpha}\gamma_{pq}^{ij} = - {1\over 2} \Omega^{--i}
v_p^{+\alpha} \gamma_{pq}^{i}\; , \qquad \\
\label{dv-1+q} &  Dv_q^{+\alpha} := dv_q^{+\alpha} - \Omega^{(0)} v_q^{+\alpha} +
{1\over 4} \Omega^{ij} v_p^{+\alpha}\gamma_{pq}^{ij} = - {1\over 2} \Omega^{++i}
v_p^{-\alpha} \gamma_{pq}^{i}\; . \qquad
\end{eqnarray}
The minus sign in the {\it r.h.s.} of (\ref{dv-1-q}) guaranties that, {\it e.g.}
$Dv_q^{-\alpha}\; v_{\alpha p}{}^+ = - v_q^{-\alpha} Dv_{\alpha p}{}^+$,
\begin{eqnarray}
\label{Dv-1v=}  v_q^{-\alpha} Dv_{\alpha p}{}^- = - Dv_q^{-\alpha}\;
v_{\alpha p}{}^- = {1\over 2} \Omega^{--i} \gamma_{pq}^{i}\; ,
\qquad   v_q^{+\alpha} Dv_{\alpha p}{}^+ = - Dv_q^{+\alpha}\;
v_{\alpha p}{}^+ = {1\over 2} \Omega^{++i} \gamma_{pq}^{i}\; .
\qquad
\end{eqnarray}
Actuallty, the above equations can also be written with noncovariant derivatives,
\begin{eqnarray}
\label{dv-1v=} dv_q^{-\alpha}\; v_{\alpha p}{}^- = - v_q^{-\alpha} dv_{\alpha p}{}^- =
- {1\over 2} \Omega^{--i} \gamma_{pq}^{i}\; , \qquad dv_q^{+\alpha}\; v_{\alpha p}{}^+
= - v_q^{+\alpha} dv_{\alpha p}{}^+ =- {1\over 2} \Omega^{++i} \gamma_{pq}^{i}\; .
\qquad
\end{eqnarray}
The fact that Cartan forms $\Omega^{(0)}$ and $\Omega^{ij}$ are used
as $SO(1,1)$ and $SO(9)$ connection used to define covariant
derivative (\ref{dv-1-q}), (\ref{dv-1+q}) can be expressed by
\begin{eqnarray}
\label{Dv+v-=0}  v_q^{-\alpha} Dv_{\alpha p}{}^+ = 0 \; , \qquad
 v_q^{+\alpha} Dv_{\alpha p}{}^- = 0\; . \qquad
\end{eqnarray}

 The
Cartan forms $\Omega^{++i}$ and $\Omega^{--i}$ are covariant with respect to
$SO(1,1)\otimes SO(9)$ transformations. They provide the vielbein for the coset
$SO(1,10)\over [SO(1,1)\otimes SO(9)]$ characteristic of the string rather than of a
particle model ({\it cf.} (\ref{v-inS11})), see Eq. (\ref{harmV=IIB}) and
\cite{BZ-str,BZ-strH} for the $D=10$ counterpart. Under the $K_9$ transformations
(\ref{K9-def}), which act on the vector harmonics by
\begin{eqnarray}
\label{K9:vU} K_9\; : \; &  \delta u_m^{--}=0 \; , \qquad  \delta u_m^{++}=
2k^{++i}u_m{}^{i} \; , \qquad \delta u_m{}^{i}= k^{++i}u_m^{--} \; ,  \qquad
\end{eqnarray}
the Cartan forms $\Omega^{++i}$ transform as a connection, $\delta \Omega^{++i}=
2Dk^{++i}:= 2(dk^{++i}+ k^{++j}\Omega^{ji}- 2k^{++i}\Omega^{(0)})$, while
$\Omega^{--i}$ is invariant. This indicates that the Cartan form $\Omega^{--i}$ provide
the vielbein for the coset $SO(1,10)/[(SO(1,1)\otimes SO(9))]\subset\!\!\!\!\!\! \times
K_9=S^9$.

\end{document}